\documentclass[12pt, twoside]{report}

\usepackage[utf8]{inputenc}
\usepackage{graphicx}
\usepackage{multirow}
\graphicspath{ {Images/} }
\usepackage{lipsum}
\usepackage[a4paper,width=150mm,top=25mm,bottom=25mm,bindingoffset=6mm]{geometry}
\usepackage{fancyhdr}
\usepackage{appendix}
\usepackage[hidelinks]{hyperref}

\usepackage{caption}
\usepackage{subcaption}
\usepackage{booktabs}   % for \toprule, \midrule, \bottomrule
\usepackage{amsmath}    % for math mode symbols like \pm
\usepackage{natbib}
\setcitestyle{authoryear} %Citation-related commands

\usepackage{setspace}
\usepackage{chngcntr}

\begin{document}

\begin{titlepage}
    \begin{center}
        %\vspace*{0.5cm}
        
        \Huge
        \textbf{Machine learning approaches to
    uncover the neural mechanisms of
    motivated behaviour: from ADHD
    to individual differences in effort
    and reward sensitivity}
        
        \vspace{0.5cm}
        
        \LARGE
        Thesis submitted in part fulfilment of the requirement for the award of Doctor of Philosophy
        
        \vspace{1.5cm}
        
        \Large
        \textbf{Nam Trinh}\\
        
        \vspace{0.25cm}
        
        Supervised by Prof. Tomas Ward and Dr. Gerard Derosiere\\
         %\vspace{0.25cm}
        \vfill

        \begin{figure}[htbp]
    \centering
    \includegraphics[width=0.25\textwidth]{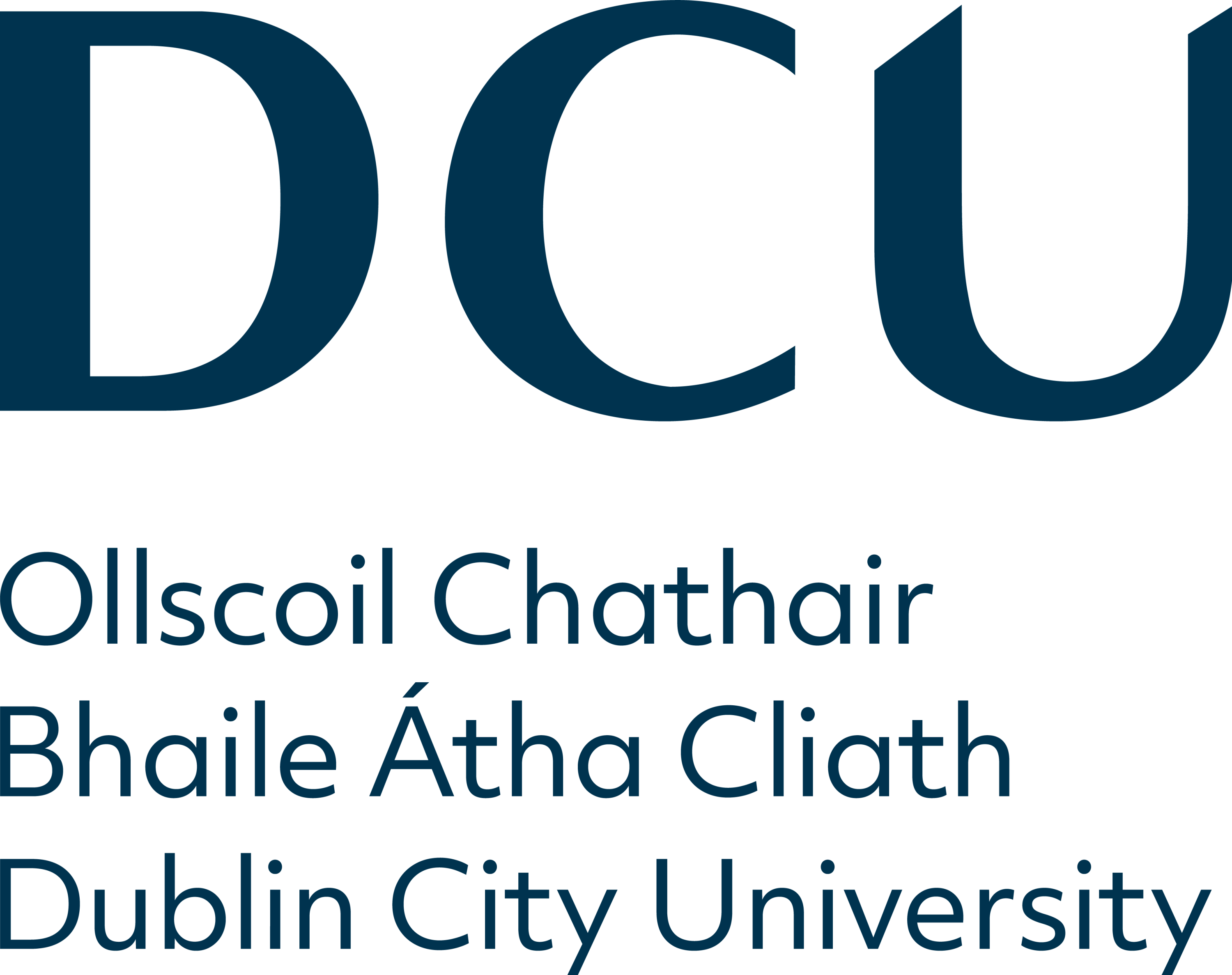}
    \hspace{2cm}
    \includegraphics[width=0.3\textwidth]{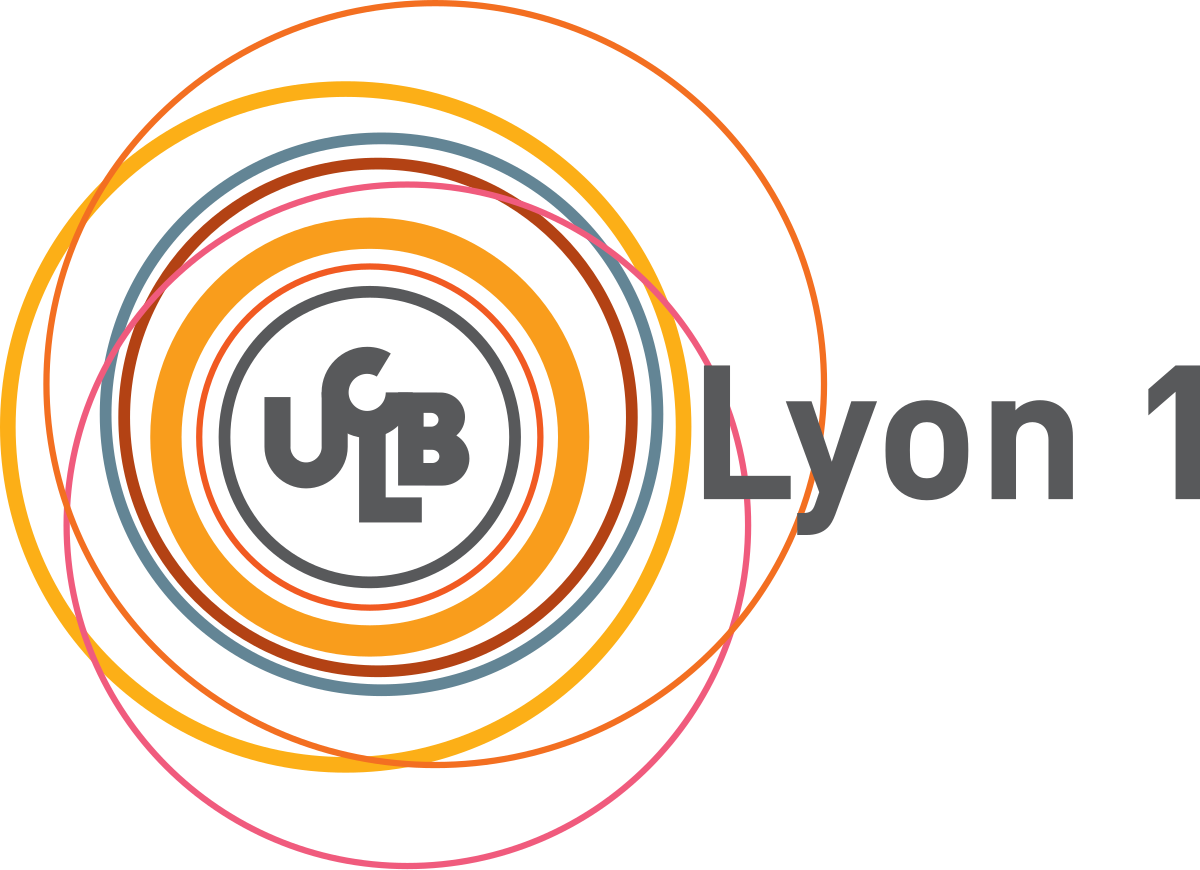}
    %\caption{Two images side by side}
    
\end{figure}

        \vspace{0.5cm}
        
        \LARGE
	    \textsc{School of Computing\\
	    Dublin City University}
	    
	    \begin{flushright}
	
	    \Large
	    December 2025 
	
	    \end{flushright}
        
    \end{center}
\end{titlepage}

\newpage

\chapter*{Declaration}
I hereby certify that this material, which I now submit for assessment on the programme of study leading to the award of Doctor of Philosophy is entirely my own work, and that I have exercised reasonable care to ensure that the work is original and have conformed to the regulations on the use and declaration of Generative AI, and does not to the best of my knowledge breach any law of copyright, and has not been taken from the work of others and to the extent that such work has been cited and acknowledged within the text of my work. 

\begin{center}
\begin{tabular}{ccc}
Sign:
\begin{minipage}{0.25\textwidth}
    \centering
    \includegraphics[width=1\linewidth]{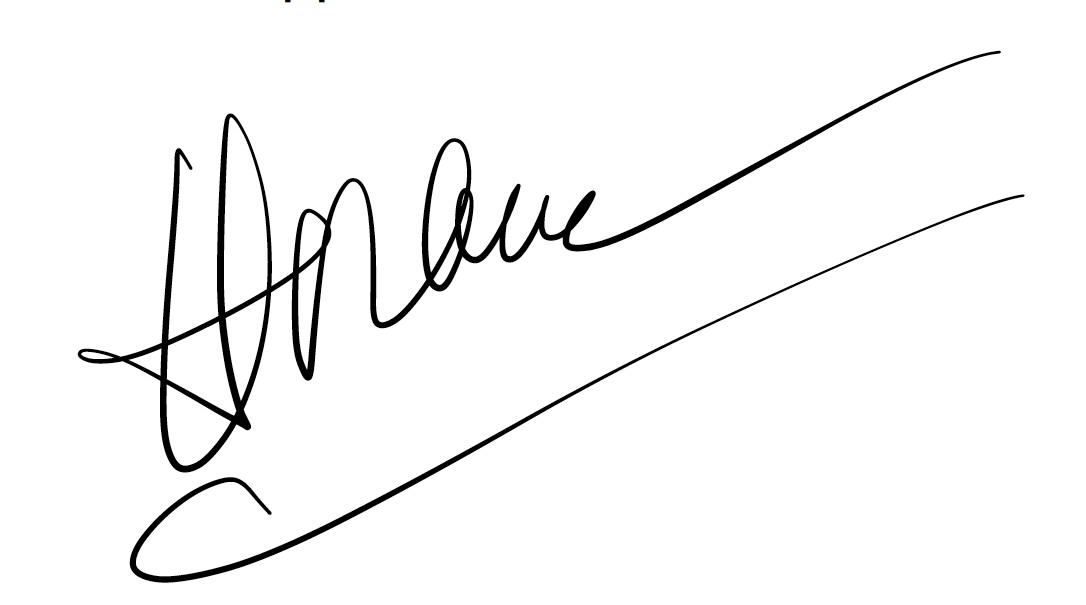}\\[2mm]
    Nam Trinh
\end{minipage}
&
\begin{minipage}{0.27\textwidth}
    \centering
    ID No.: 18215144
\end{minipage}
&
\begin{minipage}{0.3\textwidth}
    \centering
    Date: 02 December 2025
\end{minipage}
\end{tabular}
\end{center}

\chapter*{Generative AI usage declaration}

\noindent
This declaration outlines the use of generative AI tools during the preparation of my thesis, including the tools used, the outputs, example prompts, and a brief reflection on their role.

\section*{Tools Used}
\begin{itemize}
    \item \textbf{OpenAI ChatGPT}, GPT-5.1 model (used between December 2024 and December 2025).
    \item \textbf{Anthropic Claude}, Claude 3.5 Sonnet (used between December 2023 and December 2025).

\end{itemize}

\section*{Outputs}
\begin{itemize}
    \item ChatGPT was used to generate brief summaries of academic articles to assist in assessing their potential relevance to my study. I independently reviewed each article’s abstract, section headings, and figures before deciding whether to read the full text in detail. ChatGPT’s output helps me confirm again that my interpretation of the paper is correct.
    \item Claude was used to refactor Python code that I had written, improving readability and consistency. All experimental concepts, implementation decisions, and analytical work are entirely my own; no novel algorithms, designs, or interpretations were produced by Claude.
\end{itemize}

\section*{Prompts}
\begin{itemize}
    \item For ChatGPT: ``Please summarise this article for me and highlight the main takeaway messages.''
    \item For Claude: ``Please refactor this Python code to make it more readable. Do not change the logic.''

\end{itemize}

\section*{Critical reflection}

My use of GenAI tools was supportive rather than substantive. They did not generate original research ideas, experimental designs, results, or interpretations. Instead, they improve my  efficiency by summarising papers and making my code more readable.
\chapter*{Acknowledgements}
I would like to express my deepest gratitude to my two amazing supervisors, \textbf{Prof. Tomas Ward} at Dublin City University and \textbf{Dr. Gerard Derosiere} at the Lyon Neuroscience Research Centre. Despite the challenges I have encountered in my PhD research, both of you have shown great empathy and provided meaningful advice to help me overcome these hardships. Your guidance, encouragement, and mentorship have profoundly shaped both my research and personal growth. I have learned so much from each of you, not only about science, but also about perseverance, curiosity, and integrity. 

This work was made possible through the financial support of Science Foundation Ireland (SFI) via the Centre for Research Training in Machine Learning (ML-Labs) at Dublin City University under Grant No. 18/CRT/6183. I am especially thankful to the ML-Labs team for their support in facilitating research resources and travel arrangements throughout my PhD.

To my family: \textbf{Dad and Mom}, thank you for your unconditional love, and for instilling in me the values of learning and exploration. \textbf{Thao Tran} -- my lovely wife, thank you for being by my side throughout this journey — your encouragement and wisdom have meant the world to me and thank you for believing in me  when even I did not believe in myself. 

Finally, I wish to thank my friends and housemates for their emotional support and companionship, especially during the difficult months of the COVID-19 pandemic. Your presence made the challenges of this journey lighter and the milestones more joyful. I could not have done this without you.

\tableofcontents

\listoffigures

\listoftables

\chapter*{List of abbreviations}
\begin{tabular}{ll}
ADHD & Attention-deficit/hyperactivity disorder \\
EBDM & Effort-based decision making \\
SMA & Supplementary motor area \\
ACC & Anterior cingulate cortex \\
OFC & Orbitofrontal cortex \\
dlPFC & Dorsolateral prefrontal cortex \\
vmPFC & Ventromedial prefrontal cortex \\
vmPu & Ventromedial putamen \\
DMN & Default mode network \\
ECN & Executive control network \\
SN & Salience network \\
EEG & Electroencephalography \\
ERP & Event-related potential \\
TBR & Theta/beta ratio \\
MRI & Magnetic resonance imaging \\
fMRI & Functional magnetic resonance imaging \\
dMRI & Diffusion magnetic resonance imaging \\
DTI & Diffusion tensor imaging \\
DWI & Diffusion-weighted imaging \\
FA & Fractional anisotropy \\
MD & Mean diffusivity \\

\end{tabular}
\newpage
\begin{tabular}{ll}
PPC & Posterior parietal cortex \\
SPL & Superior parietal lobule \\
SEF & Supplementary eye field \\
GMV & Grey matter volume \\
MNI & Montreal neurological institute (coordinate space) \\
MVC & Maximal voluntary contraction \\
$\beta_\mathrm{Effort}$ & Effort sensitivity parameter \\
$\beta_\mathrm{Reward}$ & Reward sensitivity parameter \\
ML & Machine learning \\
LR & Logistic regression \\
SVM & Support vector machine \\
RF & Random forest \\
GB & Gradient boosting \\
ET & Extra trees classifier \\
AdaB & Adaboost \\
DT & Decision tree \\
MLP & Multi-layer perceptron \\
KNN & K-nearest neighbors \\
GNB & Gaussian naïve Bayes \\
LDA & Linear discriminant analysis \\
QDA & Quadratic discriminant analysis \\
AUC & Area under the receiver operating characteristic curve \\
TE & Echo time \\
TR & Repetition time \\
TI & Inversion time \\
FOV & Field of view \\
MATLAB & Matrix laboratory (software) \\
\end{tabular}
\newpage
\thispagestyle{plain}
\begin{singlespace}

\begin{center}
    \Large
    \textbf{Machine learning approaches to
uncover the neural mechanisms of
motivated behaviour: \\from ADHD
to individual differences \\in effort
and reward sensitivity}
    
    \vspace{0.4cm}
    \large

    \vspace{0.4cm}
    \textbf{Nam Trinh}
    
    \vspace{0.9cm}
    \textbf{Abstract}

\end{center}

Motivated behaviour relies on the brain’s capacity to evaluate effort and reward. Dysregulation within these processes contributes to a spectrum of conditions, from hyperactivity in attention-deficit/hyperactivity disorder (ADHD) to diminished goal-directed behaviour in apathy -- characterised by disrupted effort sensitivity and/or reward sensitivity. This thesis investigates the neural mechanisms underlying ADHD using electroencephalography (EEG) and examines individual differences in effort and reward sensitivity using neuroimaging, applying machine learning approaches through three main studies. 

In Study 1, task-based and resting-state EEG were employed with machine learning models to classify adult individuals with ADHD and healthy controls. Machine learning classifiers trained on task-based EEG during a stop signal task outperformed those trained on resting-state EEG data, with the strongest predictive features arising from gamma-band spectral power over fronto-central and parietal regions. These results demonstrate that ADHD-related neural alterations become most evident when inhibitory control is actively engaged. This also aligns with evidence of reduced gamma band activity in ADHD, potentially reflecting disrupted dopamine-mediated reward circuits.

In Study 2, to investigate the structural basis of motivated behaviour, diffusion MRI and whole-brain data-driven permutation-based analyses were used to identify the association between white matter integrity and two computationally modelled parameters reflecting effort and reward sensitivity. Effort sensitivity related to SMA-, dACC- and OFC-connected white matter clusters, and reward sensitivity additionally involved fronto-parietal and sensorimotor clusters, with tracts connected to the SMA emerging as a central hub for both of these parameters. Using machine learning approaches, individual differences in these sensitivity parameters were reliably predicted by microstructural integrity at identified clusters.

Finally in Study 3, grey matter morphometry, specifically with regional grey matter volumes extracted from structural T1-weighted MRI, was used to examine the grey matter correlates of individual differences in effort sensitivity, reward sensitivity, and subclinical apathy. Effort sensitivity was related to volumes in the dorsal insula, primary motor cortex, and superior parietal cortex, whereas reward sensitivity was associated with volumes in the ventromedial putamen and posterior parietal cortex. Reduced grey matter volume in the medial prefrontal cortex, supplementary motor areas, and ventral temporal regions was also associated with higher apathy scores.  Machine learning analyses confirmed that grey matter volumes allowed robust decoding of reward sensitivity and apathy levels.

Through these three studies, the thesis advances understanding of the neural mechanisms underlying motivated behaviour by identifying neural predictors of ADHD and individual differences in effort and reward sensitivity. Across three studies, common neural circuits emerged, most notably the fronto-parietal circuits, indicating its central role in effort valuation and reward processing to support motivated behaviour. The findings in this thesis may potentially serve as clinically neural biomarkers not only for improving diagnostic accuracy in conditions such as ADHD and motivational impairments, but also for guiding the development of personalised and targeted neurotechnological interventions for individuals with neuropsychiatric disorders.
\end{singlespace}

\chapter{Introduction}
% Introduction
Why do some people struggle to focus, while others find it hard to care? In classrooms, workplaces, and homes, we encounter individuals whose challenges seem to lie not in ability but in will: one individual exhibits restlessness and impulsive responding; another lacks the drive to initiate even rewarding tasks. Though these difficulties appear distinct with one rooted in impulsivity and the other in apathy, they may share partially overlapping neural mechanisms. In this thesis, I investigate how neural mechanisms relate to ADHD and individual differences in effort sensitivity and reward sensitivity.

%Research gap
\section{Introduction}
Human behaviour is shaped by complex decisions that require evaluating whether the potential benefits of an action outweigh its associated costs (\citealp{croxson2009effort}). From everyday situations, like whether to cook dinner or order takeout, to important life choices such as career changes, our brains continuously assess the values of effort and reward (\citealp{hogan2020neural}). These computations form the basis of what is broadly referred to as \textit{motivated behaviour}, the capacity to pursue goals based on internal valuation mechanisms. Disruptions in this process contribute to a range of psychiatric and neurological disorders, including apathy, depression, and ADHD, in which individuals exhibit altered sensitivity to effort or reward (\citealp{husain2018neuroscience, hogan2020neural}).

Theoretical frameworks propose that motivated behaviour depends on the appropriate activation of neural circuits that regulate the energisation of behaviour (\citealp{turrigiano2008self}). In this view, the brain must generate sufficient activation to initiate and pursue goal-directed actions (\citealp{husain2018neuroscience}), while also modulating this activation to prevent impulsive or poorly controlled behaviours (\citealp{botvinick2001conflict}). Disruptions to this process can lead to hyperactivity or hypoactivity in motivational circuits:
\begin{itemize}
    \item Hyperactivity, involving excessive or poorly regulated activation, has been linked to impulsivity, difficulty in inhibiting responses, and challenges in sustaining controlled behaviour along with emerging evidence of altered reward processing and motivational valuation (\citealp{volkow2011motivation, koirala2024neurobiology}).
    \item Hypoactivity, reflected in disrupted neural activation to drive behaviour, is commonly associated with apathy, characterised by increased sensitivity to effort and/or decreased sensitivity to reward (\citealp{spalletta2013brain}).

\end{itemize}

From this perspective, ADHD and apathy, while often treated as distinct or even opposite conditions, may reflect different forms of motivational dysregulation arising from disrupted motivation-related circuits. ADHD is widely conceptualised as a disorder of attention and inhibitory control, supported by robust evidence of impaired response inhibition (\citealp{morsink2022studying}). Beyond attentional and inhibitory deficits, emerging evidence also highlights that ADHD involves motivational deficits and reward processing (\citealp{grimm2021effects, tegelbeckers2018orbitofrontal, plichta2014ventral, cubillo2012review}). However, the neural circuits of motivational deficits in ADHD remain poorly understood compared to those associated with attentional and inhibitory deficits.

In parallel, apathy, typically characterised by reduced motivation and goal-directed behaviour, has been linked to structural differences in frontal, temporal, and striatal regions, even in subclinical populations (\citealp{spalletta2013brain}). In healthy individuals, several studies show that people differ substantially in how sensitive they are to effort and reward (\citealp{treadway2012dopaminergic, chong2017neurocomputational}). Some individuals are highly reluctant to exert effort and reduce the perceived value of rewards when effort is required, while others are far more willing to invest effort even for relatively modest rewards (\citealp{westbrook2013subjective, apps2015role, jurgelis2021heightened}). These differences are not merely behavioural variations; they also reflect underlying variability in the brain motivation-related circuits that evaluate costs and benefits, particularly circuits involving the frontal cortex, striatum, and cingulate regions (\citealp{bartra2013valuation, salamone2016activational}). Recent work also highlights the role of mesocorticolimbic dopamine systems, along with fronto-striatal and cingulate networks, in shaping effort-based decisions (\citealp{salamone2024neurobiology, wang2024neural, suzuki2021distinct}). Yet, the neural correlates of individual differences in effort and reward sensitivity among healthy individuals remain incompletely characterised. 

%Understanding these individual differences is crucial for building a more complete picture of motivated behaviour and for distinguishing typical variability from dysregulation.

Machine learning provides a powerful framework for neuroscience research particularly when analysing high-dimensional physiological and neuroimaging data. Machine learning methods are well suited for detecting subtle, multivariate patterns that may differentiate clinical populations from neurotypical controls (\citealp{bishop2006pattern, murphy2012machine}). Within neuroscience, predictive modelling has become increasingly important for identifying biomarkers of psychiatric and neurological conditions (\citealp{badrulhisham2024machine, woo2017building}). In EEG research, machine learning has been widely used for classifying cognitive states and clinical conditions by leveraging spectral features, network dynamics, and event-related activity (\citealp{derosiere2014towards, lotte2018review, craik2019deep}). Machine learning approaches have also been applied to neuroimaging data, enabling the prediction of behavioural measures and diagnostic status from grey matter morphology and white matter connectivity (\citealp{arbabshirani2017single, vieira2017using}). Together, these studies support the use of machine learning to examine the structural correlates of behavioural measures in terms of prediction capability, thereby motivating the use of machine learning analyses in this study.

To advance understanding of motivated behaviour, this thesis uses different types of data modalities and multiple machine learning approaches to investigate how ADHD and individual differences in effort and reward sensitivity map onto the neural systems that support motivation. First, I investigate how spectral power bands during the stop signal task and resting-state predict ADHD conditions. Second, I explore how individual differences in effort and reward sensitivity are predicted by white matter integrity in healthy individuals. Third, I examine how variations in effort and reward sensitivity are associated with regional grey matter volume measures.

\section{Research questions}
This thesis investigates the neural and physiological correlates of motivated behaviour by analysing EEG data in individuals with ADHD and neuroimaging data in healthy controls. Specifically, I will analyse electroencephalography (EEG) recordings from a cohort of adult individuals with ADHD and neurotypical controls, alongside with diffusion and T1-weighted magnetic resonance imaging (MRI) data collected from a cohort of healthy control participants. The present thesis addresses three Research Questions as follows:

\begin{enumerate}
    \item \textbf{EEG-based adult ADHD classification:} How can EEG-derived spectral band powers from resting-state and task-related activity differentiate individuals with ADHD from neurotypical controls, and what do these features reveal about neural dynamics in motivation-related circuits?
    
    \item \textbf{White matter integrity correlates of effort and reward sensitivity:}  How does white matter integrity predict individual differences in effort and reward sensitivity? 
    
    \item \textbf{Grey matter correlates of effort sensitivity, reward sensitivity and subclinical apathy:} How does variability in cortical and subcortical grey matter volumes predict individual differences in effort and reward sensitivity?
\end{enumerate}

\section{Thesis structure}

The structure of this thesis is organised as follows:

\begin{itemize}
    \item \textbf{Chapter 2} presents the conceptual and theoretical background of ADHD, including its neural mechanisms and the need for reliable biomarkers. The chapter also reviews how individual differences in effort and reward sensitivity relate to the neural circuits involved in effort-based decision-making. Finally, it introduces work using machine learning in neuroscience, outlining how these approaches have been applied to EEG and neuroimaging data.

    \item \textbf{Chapter 3} investigates the classification of ADHD using EEG data collected during both resting-state and task-related (stop-signal) paradigms. Machine learning approaches are applied to classify individuals with ADHD from healthy controls with spectral powers across different frequency bands as input features.

    \item \textbf{Chapter 4} examines the relationship between white matter integrity measures and individual variability in effort and reward sensitivity in healthy controls. Using diffusion MRI and effort and reward sensitivity parameters derived from the computational modelling of an effort-based decision-making task, this chapter characterises the white matter integrity correlates of motivational regulation parameters.

    \item \textbf{Chapter 5} extends the investigation of structural correlates in Chapter 4 to grey matter morphology, studying associations between regional grey matter volumes and motivated behaviour parameters including effort sensitivity, reward sensitivity and apathy scores.

    \item \textbf{Chapter 6} summarises the findings in preceding main chapters (3, 4 and 5) and concludes the thesis. 
\end{itemize}

Supplementary materials accompanying the main analyses are presented in the \textit{Appendices}, providing additional methodological details, supporting results, and extended discussion of key findings.

\section{Related publications}
The work presented in this thesis has led to several related publications and conference presentations, as outlined below.
\begin{itemize}
    \item Chapter 3 was based on our previous publication: ``Trinh et al. Task-related and resting-state EEG classification of adult patients with ADHD using machine learning. 2023 IEEE 19th International Conference on Body Sensor Networks (BSN). IEEE, 2023." and was significantly extended based on the comments and feedback from reviewers of the conferences and DCU's annual research progress report (PGR2) examiners. 
    \item Chapter 4 was previously published as a preprint (doi: \href{https://doi.org/10.1101/2025.08.19.671080}{10.1101/2025.08.19.671080}). A condensed version of this study was also presented as an oral presentation at the International Conference on Cognitive Neuroscience (ICON 2025) in Porto, Portugal. At the time of writing this thesis, this preprint has passed the first round of journal peer reviews and is in major revision with NeuroImage.
    \item Chapter 5 was accepted for publication as a poster in the Society for Neuroscience (SfN) Annual meeting 2024, held in Chicago, Illinois, USA, in September 2024.
    \item The implementation of the machine learning analyses in this thesis was inspired by the framework presented in this journal paper by \citealp{zhang2022comparison}, titled ``A comparison of distributed machine learning methods for the support of ``many labs" collaborations in computational modeling of decision making", Frontiers in Psychology, in which as a co-author, I contributed to the implementation of a recurrent neural network in predicting behavioural parameters in a gambling task.
\end{itemize}

\chapter{Related work and background}
% Use this paper to write more Associations between ADHD symptoms, executive function and frontal EEG in college students
Human behaviour arises from the brain’s ability to dynamically modulate neural circuits involved in motivation. When such regulation is perturbed, behaviour can turn toward hyperactivity or hypoactivity. In this chapter, I would like to provide a literature review on ADHD and the underlying neural mechanisms of this condition. I then provide an overview of the neural circuits underlying effort-based decision making in motivated behaviour. Finally, I summarise the use of machine learning in neuroscience research with input data from EEG and neuroimaging. 

\section{Attention-deficit hyperactivity disorder}
ADHD is a prevalent psychological disorder characterized by attention deficits and high impulsivity, impacting about 5\% of children globally (\citealp{polanczyk2007worldwide, wolraich2019clinical}), with a possibility of continuing into adulthood observed in 40\% - 67\% of cases (\citealp{kiiski2020eeg}, \citealp{kessler2010structure}). Current diagnostic approaches predominantly rely on subjective behavioural assessments, conducted through structured clinical interviews and standardised questionnaires, such as the Diagnostic and Statistical Manual of Mental Disorders (DSM-5) criteria (\citealp{diagnostic2013statistical}) and Conner's ADHD scales (\citealp{conners1998rating}). In parallel, there has been a considerable effort to develop objective neuro-physiological markers through physiological measurements such as EEG and fMRI (\citealp{alim2023automatic}, \citealp{maniruzzaman2023optimal}, \citealp{anchana2023comparative}, \citealp{zhang2023machine}, \citealp{parlatini2023white}).
\subsection{Neural mechanism underlying ADHD}
The neurobiological underpinnings of ADHD have been extensively investigated over the past several decades, revealing a complex interplay of structural, functional, and neurochemical abnormalities that contribute to the disorder's characteristics. Understanding these neural mechanisms is crucial for developing more precise diagnostic tools and targeted interventions.
\subsubsection{Structural correlates of ADHD}
Neuroimaging studies have consistently demonstrated structural brain differences in individuals with ADHD and healthy controls. Meta-analyses of MRI studies in ADHD have reported reduced total brain volume in children with ADHD, with particular reductions in the prefrontal cortex (PFC), basal ganglia, corpus callosum and cerebellum (\citealp{valera2007meta, nakao2011gray}). 

Specifically, the PFC, which plays an essential role in attention, executive control and working memory, has been reported with delayed cortical maturation in children with ADHD, with peak cortical thickness occurring approximately three years later than in typically developing children (\citealp{shaw2007attention}). Volumetric reductions have been also observed to be significantly associated with ADHD across multiple PFC subregions, including the dorsolateral prefrontal cortex (dlPFC) (\citealp{cubillo2012review}), and the ventrolateral prefrontal cortex (vlPFC), which contributed to response inhibition and attentional set-shifting (\citealp{rubia2014effects}). \citealp{makris2007cortical} reported a cortical thinning in the PFC in the right hemisphere, particularly in regions associated with attention networks including the inferior parietal lobule, the anterior cingulate cortex and the dlPFC. \citealp{hai2022right} reported a significantly thinner right superior frontal gyrus (SFG) in children with ADHD compared with typically developing controls. Additionally, reduced gray matter volume in the lateral prefrontal gyrus has been associated with greater symptom severity and impaired inhibitory control in ADHD populations (\citealp{depue2010behavioral}). Longitudinal studies have further demonstrated that while some PFC structural associations with ADHD may normalise with age, persistent deficits in frontal-striatal connectivity remain into adulthood, suggesting that developmental trajectories in ADHD differ fundamentally from typical neurodevelopment (\citealp{proal2011brain, hoogman2017subcortical}).

The basal ganglia, particularly the caudate nucleus and putamen, also showed 
reductions in gray matter volumes among ADHD populations (\citealp{ellison2008structural}). Another work by \citealp{qiu2009basal} reported that boys with ADHD showed significantly reduced basal ganglia volumes and localized shape compressions in the caudate, putamen, and globus pallidus, indicating atypical development of frontal–subcortical circuits. These subcortical structures are integral components of fronto-striatal circuits that mediate reward processing, motor control, and executive functions (\citealp{haruno2006different, grahn2008cognitive}). The observed structural abnormalities in these regions align with the dopaminergic dysfunction hypothesis of ADHD, as the basal ganglia are rich in dopamine receptors and serve as key nodes in dopaminergic pathways (\citealp{volkow2009evaluating}). 

Recent advancements in diffusion MRI have further contributed to the exploration of white matter abnormalities in ADHD, revealing that fractional anisotropy reductions in major white matter tracts, including the corpus callosum, corona radiata, cerebellum and bilateral internal capsule (\citealp{van2012diffusion}). 
\subsubsection{Functional brain networks and connectivity}
Beyond structural differences, functional neuroimaging studies have revealed alternation patterns of brain activation and connectivity in individuals with ADHD. Resting-state functional MRI (rs-fMRI) studies have consistently demonstrated alterations in large-scale brain networks, particularly the default mode network, the executive control network, and the salience network (\citealp{castellanos2012large, cortese2012toward}).

The default mode network, which typically showed increased activity during rest and decreased activity during task engagement, exhibits abnormal patterns in ADHD (\citealp{norman2023evidence}). Specifically, individuals with ADHD often failed to adequately suppress the default mode network's activity during attention-demanding tasks, leading to interference with task-relevant processing (\citealp{sonuga2007spontaneous}).

\subsection{Diagnostic Challenges in ADHD}
Despite decades of research and clinical experiments, the diagnosis of ADHD remains challenging due to the disorder's heterogeneous presentation, developmental trajectory, and reliance on subjective assessment methods.

\subsubsection{Subjectivity and variability in behavioural assessment}
The current gold standard for ADHD diagnosis relies heavily on behavioural observations and subjective reports from multiple sources, including parents, teachers, and the individuals themselves. The DSM-5 criteria require the presence of at least six symptoms of inattention and/or hyperactivity-impulsivity (or five for individuals aged 17 years and older) that are present in multiple settings, persist for at least six months, and cause significant functional impairment (American Psychiatric Association, 2013). However, the interpretation of these criteria is inherently subjective and can be influenced by numerous factors, including informant biases, cultural expectations, environmental context, and the observer's familiarity with typical developmental behaviour (\citealp{eng2024evidence, tam2025individual}). 

Rating scales such as the Conners' Rating Scales, the ADHD Rating Scale-IV, and the Vanderbilt ADHD Diagnostic Rating Scale are widely used to quantify symptom severity and aid in diagnosis (\citealp{conners1998rating, dupaul1998parent}). While these instruments have demonstrated reliability and validity, they remain susceptible to rater bias and may not capture the full complexity of ADHD symptomatology. Discrepancies between parent and teacher ratings are common, with correlations typically ranging from 0.3 to 0.5, reflecting genuine situational variability in symptom expression as well as differences in rater perspectives and expectations (\citealp{achenbach1987child}).

\subsubsection{Heterogeneity and comorbidity}
ADHD is a highly heterogeneous disorder, with substantial variability in symptom profiles, severity, and associated impairments across individuals (\citealp{ging2025symptom}). The DSM-5 recognizes three presentations: predominantly inattentive, predominantly hyperactive-impulsive, and combined presentation. However, even within these subtypes, there is considerable heterogeneity in the specific symptoms present, their severity, and their functional impact (\citealp{epstein2013changes}).

This heterogeneity is further complicated by high rates of psychiatric comorbidity. Approximately 60-80\% of individuals with ADHD have at least one comorbid psychiatric disorder (\citealp{gnanavel2019attention}). Common comorbid conditions include oppositional defiant disorder (ODD), conduct disorder (CD), anxiety disorders, mood disorders, learning disabilities, and autism spectrum disorder (ASD) (\citealp{gnanavel2019attention}). The presence of comorbid conditions can complicate diagnosis, as symptom overlap between disorders can make it difficult to determine which symptoms are attributable to ADHD versus other conditions. For example, inattention can be a feature of anxiety disorders, depression, and learning disabilities, while impulsivity may be present in bipolar disorder and disruptive behaviour disorders (\citealp{friesen2021diagnosis}).

The high comorbidity rates raise important questions about the validity of ADHD as a distinct diagnostic entity versus a dimensional trait that co-occurs with other forms of psychopathology. Some researchers have proposed that ADHD may be better conceptualized as a spectrum of executive function deficits that cut across traditional diagnostic boundaries (\citealp{sonuga2003dual}). This perspective is supported by evidence of shared genetic risk factors and overlapping neural substrates across ADHD and other neurodevelopmental and psychiatric disorders (\citealp{cross2013identification}).
\subsubsection{Developmental considerations}
ADHD symptoms and their manifestations change across development, presenting additional diagnostic challenges. Hyperactivity and impulsivity tend to be most prominent in early childhood, while attention deficits may become more apparent as academic and organizational demands increase in later childhood and adolescence (\citealp{cherkasova2013developmental}). In adulthood, overt hyperactivity often diminishes, but internal restlessness, difficulty with sustained attention, and executive function deficits typically persist (\citealp{faraone2006age}).

The developmental trajectory of ADHD symptoms indicates that diagnostic criteria appropriate for children may not adequately capture the disorder's presentation in adolescents and adults. The DSM-5 made some accommodations for this by reducing the symptom threshold for adults and providing examples of how symptoms might manifest in older individuals. However, concerns remain about whether current criteria are sufficiently sensitive to adult ADHD, particularly in individuals who have developed compensatory strategies or whose symptoms are primarily internal rather than behavioural.

Furthermore, the requirement that symptoms be present before age 12 (as specified in DSM-5) can be problematic for adults seeking diagnosis, as retrospective recall of childhood symptoms is often unreliable and corroborating informants may not be available (\citealp{sibley2012diagnosing}). This age-of-onset criterion has been debated, with some researchers arguing that late-onset ADHD (with symptom onset after age 12) may represent a valid variant of the disorder (\citealp{moffitt2015adult}).
\subsubsection{The need for objective biomarkers}
The challenges inherent in current diagnostic approaches have motivated extensive research into objective biomarkers that could complement behavioural assessments and improve diagnostic accuracy (\citealp{hurjui2025biomarkers}). Ideal biomarkers would be reliable, valid, sensitive, specific, and feasible to implement in clinical settings (\citealp{thome2012biomarkers}). While no single biomarker has yet achieved sufficient diagnostic utility to replace clinical assessment, several promising candidates have emerged from neuroimaging, neurophysiological, and genetic research (\citealp{michelini2022treatment}).

Neuroimaging biomarkers, including structural MRI measures of brain volume and cortical thickness, functional MRI measures of brain activation and connectivity, and DTI measures of white matter integrity, have shown group-level differences between individuals with ADHD and controls (\citealp{narr2009widespread, hart2013meta, parlatini2023white}). However, the substantial overlap in distributions between groups, along with the high cost and limited accessibility of neuroimaging, has precluded their use as a standalone diagnostic tool (\citealp{hoogman2019brain}).

Electroencephalography (EEG) has emerged as a particularly promising modality for biomarker development due to its non-invasive nature, relatively low cost, high temporal resolution, and direct measurement of neural activity (\citealp{arns2013decade}). The following sections will explore the extensive literature on EEG markers of ADHD, with particular focus on spectral power abnormalities that form the basis for the predictive modeling approach employed in this thesis.
\subsection{EEG markers of ADHD}
EEG has been used to study ADHD for over half a century, with early studies in the 1970s documenting differences in brain electrical activity between children with ADHD and typically developing controls (\citealp{satterfield1973response}). Since then, a substantial body of research has accumulated, revealing consistent patterns of EEG abnormalities in ADHD that reflect the underlying neural dysfunction associated with the disorder (\citealp{loo2012clinical, arns2013decade, lenartowicz2014electroencephalography, arns2020neurofeedback}).

EEG measures the electrical activity of the brain by recording voltage fluctuations at the scalp surface, which reflect the summated postsynaptic potentials of large populations of cortical pyramidal neurons firing in synchrony. The continuous EEG signal can be decomposed into frequency components through spectral analysis, typically using Fourier transform methods. The resulting power spectrum reveals the distribution of signal power across different frequency bands, which are conventionally divided into delta (0.5-4 Hz), theta (4-8 Hz), alpha (8-13 Hz), beta (13-30 Hz), and gamma ($>$30 Hz) ranges.

Each frequency band is associated with distinct functional states and neural processes. Delta activity is prominent during deep sleep and may reflect cortical-subcortical interactions (\citealp{steriade1993thalamocortical}). Theta activity is associated with drowsiness, memory processes, and cognitive control, with frontal midline theta particularly implicated in executive functions (\citealp{cavanagh2014frontal}). Alpha activity is thought to reflect cortical idling or active inhibition of task-irrelevant regions (\citealp{klimesch2012alpha}). Beta activity is associated with active cognitive processing, motor activity, and arousal (\citealp{engel2010beta}). Gamma activity is linked to sensory processing, attention, and consciousness, though it is more difficult to reliably measure from scalp EEG due to its lower amplitude and susceptibility to muscle artifact (\citealp{fries2009neuronal}).

In ADHD research, quantitative EEG analysis has been extensively employed to characterize spectral power differences between individuals with ADHD and controls (\citealp{arns2013decade}). These analyses typically involve recording EEG during resting states (eyes open and/or eyes closed) or during cognitive tasks, followed by artifact removal, spectral decomposition, and statistical comparison of power in different frequency bands across groups (\citealp{arns2013decade}). In this study we focus on spectral powers extracted from EEG data for diagnosis of ADHD using machine learning techniques.

\subsubsection{The theta/beta ratio}
The most widely studied EEG marker of ADHD is the theta/beta ratio (TBR), which represents the ratio of power in the theta band (typically 4-8 Hz) to power in the beta band (typically 13-30 Hz), usually measured at frontal or central electrode sites (\citealp{arns2013decade}). The TBR gained prominence following studies in the 1990s and early 2000s that reported elevated TBR in children with ADHD compared to controls, with effect sizes often in the medium to large range (\citealp{monastra1999asessing, clarke2001age}).

The theoretical rationale for the TBR as an ADHD marker comes from the interpretation of elevated theta as reflecting cortical hypoarousal or maturational lag, while reduced beta reflects decreased cortical activation or cognitive processing efficiency. The combination of these two abnormalities, captured by the TBR, was proposed to reflect the fundamental neural dysfunction in ADHD (\citealp{barry2003review}).

The TBR gained considerable attention as a potential diagnostic biomarker, with some researchers and clinicians advocating for its use in clinical practice. In 2013, the U.S. Food and Drug Administration (FDA) approved the Neuropsychiatric EEG-Based Assessment Aid (NEBA) system, which uses the TBR to aid in ADHD diagnosis in children and adolescents aged 6-17 years. However, this approval was controversial and sparked considerable debate among the scientific community (\citealp{gloss2016practice, kiiski2020eeg}). 

Subsequent research has revealed significant limitations and inconsistencies regarding the TBR as an ADHD biomarker. Meta-analyses have shown that while the TBR is elevated in ADHD on average, the effect size has decreased over time, with more recent studies reporting smaller differences than earlier studies (\citealp{boxum2024challenging, arns2013decade}). Furthermore, the TBR shows substantial developmental changes, decreasing with age throughout childhood and adolescence, which complicates its interpretation and necessitates age-appropriate normative data (\citealp{finley2022periodic, lenartowicz2014electroencephalography, monastra1999asessing}). The TBR also exhibits considerable variability across individuals with ADHD, with many individuals falling within the normal range, limiting its sensitivity and specificity as a diagnostic marker (\citealp{loo2013characterization}).

The controversy surrounding the TBR highlights the challenges in translating group-level EEG findings into clinically useful biomarkers. While the TBR may capture some aspects of neural dysfunction in subgroups of individuals with ADHD, it lacks the sensitivity and specificity required for diagnostic application at the individual level (\citealp{saad2018theta}). Nevertheless, the extensive research on the TBR has advanced our understanding of spectral power abnormalities in ADHD and has motivated more sophisticated approaches to EEG-based classification.

\subsubsection{Spectral band abnormalities across frequency bands}
Beyond the TBR, studies have revealed abnormalities in absolute and relative power across multiple frequency bands in individuals with ADHD. These findings provide a more detailed literature of the spectral characteristics of ADHD and suggest that multiple frequency bands may contribute to the neural signature of the disorder.

\textbf{Delta band abnormalities. }Delta activity, which is prominent during deep sleep but typically minimal during wakefulness (\citealp{long2021sleep}), has received less attention in ADHD research. Some studies have reported elevated absolute delta power during waking states in individuals with ADHD (\citealp{kiiski2020eeg}), which may reflect excessive drowsiness, cortical hypoarousal, or intrusion of sleep-like states during wakefulness (\citealp{markovska2017quantitative, rommel2017altered, bresnahan2002specificity}). For example, \citealp{yoon2023differences} found significantly higher delta power in adults with ADHD during eyes-open resting state. Similarly, \citealp{rommel2017altered} reported greater delta power in adolescents with ADHD during both resting and task conditions (Continuous Performance Test–OX), and \citealp{kamida2016eeg} observed increased frontal and parietal delta activity in children with ADHD under eyes-closed conditions.

\textbf{Theta band abnormalities. }Elevated theta power, particularly in frontal and central regions, is one of the most consistently reported EEG findings in ADHD (\citealp{barry2003review, snyder2006meta}).
The interpretation of elevated theta in ADHD has evolved over time. Early theories emphasized cortical hypoarousal, suggesting that individuals with ADHD have insufficient cortical activation to support sustained attention and cognitive control (\citealp{satterfield1973response}). More recent perspectives have emphasized the role of theta oscillations in cognitive control and working memory processes (\citealp{cavanagh2014frontal, kim2025resolving}). Frontal midline theta, which increases during tasks requiring cognitive control, executive functions, and working memory, has been found to be altered in ADHD, with some studies reporting reduced task-related theta increases in individuals with ADHD (\citealp{missonnier2013eeg, koehler2009increased}). This suggests that the relationship between theta activity and ADHD may be more complex than simple hypoarousal, potentially reflecting dysregulation of theta-mediated cognitive control mechanisms.

\textbf{Beta band abnormalities. }Reduced beta power in ADHD has been reported in numerous studies, though findings have been somewhat less consistent than for theta abnormalities (\citealp{barry2003review}). Because beta activity is linked to cognitive processing, motor control and cortical arousal (\citealp{lundqvist2024beta}), its reduction in ADHD may indicate decreased cortical activation, inefficient cognitive processing, or altered sensorimotor function (\citealp{tzagarakis2010beta}). 

\textbf{Alpha band abnormalities. }Compared to more consistent findings in the theta and beta bands, studies of resting‐state alpha (8-13 Hz) power in ADHD have produced mixed results. Some investigations (especially in children) report reduced alpha power, while others (particularly in adult samples) report increased or no difference in alpha. For example, one study of adult ADHD found elevated alpha power (\citealp{koehler2009increased}) whereas another found reduced alpha in college-student ADHD participants (\citealp{woltering2012resting}). Alpha oscillations are generally interpreted as reflecting cortical idling or active inhibition of task-irrelevant regions, and alpha desynchronization (power decrease) occurs during engagement in cognitive tasks in healthy subjects (\citealp{trondle2022decomposing}). Thus, alpha abnormalities in ADHD may relate to altered regulation of cortical excitability or suppression of irrelevant processing — but the precise functional significance remains unclear and may vary by age, subtype and brain regions.

\textbf{Gamma band abnormalities. }Gamma oscillations (typically $>$30 Hz) have been less extensively investigated in ADHD compared to lower-frequency EEG bands such as theta, alpha, and beta. This is partly due to technical challenges in reliably measuring gamma activity from EEG, since gamma signals are of lower amplitude and are easily contaminated by muscle artifacts and other high-frequency noise. Nevertheless, the limited available evidence indicates that reduced gamma activity is associated in individuals with ADHD reported in both resting-state and task-related contexts (\citealp{tombor2021atypical, barry2010resting, prehn2015early}). Gamma oscillations are thought to reflect local cortical processing and the binding of distributed neural representations into coherent percepts and cognitive states (\citealp{herrmann2010human, guan2022role}). Abnormalities in gamma activity in ADHD could possibly contribute to difficulties in integrating sensory information, maintaining focused attention, and coordinating distributed neural networks.

\section{Effort-based decision making and individual differences in effort and reward sensitivity}
While ADHD can be viewed as a state of dysregulated motivation and characterised by heightened impulsivity and hyperactivity, apathy is characterised by diminished initiation and engagement in goal-directed behaviour (\citealp{costello2024apathy}). Effort-based decision making (EBDM) represents a fundamental aspect of motivated behaviour, where individuals must weigh the costs associated with physical or cognitive exertion against the potential rewards to be gained. Here, I provide a literature review of the conceptual and empirical foundations of EBDM and how disruptions in this process contribute to apathy.

\subsection{Neural mechanisms underlying EBDM}
EBDM is a core component of motivated behaviour, in which individuals assess whether the anticipated rewards (considering their magnitude, probability and delay) are sufficient to justify the physical or cognitive effort required (\citealp{botvinick2009effort}). This cost-benefit computation is regulated by distributed neural circuits that represents information about the cost of effort, the benefit/valuation of reward, and the integration of those cost-benefit signals into a net subjective value (\citealp{bailey2016neural, salamone2007effort, walton2006weighing}). Understanding the neural substrates of effort-based decision making has become increasingly important, as disruptions in this system appear to underlie motivational deficits across numerous neuropsychiatric conditions. Below I provide an overview of the neural circuits that support the process of cost-benefit computation.

In EBDM, effort valuation represents the process of computing the effort-related costs in potential actions. The anterior cingulate cortex (ACC) has been implicated in encoding the value of effort expenditure and in determining whether effort costs are worthwhile given expected rewards (\citealp{walton2003functional, rushworth2004action}). Neuroimaging studies in humans have shown ACC activation during EBDM tasks, with activity scaling according to the effort requirements of chosen actions (\citealp{croxson2009effort, prevost2010separate}). Lesion studies in rodents and non-human primates have revealed that ACC damage impairs the ability to select high-effort options even when they are associated with greater rewards, suggesting that the ACC is essential for integrating effort costs into value-based decisions (\citealp{walton2003functional, rudebeck2006separate}).

The mesolimbic dopamine system, particularly projections from the ventral tegmental area (VTA) to the nucleus accumbens (NAc), plays a critical role in EBDM (\citealp{salamone2024neurobiology}). Early theories posited that dopamine primarily encoded reward prediction and hedonic pleasure. However, recent evidence suggests that dopamine is more accurately characterised as representing the motivation to overcome effort costs and to enable reward-seeking behaviour rather than directly mediating pleasure itself (\citealp{berridge1998role, salamone2007effort}). Pharmacological studies in rodents have demonstrated that dopamine depletion in the NAc shifts animals' preferences away from high-effort, high-reward options toward low-effort, low-reward alternatives, without affecting their hedonic responses to freely available rewards (\citealp{salamone1991haloperidol, cousins1996nucleus}).

Reward valuation relies on brain regions that represent the subjective value of outcomes. The dorsal striatum, including the caudate nucleus and putamen, contributes to the process of reward valuation, particularly in the context of learning action-outcome associations and habit formation (\citealp{balleine2010human}). The dorsal striatum receives dopaminergic input from the substantia nigra and glutamatergic input from cortical regions, positioning it to integrate information about effort requirements with learned action values (\citealp{haber2010reward}). Functional neuroimaging studies have demonstrated that dorsal striatal activity during effort-based tasks predicts individual differences in effort sensitivity and willingness to work for rewards (\citealp{treadway2012dopaminergic}).

The ventral striatum (VS), particularly the nucleus accumbens (NAc), contributes to EBDM through reward anticipation and detection (\citealp{haber2011neuroanatomy}). Several studies report that the VS encodes the expected value of options during EBDM tasks in which ventral striatal activity scales with reward magnitude and subjective value, even when choices involve effort costs (\citealp{croxson2009effort, prevost2010separate}). An fMRI study by \citealp{pagnoni2002activity} reveals that the VS contributes to representing errors in predictions of rewards. Lesion and pharmacological studies in rodents have further established the causal importance of the NAc, showing that dopamine depletion in this region shifts animals' preferences away from high-effort options toward smaller, easily obtainable rewards, even when the effort-discounted value favors the larger reward (\cite{salamone2007effort, floresco2008cortico}). The NAc core and shell subregions appear to have different but complementary roles in EBDM, with the core being particularly important for selecting actions based on effort-benefit calculations, whereas the shell seems more involved in modulating reward valuation and the refinement of reward-seeking behaviour (\cite{ghods2010differential, floresco2018differential}). 

EBDM also depends on neural systems that integrate effort costs with reward benefits to compute the overall subjective value of an option. The ventromedial prefrontal cortex (vmPFC) and orbitofrontal cortex (OFC) play a central role in representing the reward value of different options after integrating multiple decision variables, including effort costs (\citealp{rangel2010neural, levy2012root}). Neuroimaging studies have shown that activity in these regions correlates with the subjective value of choices during effort-based tasks, suggesting they serve as a common currency for comparing options with different cost-benefit profiles (\citealp{prevost2010separate, bonnelle2016individual}). The vmPFC appears particularly important for translating motivational signals into goal-directed behaviour, with damage to this region associated with apathy and reduced willingness to exert effort (\citealp{levy2006apathy}). Integration also relies on distributed corticostriatal pathways. White matter connectivity between these regions is essential for coordinating the process of effort/reward integration. Diffusion imaging studies have revealed that the integrity of white matter tracts connecting frontal cortical regions with striatal structures predicts individual differences in effort-based decision making (\citealp{bonnelle2015characterization, le2018anatomy}). Specifically, reduced fractional anisotropy (FA) in tracts connecting the ACC to the striatum has been associated with decreased willingness to exert effort for rewards, suggesting that efficient communication between these regions is necessary for optimal motivated behaviour (\citealp{bonnelle2015characterization}).

Recent computational modeling approaches have provided additional insights into the neural mechanisms of EBDM. These models typically assume that individuals discount the value of rewards based on the effort required to obtain them, with the degree of discounting varying across individuals (\citealp{hartmann2013parabolic, chong2017neurocomputational}). Neuroimaging studies using these computational models have identified specific brain regions whose activity correlates with model-derived parameters such as effort sensitivity and reward sensitivity (\citealp{prevost2010separate, skvortsova2014learning}). These findings suggest that individual differences in EBDM reflect variations in how the brain computes and integrates effort costs with reward values.

\subsection{Disruption of EBDM and apathy}
Apathy, defined as a quantitative reduction in goal-directed behaviour and cognition (\citealp{levy2006apathy}), represents a common consequence of disrupted effort-based decision making. While apathy has traditionally been conceptualized as a lack of motivation, contemporary frameworks recognize it as a multidimensional syndrome that can arise from disruptions at various stages of goal-directed behaviour, including goal generation, planning, and execution (\citealp{levy2006apathy, husain2018neuroscience}).

The relationship between disrupted EBDM and apathy is supported by converging evidence from clinical populations. Patients with Parkinson's disease, a condition characterized by dopamine depletion, frequently exhibit both apathy and impaired performance on EBDM tasks (\citealp{chong2015dopamine, le2018anatomy}). These patients show reduced willingness to exert physical effort for rewards, with the severity of this deficit correlating with apathy ratings (\citealp{chong2015dopamine}). Importantly, dopaminergic medication can partially restore normal effort-based decision making in these patients, providing causal evidence for the role of dopamine in motivated behaviour (\citealp{chong2015dopamine}).

Apathy is also prevalent in patients with focal brain lesions affecting the EBDM-related neural circuits. Lesions to the ACC, vmPFC, and basal ganglia are particularly associated with apathy (\citealp{levy2006apathy, stuss2000executive}). \citealp{bonnelle2015characterization} demonstrated that traumatic brain injury patients with apathy showed reduced white matter integrity in tracts connecting the ACC to the ventral striatum, and that this structural disconnection predicted impaired performance on effort-based decision-making tasks. This study provided critical evidence that disrupted communication between cortical and subcortical regions involved in effort-based decision making can lead to apathy.

Schizophrenia is also characterised by prominent motivational deficits and apathy, particularly in the negative symptom domain (\citealp{foussias2010negative}). Patients with schizophrenia show impaired EBDM, with reduced willingness to exert effort for rewards (\citealp{gold2013negative, reddy2015effort}). These deficits appear to reflect abnormalities in the neural systems that represent reward value and compute effort costs, including the striatum and prefrontal cortex (\citealp{barch2010goal}). Importantly, motivational deficits in schizophrenia predict functional outcomes, including employment status and community functioning, highlighting the real-world significance of these impairments (\citealp{reddy2015effort}).

Depression represents another condition in which disrupted EBDM may contribute to motivational deficits. Depressed individuals show reduced willingness to exert effort for rewards, even when their ability to experience pleasure from rewards remains intact (\citealp{treadway2012dopaminergic, yang2014motivational}). Neuroimaging studies have revealed that depressed individuals show altered activity in the ACC and striatum during effort-based decision making, with the magnitude of these alterations correlating with symptom severity (\citealp{treadway2012dopaminergic}).  In this thesis we focused mainly on the motivational deficits with apathy and considered depression as a covariate in our analyses. 

\subsection{Neuroimaging and apathy}
Neuroimaging studies have provided important insights into the neural mechanisms underlying apathy, revealing how disruptions in brain structure and connectivity contribute to motivational deficits across a wide range of neurological and psychiatric disorders. Both grey and white matter abnormalities have been reported, reflecting the complex and distributed nature of the neural systems that support goal-directed behaviour. 

\subsubsection{Grey matter correlates of apathy}
Structural MRI has been used to investigate the neural correlates of apathy and EBDM. These studies have examined both regional brain volumes and cortical thickness, revealing consistent associations between structural alterations in specific brain regions and motivational deficits.

Structural imaging studies in multiple clinical populations implicate the cingulate cortex (especially the anterior cingulate/medial frontal region) in apathy. A study in Alzheimer's disease (AD) revealed that apathetic patients showed greater cortical thinning in the left ACC and left OFC compared to non‐apathetic patients (\citealp{tunnard2011apathy}). Similarly, in a voxel‐based morphometry study of mild AD patients, grey matter density loss in the ACC (and frontal cortex) was associated with apathy severity (\citealp{bruen2008neuroanatomical}). In Parkinson’s disease, high apathy scores correlate with reduced grey matter density in regions including the cingulate gyrus and inferior frontal cortex (\citealp{maggi2024anatomical, reijnders2010neuroanatomical}). In patients with behavioural-variant frontotemporal dementia (bvFTD), atrophy of the ACC has been shown to be associated with apathy (\citealp{rouse2025behavioural}). These findings are consistent with models of apathy that propose disruption of decision making systems (involving effort and reward integration) mediated by fronto‐cingulate networks and atrophy of the ACC may impair evaluation of effort costs and reward benefits, leading to reduced motivation.

The vmPFC and OFC have also been consistently implicated in structural MRI studies of apathy. Reduced volume in these regions has been associated with apathy in traumatic brain injury, stroke, and neurodegenerative diseases (\citealp{jenkins2022transdiagnostic, sheelakumari2020neuroanatomical, ducharme2018apathy}). Given the role of these regions in representing subjective value (\citealp{le2018anatomy, levy2006apathy}), structural alterations may impair the ability to appropriately value potential goals and outcomes, contributing to motivational deficits.

Striatal volume has been linked to both apathy and effort-based decision making. Studies in Parkinson's disease have revealed associations between reduced striatal's grey matter density and apathy severity (\citealp{reijnders2010neuroanatomical}). In individuals with schizophrenia, striatal volume has been shown to correlate with apathy (\citealp{roth2016apathy}). These findings are consistent with the critical role of the striatum in motivated behaviour and reinforcement learning.

\subsubsection{White matter integrity correlates of apathy}

While structural MRI studies have focused primarily on grey matter alterations, diffusion MRI has enabled investigation of white matter microstructure and connectivity in relation to apathy and effort-based decision making. The advent of DTI has enabled researchers to quantify white matter microstructural integrity in vivo through metrics such as fractional anisotropy (FA), which reflects the directional coherence of water diffusion along axonal tracts, and mean diffusivity (MD), which captures overall tissue organization (\citealp{tournier2011diffusion, pierpaoli1996diffusion}). Advanced analytical approaches including tract-based spatial statistics (TBSS) and tractography have facilitated systematic investigation of the relationship between white matter disruption and apathy severity across diverse clinical populations (\citealp{tournier2011diffusion, smith2006tract}).

Converging evidence from multiple neurological conditions supports the association between white matter integrity loss and apathy symptomatology. In Alzheimer's disease, \citealp{hahn2013selectively} demonstrated that apathy severity associated with reduced FA in the intrinsic brain network, suggesting that interhemispheric and frontoparietal disconnection contributes to motivational deficits in dementia. Similarly, investigations in Parkinson's disease have revealed that apathetic patients exhibit significantly lower FA values in the corpus callosum, anterior corona radiata, and superior corona radiata and left cingulum compared to non-apathetic participants (\citealp{zhang2018reduction}). In cerebral small vessel disease, \citealp{le2018dysfunctional} reported that apathy was associated with reduced FA in frontostriatal pathways including the corpus callosum, anterior cingulum and orbitofrontal-anterior cingulate. In mild cognitive impairment, widespread white matter abnormalities involving the anterior thalamic radiation, the corpus callosum, the inferior fronto-occipital fasciculus/uncinate fasciculus, and the anterior corona radiata have been associated with greater apathy severity, suggesting that motivational impairment may serve as an early marker of network disruption preceding overt dementia (\citealp{setiadi2021widespread}). In another study with traumatic brain injury, FA in the superior longitudinal fasciculus, inferior longitudinal fasciculus, and internal capsule was negatively correlated with apathy severity (\citealp{navarro2021apathetic}).

These findings collectively support a disconnection hypothesis of apathy, in which disruption of frontostriatal and frontotemporal white matter networks compromises the neural circuitry underlying motivation, reward anticipation, and goal-directed behaviour. The consistent involvement of the anterior cingulum, which connects the anterior cingulate cortex to subcortical structures including the ventral striatum, suggests that apathy emerges when pathways mediating reward valuation and effort-based decision-making are compromised (\citealp{bonnelle2016individual}). Similarly, disruption of the corpus callosum may impair the interhemispheric coordination necessary for complex motivational states (\citealp{schutter2013corpus}). Understanding apathy as a network-level phenomenon arising from structural disconnection opens avenues for targeted interventions, including neuromodulation approaches that aim to restore functional connectivity within affected circuits and pharmacological strategies that enhance dopaminergic transmission along compromised frontostriatal pathways (\citealp{husain2018neuroscience}). 

\subsubsection{Effort and reward sensitivity in computational modelling}
Using computational modelling approaches, apathy can be described as emerging from distinct motivational deficits, specifically \textit{hypersensitivity to effort costs and/or hyposensitivity to reward value} (\citealp{costello2024apathy, bonnelle2015characterization}). These computational frameworks have proven valuable in decomposing motivated behaviour by quantifying how individuals weigh effort (effort sensitivity) against reward (reward sensitivity) when making goal-directed decisions (\citealp{chong2015dopamine, le2018anatomy}). While neuroimaging studies have implicated frontal cortical regions, anterior cingulate cortex and ventral striatum in effort valuation, reward processing and effort--reward integration (\citealp{le2018anatomy, croxson2009effort, prevost2010separate}), the specific neural correlates associated with individual differences in effort and reward sensitivity remain subject to ongoing investigation.

\section{Machine learning in neuroscience research}
In this thesis, I aim to evaluate the predictive powers of brain physiological signals and neuroimaging measures in predicting behavioural measures and computational modelling parameters using machine learning models and feature importance analysis. Here, I would like to provide a brief literature in how machine learning techniques have been employed in neuroscience, specifically with EEG and neuroimaging data. The application of machine learning techniques to EEG and neuroimaging data represents a paradigm shift from traditional group-level statistical analyses toward individual-level predictions, offering the potential to capture subtle, spatially distributed patterns that may be imperceptible to conventional statistical approaches (\citealp{arbabshirani2018advanced}).

Machine learning consists of algorithms capable of generalising rules or patterns from a labeled set of input data, and using that knowledge to generate predictions or classifications on data not seen before (\citealp{kotsiantis2007supervised}). The field of neuroscience has greatly benefited from machine learning, as machine learning algorithms have been widely used to build classifiers for a wide range of diseases using magnetic resonance imaging (MRI) information as input features (\citealp{arbabshirani2017single}). These input features can encompass diverse neuroimaging modalities, including structural measures such as grey matter volumes from voxel-based morphometry (\citealp{ashburner2000voxel}), cortical thickness measures (\citealp{fischl2000measuring}), and white matter microstructural measures derived from diffusion MRI (e.g. fractional anisotropy and mean diffusivity) (\citealp{mandl2008functional}), as well as functional connectivity patterns (\citealp{iturria2008studying}).

\subsection{Machine learning approaches for EEG-based prediction}
The application of machine learning to EEG data has shown considerable promise across multiple domains of neuroscience research. Machine learning is a subset of artificial intelligence focusing on developing computer algorithms and statistical models to enable computer systems to make predictions from learning data (\citealp{jordan2015machine}). The capability of ML to handle high-dimensional data has been demonstrated in many fields of studies, and may be applied in an automated process to increase predictive accuracy (\citealp{papoutsoglou2021automated, amini2018review}).

Machine learning models applied to EEG data can be broadly categorized into conventional ML and deep neural networks (DNN). The main difference between the two categories is the input to the model: the input of conventional ML is typically handcrafted features, such as demographics and quantitative EEG features (\citealp{chen2024electroencephalogram}). On the other hand, deep neural networks automatically extract features from raw data (\citealp{chen2024electroencephalogram}). Deep learning can automatically capture essential characteristics from a large volume of data by optimizing its parameters through back-propagation and stochastic gradient descent (\citealp{cui2023towards}), and it is reported that deep learning has achieved better performance than conventional methods in many BCI domains such as identifying attentive mental state, movement-related cortical potential recognition, and detection of driver drowsiness (\citealp{cui2023towards}).

Recent advances have demonstrated impressive performance in clinical applications. Random forest classifiers have been used to predict individual major depressive disorder status using baseline quantitative EEG coherence (\citealp{mcvoy2025assessing}). Furthermore, EEG analysis is considered a good, least invasive, and reliable way for identifying any neurological disorder  and with promising new advancements in machine learning-based algorithms, early and precise prediction might induce a radical shift (\citealp{zhang2023applied}), with cutting-edge AI methods being applied for exploiting EEG data for early warning symptoms detection across various conditions including Parkinson's disease, sleep apnoea, drowsiness, schizophrenia, motor imagery classification, and emotion recognition (\citealp{li2025machine}).

\subsection{Strutural neuroimaging and behavioural prediction}
Beyond EEG, structural neuroimaging combined with machine learning offers powerful tools for predicting individual differences in behaviour and cognition. ML algorithms use MRI information as input features, including structural gray matter readings obtained from cortical thickness (\citealp{ad2006civet, fischl2000measuring}) or GM density values from voxel-based morphometry (\citealp{ashburner2000voxel}), microstructural changes in the white matter from diffusion-weighted imaging (fractional anisotropy) (\citealp{mandl2008functional}), connectivity matrices (\citealp{iturria2008studying}), or parameters derived from network analyses (\citealp{iturria2013anatomical, rubinov2010complex}).

The predictive utility of structural brain features has been demonstrated across multiple domains. For schizophrenia patients and healthy controls, features derived from resting-state fMRI, structural MRI gray matter, and DTI fractional anisotropy were extracted and fed into a regression model, achieving high prediction for both cognitive scores and symptomatic scores (\citealp{sui2018multimodal}). With high accuracy, future literacy was predicted predominantly based on gray matter volume in the left occipito-temporal cortex and local gyrification in the left insular, inferior frontal, and supramarginal gyri, with phonological awareness significantly predicting future literacy (\citealp{beyer2022structural}), indicating that brain morphology of the large-scale reading network at a preliterate age can predict how well children learn to read (\citealp{beyer2022structural}).

White matter integrity has emerged as a particularly important predictor of cognitive and behavioural outcomes. A study with white matter connectivity from diffusion MRI as input to machine learning models has been shown to identify individuals with autism and achieve 78\% classification accuracy (\citealp{zhang2018whole}). Using a linear support vector machine on white matter integrity measures (white matter volume, fractional anisotropy and mean diffusivity) from structural and diffusion MRI, a study showed that white matter features can distinguish children with developmental dyslexia from controls with 83.6\% accuracy, highlighting widespread alterations across reading-related, limbic, and motor white matter pathways (\citealp{cui2016disrupted}). Using an adaptive-boosting framework applied to diffusion, structural, and functional magnetic resonance imaging measures, the study showed that integrating superficial and deep white-matter properties with cortical thickness and resting-state connectivity yields the strongest ability to distinguish individuals with early multiple sclerosis from healthy controls (\citealp{buyukturkoglu2022machine}).

Together, structural neuroimaging paired with modern machine learning approaches provides a robust framework for capturing individual variation in cognition and behaviour (\citealp{sui2020neuroimaging, tejavibulya2022predicting}). By leveraging microstructural and network level measures, these models consistently achieve high predictive performance across clinical and developmental domains (\citealp{karimi2024diffusion, tejavibulya2022predicting}). 

In this thesis, I leverage machine learning methodologies across multiple levels of analysis. First, we employ classification algorithms to predict ADHD diagnosis from EEG signals, employing machine learning's ability to detect complex spectral power band patterns in neural activity. Then, we examine how structural brain integrity -- both white and grey matter -- relates to individual differences in effort and reward sensitivity, using a wide range of machine learning classification algorithms.

%\chapter{Methodology}
%\input{Chapters/litreview}

\chapter{Task-based versus resting-state EEG for machine learning classification of adult ADHD}
In this chapter, I aim to assess the effectiveness of task-related electroencephalography (EEG) and resting-state EEG in classifying adult patients with ADHD from healthy controls. Specifically I applied machine learning techniques to classify ADHD and control participants based on spectral band power features extracted from EEG. The primary objective was to test whether task-based EEG acquired during a stop-signal task, which engages inhibitory control processes, would provide superior classification performance compared to resting-state EEG. I hypothesised that task-based EEG would provide disorder-relevant characteristics of inhibitory control, whereas resting-state signals would offer weaker predictive power. To further characterise the neural features underlying classification, we conducted permutation-based feature importance analyses, which identified theta–beta ratios as key predictors during rest and gamma-band oscillations over fronto-central and parietal sites during task engagement. Together, these analyses provide an insight into why task-based EEG more robustly captures ADHD-related dysfunction than resting-state activity. 

\section{Introduction}
ADHD is a neurodevelopmental disorder that affects a large amount of school-aged children and adults worldwide (\citealp{biederman2006functional}, \citealp{kessler2006prevalence}). The core symptoms of ADHD include inattention, hyperactivity, and impulsivity, which can lead to impairments in academic, social, and occupational functioning (\citealp{sonuga2010developmental}). The diagnosis of ADHD is primarily based on clinical assessment and behavioural observations. While these tools are imporant, they are also inherently subjective and susceptible to rater bias and contextual influences. This has motivated increasing interest in developing objective, biologically grounded markers that can support diagnosis, prognosis and treatment stratification in ADHD (\citealp{banaschewski2007annotation}).

EEG is a non-invasive and cost-effective tool that measures the electrical activity of the brain and has been widely used in the study of ADHD (\citealp{castellanos2002neuroscience}). EEG recordings can be obtained in different states, including resting state and task-based paradigms. Resting-state EEG measures the spontaneous neural activity of the brain when the participant is at rest, whereas task-based EEG measures the neural activity associated with performing a specific cognitive task (\citealp{karamzadeh2013capturing}, \citealp{bandara2016task}). 

Early EEG studies in ADHD focused on resting state spectral power. In children, increased theta power and decreased beta power, often summarised as an elevated theta-to-beta ratio (TBR), have been reported in ADHD relative to typically developing individuals and were proposed as a diagnostic biomarkers (\citealp{snyder2006meta, markovska2017quantitative}). Some studies reported high sensitivity and specificity using single-channel TBR measures (\citealp{monastra1999asessing}) and machine learning approaches in children have yielded classification accuracies in the 75–85\% range using spectral features (\citealp{chen2019eeg}). However, in adults the literature is considerably more mixed: while some work has replicated elevated TBR or related spectral alterations, others have failed to find reliable group differences or diagnostic utility, raising concerns about the robustness and generalisability of resting state TBR as a biomarker in later life (\citealp{kiiski2020eeg, newson2019eeg}). These inconsistencies may reflect developmental changes, heterogeneity within ADHD, or sensitivity of resting-state measures to state factors such as drowsiness or mind wandering.

In parallel, a large body of work has examined EEG during cognitive tasks involving executive control. Inhibitory control paradigms such as the stop signal task and Go/No-Go tasks are particularly relevant, as deficits in response inhibition are considered a core feature of ADHD and are robustly reflected in longer stop signal reaction times and increased commission and omission errors (\citealp{senkowski2024assessing, barkley1997behavioral}). Task-based EEG studies have revealed alterations in several frequency bands, including theta, beta and gamma oscillations, as well as in event-related potentials (ERPs) such as the N2 and P3, which are linked to conflict monitoring, inhibitory control and attentional allocation (\citealp{woltering2013neurophysiological, grane2016erp, czobor2017electrophysiological}). In studies involving children, increased theta power and an elevated theta/beta ratio have been linked to ADHD (\citealp{markovska2017quantitative}). \citealp{bink2015eeg} reported spectral analyzes revealing group-specific patterns, as adolescents with ADHD showed greater theta activity than those with comorbid autism spectrum disorder (ASD) and ADHD during eyes open and task conditions, with only the ADHD group displaying an association between increased theta and poorer attention performance. Gamma-band activity ($>$30 Hz) in particular has been associated with the binding and coordination of distributed neural assemblies supporting attention and cognitive control (\citealp{herrmann2004cognitive}). Importantly, gamma oscillations have been shown to increase during the processing of task-relevant visual and auditory targets and to differ between individuals with ADHD and healthy controls (\citealp{herrmann2001human, debener2003top})..

More recently, machine learning methods have been applied to EEG data to develop predictive models that classify individuals with ADHD versus healthy controls or predict symptom severity (\citealp{dubreuil2020deep, tosun2021effects, hosseini2020review}). Most of these studies have used resting-state EEG, often focusing on spectral power and TBR. In child samples, promising classification performance has been reported, whereas in adults results are more modest and variable, mirroring the inconsistency of resting-state markers themselves. A smaller but growing number of studies have applied ML to task-based EEG, including spectral and time–frequency measures acquired during cognitive control paradigms, sometimes achieving high accuracy using, for example, convolutional neural networks on multi-channel task EEG (\citealp{dubreuil2020deep}). However, existing ML work typically examines either resting-state or task-based EEG in isolation, often with differing feature sets and analysis pipelines. As a result, it remains unclear whether task-based EEG systematically affords better diagnostic classification than resting-state EEG when both are analysed within a machine learning framework.

From a methodological standpoint, this chapter concentrates on spectral power features rather than ERPs. ERPs provide precise information about the timing of discrete cognitive events (e.g., N2/P3 components related to conflict monitoring and response inhibition) but typically require averaging across many stimulus-locked trials, which cannot be obtained with resting-state EEG. Spectral power, by contrast, can be robustly extracted using time–frequency decomposition methods from both resting-state and stop signal task EEG data, making it a practical choice for our machine learning pipelines aimed at comparing resting-state and task-based EEG with matched feature spaces.

The primary aim of this chapter is to directly compare the classification performance of machine learning models trained on resting-state versus task-based EEG in adults with ADHD and healthy controls, using a common spectral power feature framework. Methodologically, our analyses focus on spectral power features because event-related potentials cannot be derived from resting-state recordings in the absence of time-locked events. Resting-state EEG was recorded with eyes closed, and task-based EEG was acquired during a stop-signal task that robustly engages inhibitory control processes. We trained and evaluated a diverse set of supervised ML classifiers on spectral band power and theta-to-beta ratio (TBR) features from both paradigms, using identical preprocessing and nested cross-validation procedures, and assessed whether task-based EEG leads to superior classification performance relative to resting-state EEG. We further used permutation-based feature importance to identify feature contribution to classification tasks. We hypothesised that: (i) models trained on stop-signal task EEG would outperform those trained on resting-state EEG in distinguishing ADHD from controls; and (ii) task-based classification would be supported by specific frequency spectral features over fronto-central and parietal cortices, consistent with literature about altered recruitment of executive control networks in ADHD.

\section{Methods}
% Make a figure here: A. EEG data acquision (resting-state and task-based), B. Go/No-go task description, ML analysis (17.6cm)

\subsection{Participants}
EEG recordings were acquired at Trinity College Dublin using a BioSemi ActiveTwo system with 70 electrodes, of which 64 were positioned according to the 10--5 system \citealp{kiiski2020eeg}. Resting-state EEG data were obtained from 52 adults with ADHD and 98 neurotypical controls. ADHD participants met formal diagnostic criteria for ADHD and scored above the clinical threshold (T-score $>$ 65) on the Conners' Adult ADHD Rating Scale (CAARS, \citealp{conners1999conners, la2008detecting, smyth2019evaluating}) , whereas controls had no ADHD diagnosis and scored below threshold (T-score $<$ 65). EEG during the stop-signal task was recorded in a subset comprising 24 adults with ADHD and 78 controls.

\begin{figure}
    \centering
    \includegraphics[width=1\linewidth]{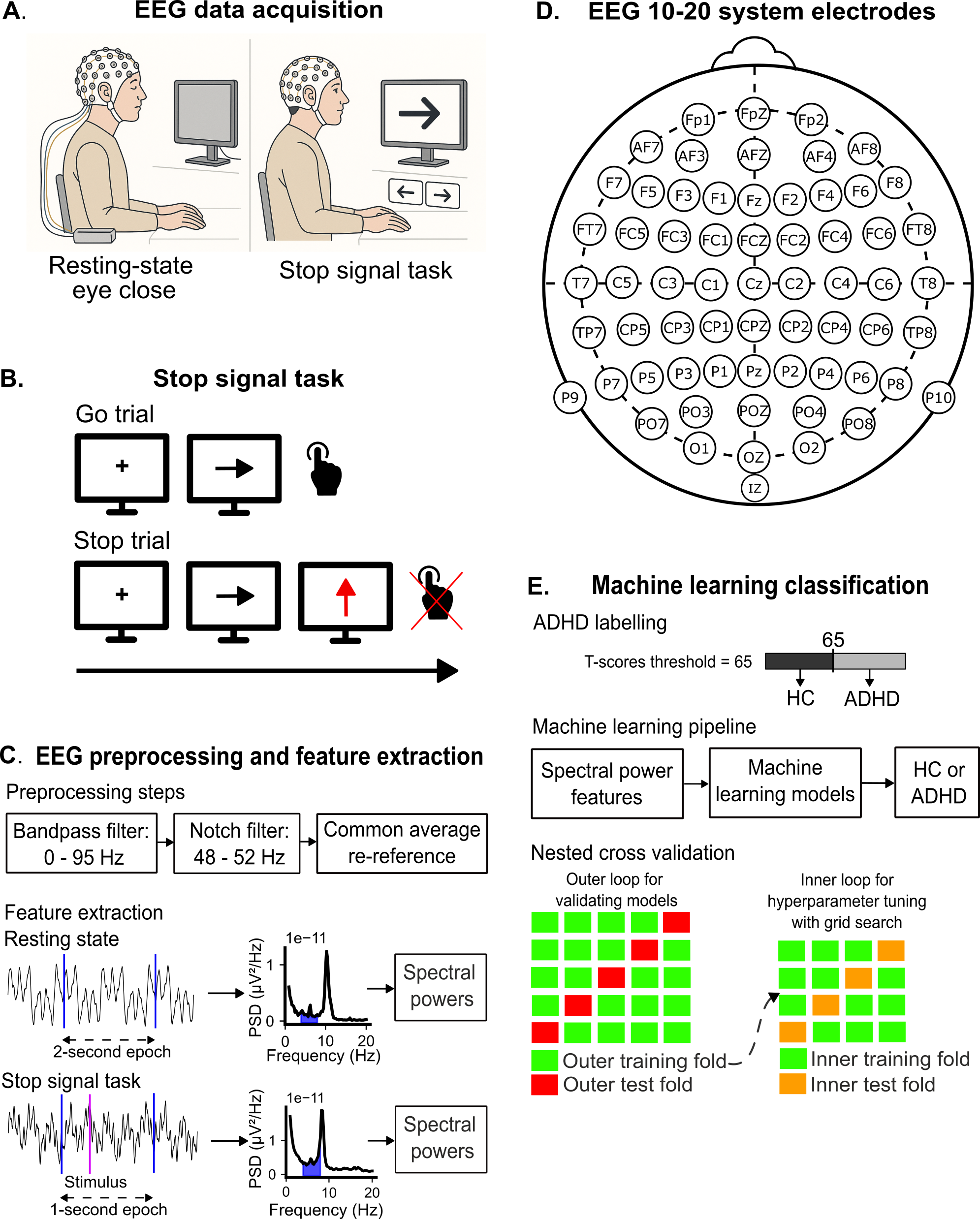}
    \caption[Experimental design, EEG data acquisition, and classification pipeline.]{\textbf{Experimental design, EEG data acquisition, preprocessing, feature extraction and classification pipeline.}
A. EEG was recorded during two conditions: resting state with eyes closed and performance of the stop-signal task. B. In the stop-signal task, participants responded to directional arrows on ``Go” trials and withheld responses when a stop signal appeared. C. EEG data were preprocessed with bandpass (0–95 Hz) and notch (48–52 Hz) filters, followed by common average re-referencing. Spectral power features were extracted from 2-s resting-state epochs and 1-s task-related epochs. D. Electrode placement followed the international 10–20 system.
E. Spectral power features were used as inputs to machine-learning classifiers to distinguish ADHD from healthy control participants. Model training employed nested cross-validation, with outer folds for model validation and inner folds for hyperparameter tuning via grid search.}
    \label{fig:chapter_3_method}
\end{figure}

\subsection{EEG data acquisition}
Resting state EEG data collection was performed in a quiet and darkened room, with participants seated in front of a computer monitor at a distance of 1.05 meters (\citealp{kiiski2020eeg}). The monitor employed a screen resolution of 1024x768 and a refresh rate of 75Hz. Data was collected with participants in both an eye-closed and eye-opened states, however, for the present study, only the eye-closed resting state EEG data was utilised in experiments. The eyes-open resting-state EEG may potentially introduce an additional confound to the experiment associated with visual processing and fixation demands as well as increase eye-movement–related artifacts in the signal.

For the stop signal task, participants were presented with left and right arrow stimuli on a computer monitor (\citealp{rueda2021brain}). They sat in front of the monitor at a standardized distance of 108 cm, with a resolution of 1024x768 pixels and a refresh rate of 75 Hz. The task involved an adaptive event-related Stop Signal Task (SST) lasting approximately 9 minutes. It consisted of 135 Go trials and 45 Stop trials, presented in three blocks of 60 trials. Each trial began with a 1000 ms fixation cross, followed by a 750 ms Go stimulus (arrow pointing left or right). Participants were instructed to respond as quickly as possible by pressing a button corresponding to the direction of the arrow in Go trials (see Figure \ref{fig:chapter_3_method} for an illustration). In Stop trials, an upward-pointing arrow (Stop signal) appeared after the Go stimulus, requiring participants to inhibit their responses.

\subsection{EEG data processing and feature extraction}
In this study, the data analysis consists of three main steps: EEG data preprocessing, EEG spectral power extraction, and machine learning modelling (see Figure \ref{fig:chapter_3_method}.C). Two types of data were collected from separate participant groups, and to ensure comparability, all parameters and hyperparameters are kept identical between the two classification experiments. The objective is to assess the potential of task-based EEG data for enhancing ADHD diagnosis accuracy and reliability, compared to resting-state EEG data. The key difference lies in the segmentation approach (Figure \ref{fig:chapter_3_method}.C): resting state EEG data is randomly divided into 2-second segments, while stop signal task EEG data is segmented into 1-second intervals centered around the stimulus signal, with a 0.3-second pre-stimulus and 0.7-second post-stimulus time window selected for each segment.

The preprocessing steps involved filtering and artifact rejection to ensure the quality of the EEG data (\citealp{derosiere2018visuomotor, derosiere2015expectations, alamia2019implicit}). In the spectral power extraction step, the relative spectral power is extracted from 12 frequency bands for 64 EEG channels using the multitaper method (\citealp{babadi2014review}). Finally, machine learning models were trained and evaluated using nested cross-validation on both resting state and task-based EEG data to compare their diagnostic performance for ADHD. Further details of each experimental stage are elaborated in the following sections.

\subsubsection{EEG data cleaning}
EEG signals were re-referenced to the common average (\citealp{yao2019reference}), followed by bandpass filtering between 0 and 95 Hz. To mitigate the presence of electricity line noise, a notch filter is further employed, with a frequency range of 48-52 Hz (\citealp{kiiski2020eeg}). All preprocessing procedures are executed using the MNE library in Python 3.10. Relative spectral power was extracted from 12 frequency bands spanning 4 Hz each across 64 channels using the multitaper method (\citealp{babadi2014review, prerau2017sleep}). The frequency bands are defined as (1,5), (5,9), (9, 13), (13, 17), (17, 21), (21, 25), (25, 29), (29, 33), (33, 37), (37, 41), (41, 45) and (45, 49). Furthermore, a theta-to-beta ratio was calculated for each channel. The resulting feature vector has a dimension of 832.

\subsubsection{Resting state EEG data}
Resting-state EEG recordings were segmented into non-overlapping 2-second epochs for spectral analysis. To estimate relative spectral power, we applied the multitaper method using a Hanning taper, which provides improved spectral concentration and reduces variance in power estimates. Relative power was computed by normalising the power in each frequency band to the total power across all bands. The theta-to-beta ratio (TBR), a commonly used biomarker in ADHD research, was calculated by dividing the absolute power in the theta frequency band (4–8 Hz) by that in the lower beta band (13–21 Hz). This ratio was computed for each epoch and subsequently averaged across all epochs for each participant.

\subsubsection{Stop signal task EEG data}
Task-related EEG data were processed using stimulus-locked segmentation aligned to the onset of the primary stimulus in the stop-signal task. Each epoch spanned 1 s, comprising 300 ms of pre-stimulus activity and 700 ms of post-stimulus activity to capture anticipatory dynamics and the evolution of task-evoked responses. Spectral decomposition was performed using the multitaper method, which provides robust estimates of oscillatory power in short analysis windows by reducing spectral leakage. For each epoch, we computed absolute and relative power across canonical frequency bands, and post-stimulus spectra were baseline-corrected by subtracting the corresponding pre-stimulus power spectrum. The theta-to-beta ratio was then calculated by dividing baseline-corrected theta power (4–8 Hz) by lower-beta power (13–21 Hz), and averaged across epochs. These baseline-normalised absolute spectral band powers were then normalised with the total powers across all bands to obtain relative frequency bands. 

\subsection{Machine learning analysis}
To construct classification models for ADHD from EEG recordings' spectral measures, we implemented nine supervised machine learning algorithms: logistic regression (LR), support vector machines (SVM), random forests (RF), gradient boosting (GB), extra trees (ET), adaptive boosting (AB), decision trees (DT), k-nearest neighbors (KNN), and Gaussian naïve Bayes (GNB) (see Figure \ref{fig:chapter_3_method}E). These models were selected based on prior evidence demonstrating strong performance in EEG-based classification tasks (\citealp{tenev2014machine}, \citealp{lenartowicz2014use}).

\subsubsection{Model training and evaluation method}
To optimize the performance of the machine learning algorithms, hyperparameter tuning was conducted using a cross-validation grid search approach. The performance and generalisability of the models for ADHD classification were assessed using a nested cross-validation method. This involved two levels of cross-validation: an outer loop and an inner loop. The outer loop utilized multiple folds, typically through k-fold cross-validation, where each fold served as a test set while the model was trained on the remaining folds. The inner loop, within the outer loop, further optimised the hyperparameters of the models using k-fold cross-validation on the training set of the current outer fold. Different hyperparameter combinations are evaluated, and the best-performing set is selected based on a predefined evaluation metric. The nested cross-validation approach prevents information leakage from the test set during hyperparameter tuning and provides a robust evaluation of the models' generalization capabilities while avoiding overfitting. The final performance of the models is determined by aggregating the results across all outer fold iterations. By employing nested cross-validation, this study ensures a reliable assessment of the models' performance and enhances the validity of the results for ADHD classification using EEG data.

The sklearn library, which is a popular machine learning library in Python, was used for implementing all the models (\citealp{pedregosa2011scikit}). The evaluation metric used in this study is the Area under the Curve (AUC) score, which measures the ability of the model to distinguish between ADHD and healthy control groups. AUC is particularly appropriate in this experiment because it is insensitive to class imbalance and provides a robust measure of overall classification independent of decision thresholds (\citealp{fawcett2006introduction}).

\subsubsection{Permutation test}
To determine whether classification performance was significantly above chance, we applied the same validation procedure separately to resting-state EEG and stop-signal task EEG data. For each dataset, we identified the best-performing classifiers and repeated the full classification procedure 1,000 times after randomly permuting class labels, thereby generating a null distribution of accuracy and AUC under the assumption of no relationship between EEG features and diagnostic group. We then evaluated whether performance in the true-label data exceeded chance expectations using three complementary tests: (1) a Monte Carlo p-value, calculated as the proportion of permuted-label runs yielding higher accuracy or AUC than the true-label model; (2) one-sided t-tests comparing the distributions of accuracy and AUC between true-label and permuted-label models; and (3) one-sided t-tests against the theoretical chance level of 0.5. Statistical significance was defined as $p < 0.05$.

\subsubsection{Feature importance analysis}
To assess the relative contribution of each feature to model predictions, we quantified feature importance using permutation importance (\citealp{altmann2010permutation, breiman2001random}). This method evaluates the decrease in predictive performance when the values of a single feature are randomly permuted while leaving all others unchanged, thereby disrupting any relationship between that feature and the target variable. The importance of a feature is estimated as the mean performance drop across multiple permutations, with larger decreases indicating greater influence on the model’s output. Unlike measures that depend on model-specific coefficients or internal heuristics, permutation importance provides a model-agnostic estimate of feature relevance that directly reflects its impact on predictive accuracy.

\section{Results}
\subsection{Classification results}
\subsubsection{With resting-state EEG data}
%% Expanding to 12 ML algorithms with robust pipelines
%% A graph of three bars should be shown: with resting-state EEG, task-based spectral power, task-based spectral power/ERP/connectivity ,etc... 
% RSEC's power performance < SST power performance
% SST power performance < SST 
%% Feature importance study with all features
%% Combination
We evaluated the performance of nine classifiers trained on spectral power features extracted from resting-state EEG to classify between individuals with ADHD and healthy controls (see Table \ref{table:ml_results_rsec}). Across models, performance was generally modest, with area under the receiver operating characteristic curve (AUC) values ranging from 0.462 ± 0.041 (SVM) to 0.594 ± 0.024 (Gaussian Naïve Bayes; GNB). Accuracy values varied between 0.542 ± 0.040 (decision tree; DT) and 0.626 ± 0.018 (logistic regression; LR), though higher accuracy was not consistently associated with improved sensitivity or precision.

Recall was lowest for LR (0.012 ± 0.024) and highest for GNB (0.690 ± 0.037), indicating that most classifiers struggled to identify ADHD cases, with the exception of GNB. Precision values were similarly variable, ranging from 0.024 ± 0.049 (LR) to 0.430 ± 0.023 (GNB). F1-scores—a balanced measure of recall and precision—were highest for GNB (0.525 ± 0.025), followed by k-nearest neighbors (KNN; 0.342 ± 0.052) and DT (0.334 ± 0.063), and lowest for LR (0.013 ± 0.025).

Among the ensemble methods, AdaBoost achieved the highest recall (0.271 ± 0.065) and F1-score (0.308 ± 0.065), whereas random forest (RF), gradient boosting (GB), and extra trees (ET) showed only marginally better-than-chance discrimination (AUC: 0.494–0.530). Notably, GNB achieved the best overall balance between sensitivity and precision, despite relatively low accuracy, suggesting that its probabilistic framework may be better suited to the distributional characteristics of the spectral power features.

\begin{figure}
    \centering
    \includegraphics[width=0.8\linewidth]{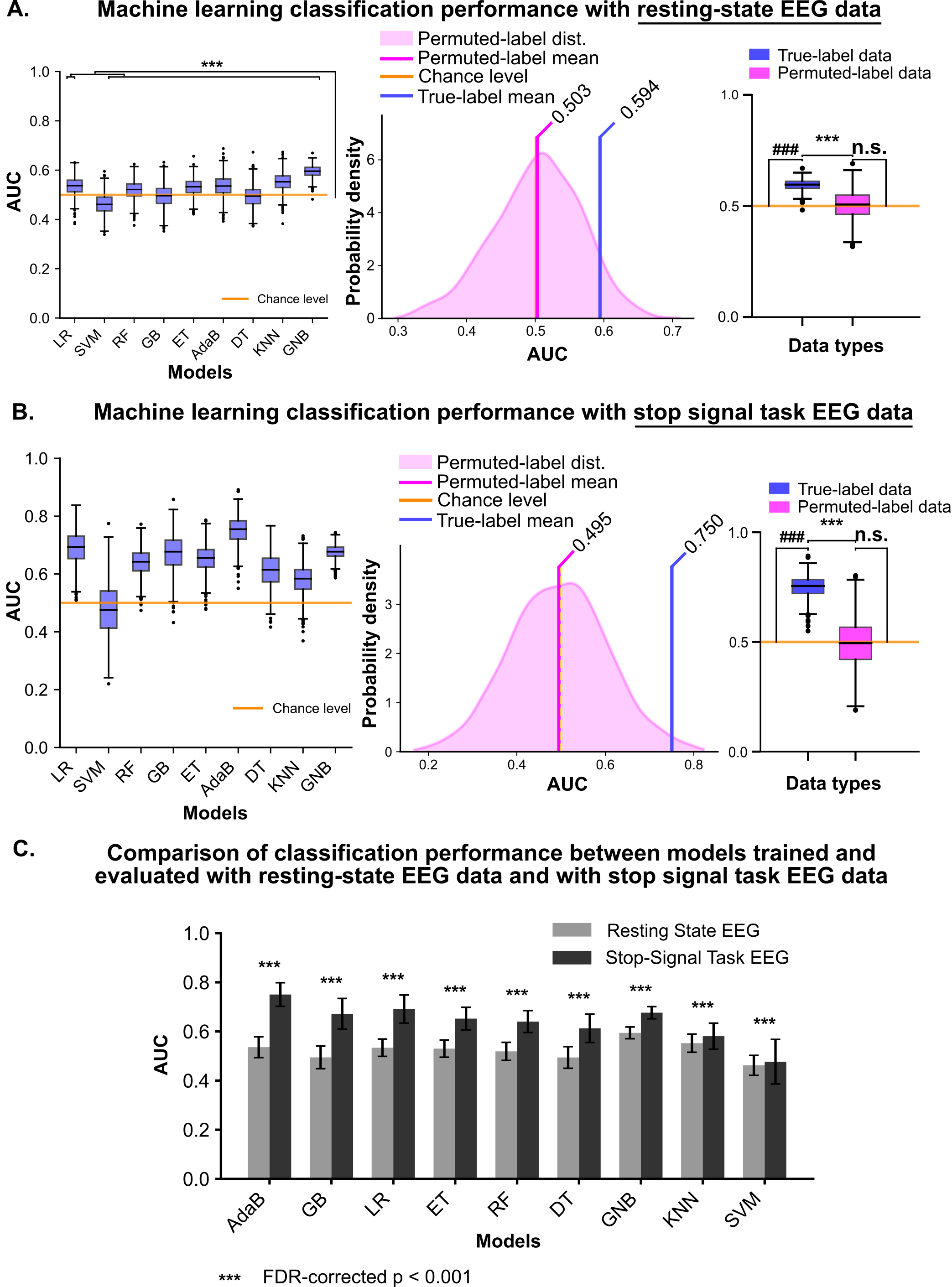}
    \caption[Statistical analysis of classification performance with resting-state and stop-signal task EEG.]{\textbf{Statistical analysis of classification performance with resting-state and stop-signal task EEG.} A. Left, distribution of AUC across models compared with chance. Middle, permutation test with 1,000 iterations showed no significant deviation from chance (Monte Carlo $p = 0.064$). Right, true-label performance was significantly greater than chance ($t = 124.99$, $p < 0.001$) and higher than permuted-label performance ($t = 42.36$, $p < 0.001$), whereas permuted-label performance did not differ from chance ($t = 1.32$, $p = 0.19$).
B. Left, distribution of AUC across models compared with chance. Middle, permutation test with 1,000 iterations indicated significant deviation from chance (Monte Carlo $p = 0.010$). Right, true-label performance was significantly greater than chance ($t = 164.5$, $p < 0.001$) and higher than permuted-label performance ($t = 67.98$, $p < 0.001$), while permuted-label performance did not differ from chance ($t = -1.41$, $p = 0.14$).
C. All classifiers trained on stop-signal task EEG achieved significantly higher AUCs than those trained on resting-state EEG (all FDR-corrected $p < 0.001$)}
    \label{fig:model_comparison}
\end{figure}
\begin{table}
\centering
\small % or \footnotesize, \scriptsize

\caption[Classification performance for ADHD versus healthy controls using resting-state EEG.]{Classification performance for ADHD versus healthy controls using resting-state EEG.
  Performance of nine classifiers: logistic regression (LR), support vector machine (SVM), random forest (RF), gradient boosting (GB), extremely randomized trees (ET), AdaBoost (AdaB), decision tree (DT), \(k\)-nearest neighbors (KNN), and Gaussian naïve Bayes (GNB). Metrics are mean \(\pm\) s.d. across cross-validation folds: area under the ROC curve (AUC), accuracy, recall (sensitivity), precision (positive predictive value) and F1 (harmonic mean of precision and recall). ADHD was treated as the positive class.}
  \label{table:ml_results_rsec}
    \resizebox{\textwidth}{!}{

\begin{tabular}{cccccc}

\toprule
Model & AUC & Accuracy & Recall & Precision & F1 \\
\midrule
LR & 0.534 ± 0.035 & 0.626 ± 0.018 & 0.012 ± 0.024 & 0.024 ± 0.049 & 0.013 ± 0.025 \\
SVM & 0.462 ± 0.041 & 0.623 ± 0.020 & 0.015 ± 0.026 & 0.032 ± 0.054 & 0.019 ± 0.029 \\
RF & 0.519 ± 0.037 & 0.590 ± 0.028 & 0.176 ± 0.048 & 0.364 ± 0.106 & 0.223 ± 0.057 \\
GB & 0.494 ± 0.046 & 0.583 ± 0.033 & 0.189 ± 0.062 & 0.346 ± 0.115 & 0.226 ± 0.064 \\
ET & 0.530 ± 0.035 & 0.600 ± 0.024 & 0.123 ± 0.047 & 0.306 ± 0.127 & 0.164 ± 0.059 \\
AdaB & 0.536 ± 0.042 & 0.593 ± 0.034 & 0.271 ± 0.065 & 0.413 ± 0.095 & 0.308 ± 0.065 \\
DT & 0.494 ± 0.044 & 0.542 ± 0.040 & 0.340 ± 0.076 & 0.356 ± 0.069 & 0.334 ± 0.063 \\
KNN & 0.552 ± 0.037 & 0.597 ± 0.032 & 0.305 ± 0.051 & 0.426 ± 0.073 & 0.342 ± 0.052 \\
\textbf{GNB} & \textbf{0.594 ± 0.024} & \textbf{0.553 ± 0.025} & \textbf{0.690 ± 0.037} & \textbf{0.430 ± 0.023} & \textbf{0.525 ± 0.025} \\
\bottomrule
\end{tabular}}
\end{table}

\subsubsection{With stop signal EEG data}

We evaluated classifier performance using spectral power features derived from the stop-signal task to classify individuals with ADHD from healthy controls ((see Table \ref{table:ml_results_sst}). Across models, classification performance was substantially better than that for spectral power–based EEG features extracted from resting-state EEG data, with most classifiers achieving high recall and precision. AUC values ranged from 0.477 ± 0.091 (SVM) to 0.750 ± 0.048 (AdaBoost), with six models exceeding 0.64 (see Table \ref{table:ml_results_sst}).

Accuracy was high across all classifiers except Gaussian Naïve Bayes (GNB; 0.631 ± 0.019), ranging from 0.786 ± 0.035 (decision tree; DT) to 0.837 ± 0.020 (AdaBoost). Sensitivity to ADHD cases was consistently strong, with recall values above 0.96 for all models except DT (0.871 ± 0.038) and GNB (0.633 ± 0.021). Precision was also robust, spanning from 0.842 ± 0.003 (k-nearest neighbors; KNN) to 0.906 ± 0.012 (GNB).

F1-scores exceeded 0.90 for most classifiers, peaking at 0.909 ± 0.010 for gradient boosting (GB) and KNN, closely followed by LR, RF, and AdaBoost (0.907–0.908). While GNB achieved the highest precision (0.906 ± 0.012), its reduced recall led to a markedly lower F1-score (0.737 ± 0.017). Among all methods, AdaBoost achieved the highest AUC while maintaining balanced sensitivity and precision.
\begin{table}[ht]
\centering
\small % or \footnotesize, \scriptsize

\caption[Classification performance for ADHD versus healthy controls using stop signal EEG data.]{Classification performance for ADHD versus healthy controls using stop signal EEG data.
  Performance of nine classifiers: logistic regression (LR), support vector machine (SVM), random forest (RF), gradient boosting (GB), extremely randomized trees (ET), AdaBoost (AdaB), decision tree (DT), \(k\)-nearest neighbors (KNN), and Gaussian naïve Bayes (GNB). Metrics are mean \(\pm\) s.d. across cross-validation folds: area under the ROC curve (AUC), accuracy, recall (sensitivity), precision (positive predictive value) and F1 (harmonic mean of precision and recall). ADHD was treated as the positive class.}
  \label{table:ml_results_sst}
  \resizebox{\textwidth}{!}{

\begin{tabular}{cccccc}
\toprule
Model & AUC & Accuracy & Recall & Precision & F1 \\
\midrule
LR & 0.691 ± 0.057 & 0.833 ± 0.017 & 0.971 ± 0.023 & 0.853 ± 0.011 & 0.907 ± 0.011 \\
SVM & 0.477 ± 0.091 & 0.833 ± 0.015 & 0.980 ± 0.022 & 0.847 ± 0.008 & 0.908 ± 0.010 \\
RF & 0.640 ± 0.045 & 0.833 ± 0.013 & 0.969 ± 0.012 & 0.854 ± 0.008 & 0.907 ± 0.008 \\
GB & 0.672 ± 0.062 & 0.835 ± 0.016 & 0.980 ± 0.021 & 0.849 ± 0.009 & 0.909 ± 0.010 \\
ET & 0.652 ± 0.046 & 0.830 ± 0.011 & 0.976 ± 0.010 & 0.846 ± 0.007 & 0.906 ± 0.006 \\
\textbf{AdaB} & \textbf{0.750 ± 0.048} & \textbf{0.837 ± 0.020} & \textbf{0.960 ± 0.019} & \textbf{0.864 ± 0.012} & \textbf{0.908 ± 0.012} \\
DT & 0.613 ± 0.058 & 0.786 ± 0.035 & 0.871 ± 0.038 & 0.877 ± 0.019 & 0.871 ± 0.024 \\
KNN & 0.581 ± 0.053 & 0.833 ± 0.013 & 0.987 ± 0.015 & 0.842 ± 0.003 & 0.909 ± 0.008 \\
GNB & 0.676 ± 0.025 & 0.631 ± 0.019 & 0.633 ± 0.021 & 0.906 ± 0.012 & 0.737 ± 0.017 \\
\bottomrule
\end{tabular}}
\end{table}

\subsubsection{Statistical analysis on model performance}
Based on the classification results from both tasks, we conducted several statistical tests to validate the statistical significance of the ML model performance. In the resting-state EEG classification task, a permutation test with 1,000 iterations yielded a Monte Carlo $p = 0.064$ (Figure \ref{fig:model_comparison}.A, middle). Classification with true labels achieved significantly greater performance than chance ($t = 124.99$, $p < 0.001$). Moreover, true-label performance was significantly higher than that obtained with permuted labels ($t = 42.36$, $p < 0.001$), indicating that the classifier captured meaningful structure in the data. In contrast, classification with permuted labels did not differ from chance ($t = 1.32$, $p = 0.19$), validating the permutation control (Figure \ref{fig:model_comparison}.A, right). In the stop-signal EEG classification task, a permutation test with 1,000 iterations yielded a Monte Carlo $p = 0.010$ (Figure \ref{fig:model_comparison}.B, middle). Classification with true labels achieved significantly greater performance than chance ($t = 164.5$, $p < 0.001$). Moreover, true-label performance was significantly higher than that obtained with permuted labels ($t = 67.98$, $p < 0.001$), indicating that the classifier captured meaningful structure in the data. In contrast, classification with permuted labels did not differ from chance ($t = -1.41$, $p = 0.14$), validating the permutation control (Figure \ref{fig:model_comparison}.B, right). 

We conducted pairwise t-tests to compare classification performance across models trained on resting-state versus stop-signal task EEG data. Across all comparisons, models trained with stop-signal task EEG exhibited significantly higher AUCs than those trained on resting-state EEG (all FDR-corrected $p < 0.001$, see Figure \ref{fig:model_comparison}.C). This pattern was consistent across model architectures (all nine trained models), indicating that task-related neural dynamics provided more predictive power for classification than resting-state activity.

\subsection{Feature importance}
\begin{figure}[!h]
    \centering
    \includegraphics[width=1\linewidth]{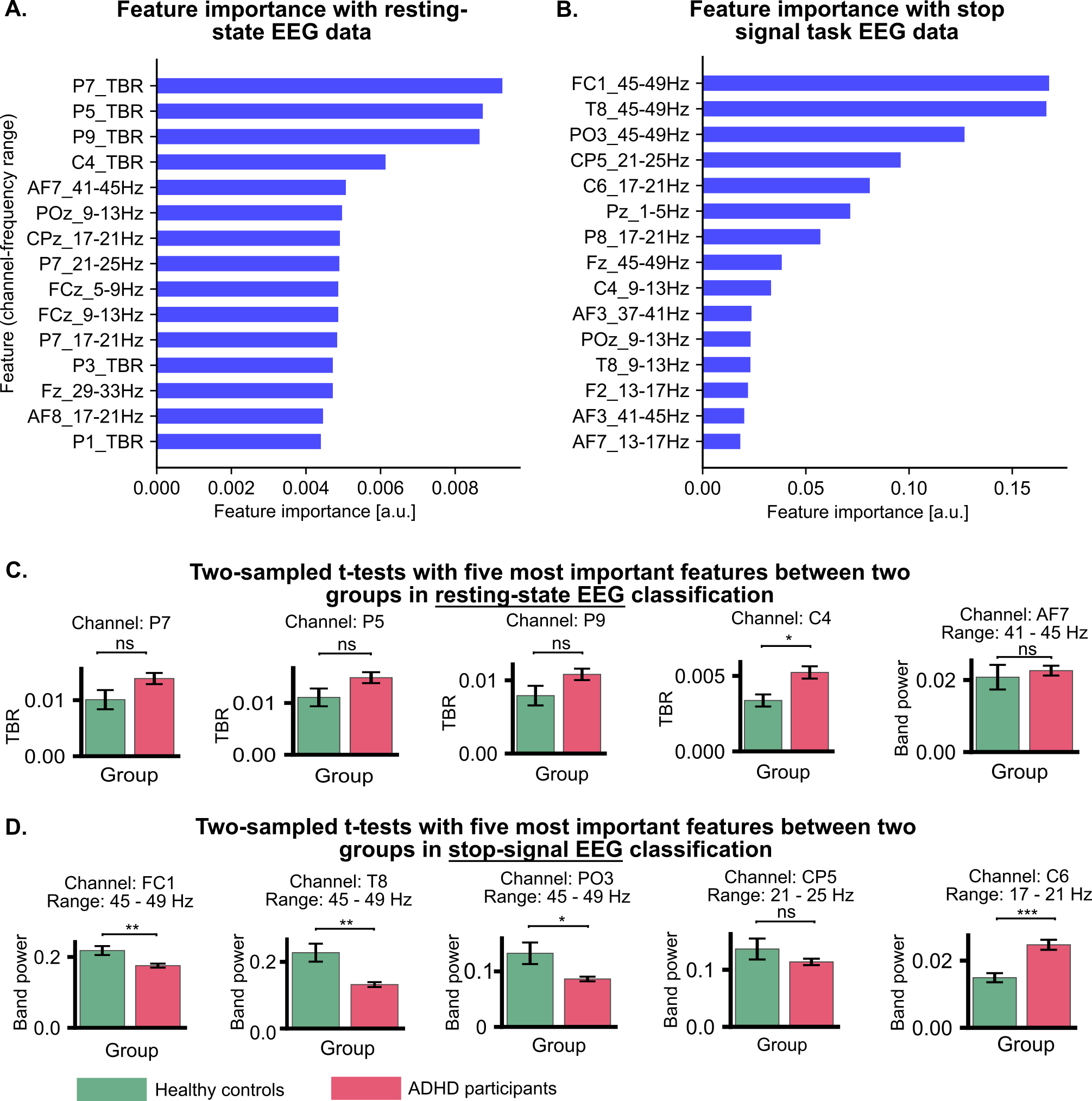}
    \caption[Feature importance of best classifiers with resting-state and stop signal task EEG data]{\textbf{Top most important EEG features for machine learning evaluated with resting-state and stop-signal task EEG data. }Permutation-based feature importance analysis revealed distinct spectral–spatial profiles across paradigms. A. In the resting-state condition, the highest-ranking predictors were theta/beta ratio (TBR) measures over parietal and central electrodes, with additional contributions from occipito-parietal beta- and alpha-band power. B. During the stop-signal task, model performance was driven primarily by relative gamma-band (45–49 Hz) power over fronto-central, temporal, and parietal sites, with secondary contributions from beta- and alpha-band activity. C. Group-level comparisons (independent-samples t-tests, FDR-corrected) of the top five resting-state features revealed a significant increase in C4 TBR in ADHD relative to controls, while parietal TBR features (P7, P5, P9) showed consistent but nonsignificant trends toward higher values in ADHD. D. For the stop-signal task, FDR-corrected tests demonstrated significantly reduced relative gamma-band power in ADHD at FC1, T8, and PO3, together with elevated relative mid-beta power at C6, whereas CP5 did not differ reliably between groups. 
}
    
    \label{fig:chapter3-feature-importance}
\end{figure}

We quantified feature importance for the two best machine learning models with two EEG datasets — resting-state and the stop-signal task — using permutation-based estimates to identify the spectral–spatial markers most predictive of model output (see Figure \ref{fig:chapter3-feature-importance}). To further explore the group-level differences of these features, we conducted independent two-samples t-tests on the top five features in each condition, with false discovery rate (FDR) correction for multiple comparisons. 

In the resting-state condition, the most influential features were predominantly theta/beta ratio (TBR) measures over parietal and central electrodes, with \texttt{P7\_TBR} emerging as the strongest contributor, followed closely by \texttt{P5\_TBR} and \texttt{P9\_TBR}. Additional high-ranking predictors included \texttt{C4\_TBR} and narrowband power features in the higher frequency ranges, such as \texttt{AF7\_41-45~Hz} and \texttt{POz\_9-13~Hz}. Notably, multiple occipito-parietal and midline sites contributed features in the beta range (17--25~Hz) and low alpha (5-13~Hz), including \texttt{CPz\_17-21~Hz}, \texttt{P7\_21-25~Hz}, \texttt{FCz\_5-9~Hz}, and \texttt{FCz\_9-13~Hz}. With t-tests, only C4 TBR remained significantly elevated in ADHD relative to controls after correction (p = 0.010, d = 0.54), whereas parietal TBR measures (P7, P5, P9) showed consistent but nonsignificant trends toward higher values in ADHD (all corrected $p \approx 0.09$).

In contrast, during the stop-signal task, the feature importance profile shifted markedly toward high-frequency (45-49~Hz) activity over fronto-central, temporal, and parietal sites. The most important feature was \texttt{FC1\_45-49~Hz}, followed by \texttt{T8\_45-49~Hz} and \texttt{PO3\_45-49~Hz}. Mid-beta activity also emerged as a key contributor, with \texttt{CP5\_21-25~Hz} and \texttt{C6\_17-21~Hz} among the top five. Lower-frequency activity played a smaller but notable role, including \texttt{Pz\_1-5~Hz} and \texttt{P8\_17-21~Hz}. Additional high-frequency predictors included \texttt{Fz\_45-49~Hz} and \texttt{AF3\_37-41~Hz}, along with midline alpha-band activity at \texttt{C4\_9-13~Hz}. FDR-corrected t-tests confirmed robust group differences for several of these features: ADHD participants showed significantly reduced gamma-band activity at FC1 at 45 - 49 Hz (p = 0.010, d = –0.83), T8 at 45 - 49 Hz (p = 0.0075, d = –1.31), and PO3 at 45 - 49 Hz (p = 0.043, d = –1.03), alongside significantly elevated mid-beta power at C6 ($p < 0.001$, d = 0.78). CP5 at range 21–25 Hz did not differ reliably between groups (p = 0.26). 

Together, these results indicate that while resting-state prediction relied heavily on parietal and central TBR measures, of which only C4 TBR was reliably altered in ADHD, successful decoding in the stop-signal task was dominated by gamma-band activity localized to fronto-central, temporal, and parietal cortices, with several of these high-frequency features also exhibiting large, statistically reliable group differences.

\section{Discussion}
\subsubsection{Superior performance of task-based EEG classification in adult ADHD}
The present chapter compared the classification performance of ML models trained on resting-state and task-based EEG data for classifying adults with ADHD versus neurotypical controls using an identical ML pipeline based on spectral power features. Classification performance using stop-signal task EEG data reliably outperformed that based on resting-state EEG, with all nine classifiers trained on stop signal task EEG achieving significantly higher AUCs (all FDR-corrected $p < 0.001$; Figure \ref{fig:model_comparison}.C). Resting-state EEG models achieved only modest performance, with AUCs close to but significantly above chance (see Figure \ref{fig:model_comparison}.A) whereas ML models trained on stop signal task EEG achieved high AUC, recall and precision, indicating robust classification between ADHD and control participants (see Figure \ref{fig:model_comparison}.B). 

This finding supports the view that inhibitory control deficits represent a core neurocognitive phenotype of ADHD (\citealp{schachar1993inhibitory, barkley1997behavioral}). The stop signal task, which requires participants to rapidly inhibit a motor response when presented with a stop signal, directly probes this fundamental capacity for response inhibition. A recent meta-analysis further confirmed inhibitory control deficits as a robust biomarker, showing that individuals with ADHD exhibit prolonged stop signal reaction times and higher omission error rates (\citealp{senkowski2024assessing}). Additionally, these findings are consistent with reports of altered executive functions associated with inhibitory control deficits in ADHD across multiple independent studies (\citealp{hadas2021right}). These investigations have reported widespread abnormalities in the neural networks involving executive control in ADHD, including alterations in prefrontal cortex activation (\citealp{arnsten2009emerging}), fronto-striatal connectivity (\citealp{konrad2010adhd}), and the coordination of distributed brain networks during cognitively demanding tasks (\citealp{firouzabadi2022neuroimaging}). The convergence of evidence from diverse methodological approaches, including functional MRI, EEG, and behavioural measures, supports the framework that executive control dysfunction, particularly in the domain of inhibitory control, represents a robust feature of ADHD neurobiology. 

\subsubsection{Modest performance of resting-state EEG classification in adult ADHD}
By contrast, the classification performance of resting-state EEG in our dataset is relatively modest with the best AUCs around 0.6. Although TBRs emerge as the most significant predictors in our feature importance analysis (see Figure \ref{fig:chapter3-feature-importance}.A), they are not significantly different between the two groups in our two-sample t-tests see Figure \ref{fig:chapter3-feature-importance}.C). This suggests that even TBR-related features may appear influential within multivariate models, their actual group-level differences are too subtle and inconsistent to support reliable classification, resulting in relatively weak classification performance. This finding aligns with recent work questioning the robustness of resting-state TBR as a diagnostic biomarker in adults (\citealp{arns2013decade, picken2020theta, loo2012clinical}). Although TBR has been widely studied and considered as a potential biomarker in ADHD diagnosis (\citealp{monastra1999asessing, snyder2006meta}), several work has proposed concerns about the reliability of TBR in adult ADHD (\citealp{kiiski2020eeg, newson2019eeg}), potentially reflecting age-related variability. The developmental trajectory of spectral characteristics is complex, with substantial changes occurring from childhood through adolescence and into adulthood (\citealp{anderson2018developmental}). Theta power decreases with age as the brain matures (\citealp{cummins2007theta}), while beta power shows more variable developmental patterns (\citealp{poil2014age}). Consequently, the TBR may not maintain the same diagnostic sensitivity and specificity in adult populations. This age-related variability represents a significant limitation for the clinical utility of resting-state TBR as a diagnostic biomarker across the lifespan (\citealp{poil2014age}). 

Moreover, resting-state EEG measures are inherently susceptible to fluctuations in arousal, attention, and vigilance that are difficult to control experimentally. Although participants were instructed to remain awake with eyes open or closed, their cognitive state can drift substantially within and across individuals. Some participants may engage in spontaneous mind-wandering, others may struggle to maintain wakefulness, and still others may actively attempt to focus their attention despite the absence of an explicit task. This heterogeneity in cognitive states during ``rest" may introduce substantial noise into resting-state EEG measurements, potentially obscuring group differences between ADHD and control populations. In contrast, task-based EEG recordings impose a more standardised cognitive context with an explicit task, reducing state-related variability and enhancing the signal-to-noise ratio for detecting ADHD-related neural abnormalities.

\subsubsection{Reduced gamma band power and motivation-related circuits}
Permutation-based feature importance analyses further indicated that task-based classification performance was strongly driven by gamma-band (45–49 Hz) power over fronto-central, temporal and parietal electrodes (Figure \ref{fig:chapter3-feature-importance}.B) and group-level comparison showed significant reductions in these gamma bands in individuals with ADHD compared to healthy controls (Figure \ref{fig:chapter3-feature-importance}.D). Early studies have reported higher gamma activity in healthy subjects when visual and auditory target stimuli were evoked since these stimuli were processed more attentively in healthy controls compared to ADHD participants (\citealp{herrmann2001human, debener2003top}). Gamma oscillations at frequencies higher than 30 Hz have been reported to involve with a wide range of cognitive processes (\citealp{herrmann2004cognitive}) including attentional selection (\citealp{muller2000modulation, sugiyama2024influence, leistiko2024effects}). Reductions in gamma-band power may be interpreted as reflecting difficulties in sustaining attention, as diminished high frequency engagement may signal reduced capacity to maintain stable task-focused processing. 

ADHD is characterised not only by core symptoms of inattention and hyperactivity but also by significant motivation deficits that impair goal-directed behaviour and reward processing (\citealp{cubillo2012review}). ADHD also involves with dysfunction of the dopamine reward pathway, with neuroimaging studies revealing reduced dopamine receptors and transporters in the nucleus accumbens and midbrain dopamine regions (\citealp{volkow2011motivation}). Neurophysiological studies have reported alterations in high-gamma activity in ADHD, including reduced frontal-central high-gamma power and increased posterior high-gamma power (\citealp{dor2021high}) and reduced gamma power in frontal-central regions in adults with ADHD (\citealp{tombor2019decreased}). Interestingly, animal studies showed that the ventral striatum and nucleus accumbens -- key regions for motivation and reward processing -- exhibit distinct gamma-band oscillations (gamma-50 and gamma-80) whose power is differentially modulated during reward delivery and approach behaviour (\citealp{van2009covert, berke2009fast}). These collections of studies indicate that reduced gamma band power may reflect disruptions in dopamine-mediated reward circuits, potentially underlying motivation deficits in ADHD. 
% limitations
\subsubsection{Methodological considerations and future work}
Several methodological considerations of the present study should be acknowledged. First, the relatively small sample size limits the generalisability of our findings and raises the possibility of overfitting in the classification models. Second, we did not conduct event-related potential (ERP) analyses, which could have provided additional insights into the temporal dynamics of inhibitory control processes in ADHD, such as alterations in N200 or P300 components that have been consistently reported in stop-signal and go/no-go paradigms. Incorporating ERP measures in future work would not only strengthen the interpretability of the machine learning results but also help clarify the neural mechanisms underlying the observed group differences. Future work could also extend these analyses to EEG datasets acquired across a broader range of cognitive tasks beyond the stop-signal task, enabling assessment of whether the observed effects generalise across different cognitive tasks.

%\textbf{Feature contribution}
A central conceptual issue concerns the extent to which EEG-based classification provides added value beyond behavioural measures alone. Behavioural performance on paradigms such as the stop-signal task reliably differentiates individuals with ADHD from neurotypical controls—for example, through prolonged stop-signal reaction times and increased intra-individual reaction-time variability (\citealp{lijffijt2005meta, lipszyc2010inhibitory}). However, reaction times and error rates represent the end product of multiple interacting neural processes and are strongly modulated by strategy, motivation, fatigue, and moment-to-moment fluctuations in attention. EEG, by contrast, offers a temporally precise window into the neural dynamics that give rise to behaviour, capturing processes such as inhibitory control, conflict monitoring, and motor preparation. These neural signatures can reveal latent dysfunction even when overt behaviour is partially compensated. Consequently, EEG markers may provide complementary mechanistic insights, enable the identification of neurophysiological subtypes, and ultimately support prognosis and treatment stratification in ways that purely behavioural or questionnaire-based assessments cannot.
%\textbf{Ablation study}

Taken together, our findings indicate that EEG acquired during an inhibitory control task provides substantially stronger diagnostic signal for adult ADHD than traditional resting-state spectral measures, and that this advantage is driven in large part by gamma-band and related oscillatory activity over fronto-central and parietal regions. These task-evoked neural markers likely reflect the efficacy with which executive control networks are recruited and coordinated under demanding conditions, and thus tap into a core neurocognitive impairment in ADHD. While behavioural measures alone remain informative and easier to obtain, EEG-based classification offers complementary mechanistic insights that could ultimately inform treatment selection and monitoring. Future work should extend the present approach by: (i) evaluating additional cognitive tasks to determine whether other paradigms similarly enhance classification performance; (ii) integrating ERPs, spectral features, behavioural measures and computational parameters within multimodal ML models; and (iii) testing whether task-based EEG markers predict clinically relevant outcomes such as treatment response or functional impairment. Such developments may bring the field closer to reliable, clinically actionable neurophysiological biomarkers for ADHD.
\section{Conclusion}
In summary, our study demonstrates that EEG collected with the stop-signal task provides superior classification performance for ADHD compared to resting-state EEG. This finding supports the view that ADHD-related neural dysfunction is most evident when executive control networks are actively involved. Feature analyses further revealed that stop-signal task classification relied heavily on gamma-band oscillations, particularly over fronto-central and parietal regions. These results suggest that resting-state features such as TBR may capture generalised cortical slowing, whereas task-based gamma power reflects more specific deficits in inhibitory control and executive functioning in ADHD.

Our findings highlight the potential of task-based EEG as a more effective biomarker source for ADHD diagnosis than traditional resting state measures. Future work should expand the task paradigms to determine whether other cognitive challenges similarly enhance classification performance. Incorporating event-related potentials alongside spectral features could provide complementary insight into the temporal dynamics of inhibitory control. Moreover, exploring advanced machine learning approaches, including deep neural networks and automated model selection (AutoML), may further optimize predictive accuracy. Finally, integrating task-based EEG measures in multimodal data (e.g., behavioural, neuroimaging, and genetic measures) may offer a path toward more robust, clinically actionable diagnostic tools for ADHD. In the next chapter, we switch our study from hyperactivity to hypoactivity aspect by investigating neuroimaging data of individuals with subclinical apathy.

\chapter{White matter integrity and computational modelling of effort and reward sensitivity}
Apathy is a common neuropsychiatric syndrome, characterised by changes in how individuals value rewards and the effort needed to obtain them. Understanding the neural basis underlying individual differences in effort and reward sensitivity is essential to explain disruptions in motivated behaviour. In this chapter (currently in major revision in the NeuroImage journal at the time of writing this thesis), I would like to examine the white matter correlates of individual differences in effort and reward sensitivity. From rodents to humans, animals constantly face a central question: is the reward worth the effort? Effort and reward sensitivity in such situations vary substantially across individuals and ultimately shape goal-directed behavior. Yet, the neural mechanisms underlying this individual differences  remain poorly understood. Here, we combined computational modelling of effort and reward sensitivity during decision-making with whole-brain diffusion MRI in 45 healthy participants to identify the white matter substrates of individual sensitivity. A data-driven, cluster-based analysis of fractional anisotropy and mean diffusivity revealed 12 clusters: five linked to effort sensitivity, all within tracts connected to major frontal valuation nodes (e.g., supplementary motor area [SMA], dorsal anterior cingulate cortex [dACC], orbitofrontal cortex [OFC]), and seven linked to reward sensitivity, spanning frontal valuation, fronto-parietal, and sensorimotor networks. The strongest associations involved two SMA-connected clusters: one shared across effort and reward sensitivity, and another consistent across both microstructural measures. Critically, microstructural measures from the five effort-related and seven reward-related clusters reliably predicted individual effort and reward sensitivity in out-of-sample machine learning analyses, respectively, whereas randomly sampled clusters did not. SMA-connected tracts were the dominant predictors in these decoding analyses, with additional contributions from fronto-parietal and sensorimotor pathways for reward sensitivity. These findings reveal a distributed white matter architecture underlying inter-individual differences in effort and reward sensitivity, with SMA pathways emerging as central hubs. They demonstrate that localized white matter microstructure can robustly predict these individual differences, offering a framework to forecast the impact of lesions or interventions on goal-directed behavior, including apathy and impulsivity. 

\section{Introduction}
Across species from rodents to humans, animals constantly confront a fundamental question: is the reward worth the effort? Everyday choices, such as choosing between doing chores or relaxing, cooking or ordering takeout, rely on evaluating effort versus reward (\citealp{hogan2020neural}). The motivation to act in these contexts depends on how sensitive an individual is to both factors. When these sensitivities become abnormal, goal-directed behaviour can break down: effort hypersensitivity leads to inaction despite potential rewards, contributing to apathy  (\citealp{husain2018neuroscience, le2018anatomy, costello2024apathy}), while reward hypersensitivity can drive action despite effort costs, fuelling impulsivity (\citealp{long2022altered, luijten2017disruption}). Even in healthy individuals, these sensitivities differ considerably (\citealp{bonnelle2016individual, fuentes2016characterizing}), influencing how people interact with their environment. Such inter-individual differences, often described as ``computational phenotypes” (\citealp{pessiglione2018not, schurr2024dynamic}), can be quantified using formal models. Yet, the neural circuits underlying this variability remain poorly understood. Revealing its anatomical basis could explain why people differ in their action tendencies and offer a baseline for identifying pathological brain--behaviour changes (\citealp{thiebaut2022emergent}), providing a framework to predict how structural brain damage might affect goal-directed behaviour.

To date, most insights into how the brain processes effort and reward come from functional neuroimaging studies examining activity across varying levels of effort and reward during decision-making tasks (\citealp{husain2018neuroscience}). These studies consistently highlight a core fronto-striatal network with partially dissociable roles: the supplementary motor area (SMA) primarily encodes effort (\citealp{bonnelle2016individual}), while the orbitofrontal cortex (OFC) tracks reward valuation (\citealp{klein2022medial}), suggesting their respective involvement in effort and reward computations. Regions including the dorsal anterior cingulate cortex (dACC; \citealp{shenhav2013expected}) and the nucleus accumbens (NAcc; \citealp{suzuki2021distinct}), encode both effort and reward, integrating them into net value signals. Although the fronto-striatal circuit forms the central hub for effort–reward evaluation, additional evidence points to the contribution of broader fronto-parietal (\citealp{etzel2016reward}) and sensorimotor structures including the primary motor cortex (M1; \citealp{derosiere2025reward}) and cerebellum (\citealp{kostadinov2022reward}) also exhibit reward-related activity during decision-making. Together, these findings indicate that evaluating effort and reward involves distributed computations across cortical and subcortical systems, anchored by the fronto-striatal network.

What remains poorly understood is how individual differences in effort and reward sensitivity arise from variability across this distributed system. While previous work has largely focused on regional activity, emerging evidence shows that white matter pathways do more than passively transmit information between regions; they actively shape communication by amplifying or dampening neural signals, thereby influencing cognition and behavior(\citealp{innocenti2022functional, thiebaut2022emergent}). Since effort and reward computations involve coordinated activity across multiple interconnected regions, inter-individual differences in structural connectivity may critically determine how these signals are integrated and acted upon—-a question that remains largely unexplored. Preliminary support for this idea comes from studies linking subclinical apathy in healthy individuals, often associated with effort hypersensitivity (\citealp{bonnelle2016individual}), to alterations in the anterior cingulum, a major tract connecting the SMA, OFC and dACC. More recently, reduced structural connectivity between SMA and NAcc has been found in individuals with increased sensitivity to effort (\citealp{derosiere2025fronto}), again implicating disrupted communication between key valuation hubs. Together, these findings suggest that white matter integrity plays an important role in shaping sensitivity to effort and reward. However, most previous studies have relied on tract-of-interest analyses, focusing on a limited set of predefined tracts. Given the extensive spatial distribution of value-related activity, there is a pressing need for whole-brain, data-driven approaches to uncover the broader structural architecture underlying individual differences in effort and reward sensitivity. 

Recent advances in diffusion imaging provide an opportunity to address this gap. Whole-brain voxel-wise metrics such as fractional anisotropy (FA) and mean diffusivity (MD) offer complementary, data-driven measures of white matter microstructural integrity (\citealp{beck2021white, song2025characterizing}). In this study, we combined computational modelling of effort and reward sensitivity in a decision making task (\citealp{gilmour2023neuropsychiatric, morris2025decision}) with diffusion imaging in 45 healthy participants to determine which brain circuits exhibit microstructural differences associated with interindividual variations in behavioral sensitivity profiles. Using a voxel-wise, cluster-based approach unconstrainted by anatomical priors, we identified multiple white matter regions where microstructure integrity strongly covaries with effort and reward sensitivity. Interestingly, these include major tracts connecting key nodes of the frontal-striatal network, as well as some fronto-parietal and sensorimotor areas. We then used microstructural metrics from these white matter regions in machine learning classifiers, demonstrating that these features can reliably predict individual phenotypes of effort and reward sensitivity.
\section{Methods}
\begin{figure}
    \centering
    \includegraphics[width=1\linewidth]{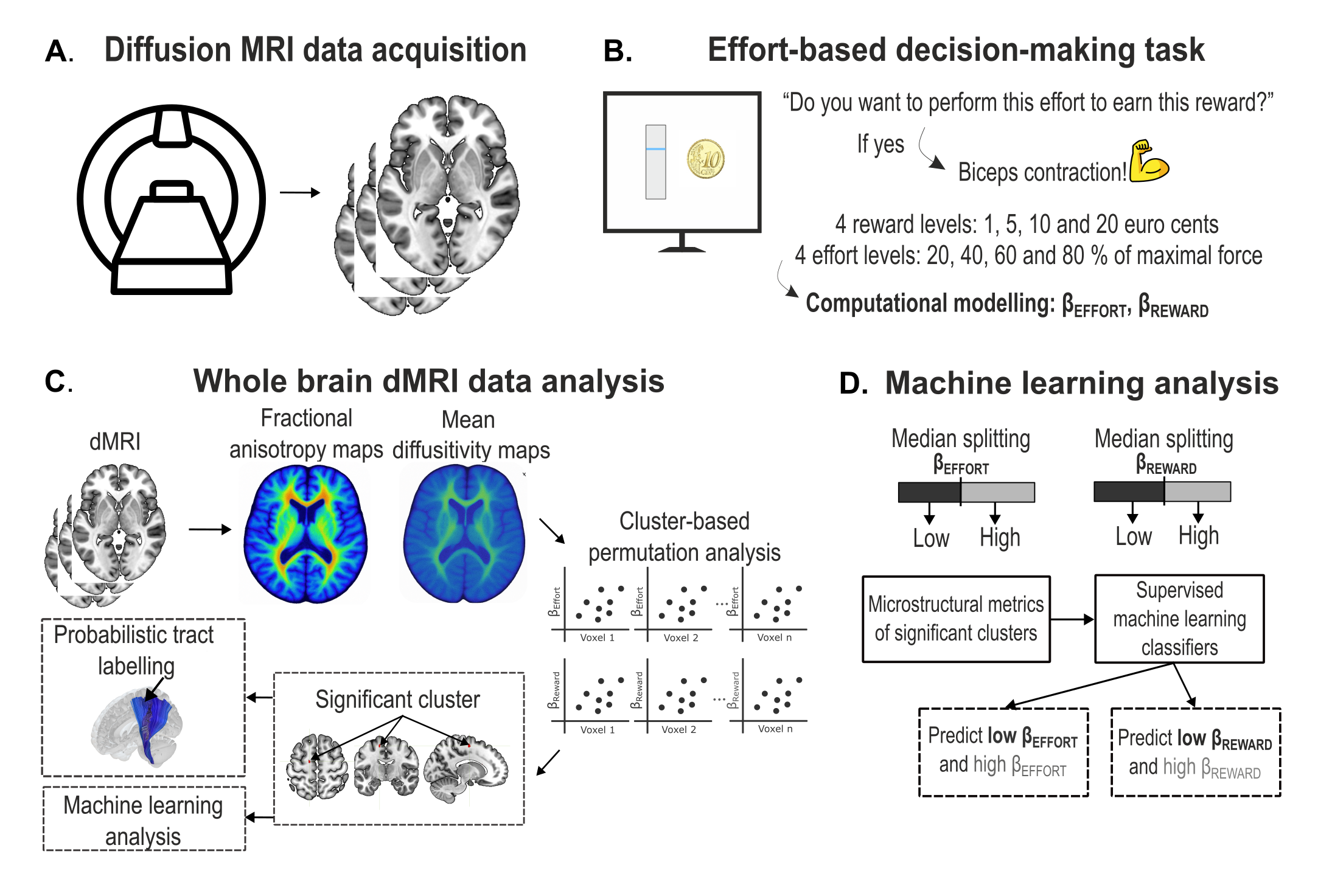}
    \caption[Experimental protocol and data analyses]{\textbf{Experimental protocol and data analyses.} A. Diffusion MRI (dMRI) data acquisition. dMRI data were acquired to assess white matter microstructure, including FA and MD. B. Effort-based decision-making task and computational modelling of behaviour. Participants completed an effort-based decision-making task, choosing whether to perform biceps contractions of varying effort levels (20\%, 40\%, 60\%, or 80\% of maximal voluntary contraction) to obtain varying monetary rewards (1, 5, 10, or 20 euro cents). Computational modelling of their acceptance rates in this task, yielded individual effort and reward sensitivity parameters ($\beta_{\text{Effort}}$ and $\beta_{\text{Reward}}$, respectively). C. Whole-brain dMRI data analysis. dMRI data were exploited to compute voxel-wise FA and MD quantification in each subject (top left panel). We then applied a cluster-based analysis to identify significant clusters where FA and MD covaried with $\beta_{\text{Effort}}$ or $\beta_{\text{Reward}}$ (left and bottom panels). Significant clusters were mapped onto white matter tracts and characterized by MNI coordinates and their metrics were used in a predictive machine learning analysis (right panel; see D.). D. Machine learning analysis. Identified microstructural features were used as inputs for machine learning classifiers predicting low and high $\beta_{\text{Effort}}$ and $\beta_{\text{Reward}}$. Classifier performance was assessed using 5-fold cross-validation, yielding average accuracy and area under the receiving operator curve metrics.}
    \label{fig:chapter4-method}
\end{figure}
\subsection{Participants}
Fifty healthy adult participants were initially recruited from the Research Participant Pool of the Institute of Neuroscience at the Université catholique de Louvain (Brussels, Belgium). Three participants did not complete the full experimental protocol, and diffusion MRI data from two participants were excluded due to corruption during acquisition. The final sample included 45 healthy individuals (mean age = 25.1 ± 0.8 years; 31 females, 14 males), whose data were included in all subsequent analyses. All subjects were right-handed according to the Edinburgh Questionnaire (\citealp{oldfield1971assessment}) and reported no history of neurological or psychiatric disorders, substance use, or medications that could affect performance. Participants received monetary compensation for their participation and could obtain additional rewards based on their task performance (see Task description section below). All procedures were approved by the institutional review board of UCLouvain, and written informed consent was obtained from all participants in accordance with the Declaration of Helsinki.

A previous study using this dataset investigated structural connectivity within a restricted set of motor-related circuits, combining streamline-based tractography with effective connectivity estimates derived from transcranial magnetic stimulation, to examine their relationship with apathy and decision-making (\citealp{derosiere2025fronto}). The present study addresses a complementary question, employing voxel-wise microstructural indices of white matter integrity (fractional anisotropy and mean diffusivity) in a whole-brain, data-driven framework to identify anatomical substrates underlying inter-individual variability in effort and reward sensitivity.

\subsection{Data acquisition}
Data were collected across two sessions separated by a minimum of 24 hours and a maximum of one week. During the first session, MRI scans were acquired at the Saint-Luc University Hospitals (Brussels, Belgium). The second session, conducted at the Institute of Neuroscience, Université catholique de Louvain (Brussels, Belgium), involved behavioral testing using a well-established effort-based decision-making task that enables computational modeling of effort and reward sensitivity (\citealp{morris2025decision, gilmour2023neuropsychiatric, pessiglione2018not}). Detailed procedures for each session are described below. 

\subsubsection{MRI data acquisition} 
Structural T1-weighted and diffusion-weighted MRI data were acquired for each participant on a 3 Tesla MRI (SIGNA\textsubscript{TM} Premier, General Electric), equipped with a 48-channel head coil (Figure \ref{fig:chapter4-method}.A). High resolution 3D T1-weighted anatomical images were obtained with the following parameters:  Echo Time (TE) = 2.96 ms, Repetition Time (TR) = 2238.93 ms, Inversion Time TI = 900 ms, 170 slices, slice thickness = 1 mm, in-plane FOV = 256 × 256 mm², matrix = 256 × 256; voxel size = 1 mm³ isotropic. Diffusion-weighted MRI (DWIs) were acquired in the axial plane with the parameters: TR = 7289 ms, TE = 57.1 ms, 70 slices, slice thickness = 2 mm, in-plane FOV = 220 × 220 mm², matrix size = 110 × 110; 2 mm isotropic voxels, with 64 gradients at b = 1000 s/mm² and one b0 reference image.

\subsubsection{Behavioural data acquisition}
Participants performed an effort-based decision-making task on a computer while seated in an ergonomic chair positioned 100 cm from a monitor (refresh rate: 100 Hz). The right arm was flexed to 90 degrees and stabilised on an armrest, with participants gripping a custom-designed handle using the right hand. The left hand remained free and was used to make choice responses via the left and right arrow keys on a standard keyboard. To minimize movement artifacts and ensure consistent biomechanical positioning during force exertion, the right forearm was secured to the armrest using an adjustable strap.

\textbf{Effort-based decision-making (EBDM) task: } The EBDM task was implemented in MATLAB (MathWorks) using Psychtoolbox (\citealp{brainard1997psychophysics}). On each trial, participants were required to decide whether to exert varying levels of physical effort in exchange for different monetary rewards (Figure \ref{fig:chapter4-method}). Effort was defined as an isometric contraction of the right biceps, elicited by instructing participants to attempt to flex their arm, bringing the fist toward the shoulder, while maintaining elbow contact with the armrest and gripping a fixed handle. The intensity of muscular contraction was continuously monitored via real-time surface electromyography. Monetary rewards were displayed in euro cents, and participants were informed that their total earnings would be paid at the end of the session.

Each trial started with the display of a vertical force gauge accompanied by a horizontal reference line indicating the required force level for that trial. Adjacent to the gauge, the monetary reward was displayed. The full scale of the gauge represented 100\% of the participant’s maximal voluntary contraction (MVC), which was individually calibrated prior to the experimental blocks (see Blocks of trials and MVC below, for details). Effort levels were set at 20\%, 40\%, 60\%, or 80\% of MVC, paired with potential rewards of 1, 5, 10, or 20 euro cents, resulting in 16 possible effort-reward combinations. Participants had 5 seconds to decide whether to accept or reject each offer, responding with their left index and middle fingers by pressing the right arrow key to accept or the left arrow key to reject.

Upon acceptance of an offer, a "Go" signal was presented on the screen following a variable delay of 1.0 to 1.2 seconds, marking the onset of the contraction period. During this period, participants received continuous visual biofeedback of their exertion level, depicted as a dynamic filling of the gauge. The extent of gauge filling was proportional to the rectified amplitude of the EMG signal recorded from the right biceps. Participants were instructed to maintain a contraction at or above the target force level — corresponding to 20\%, 40\%, 60\%, or 80\% of their MVC — for a minimum duration of 2.7 seconds within a 4-second window (i.e., $>66\%$ of the contraction period).
At the end of each trial, participants received visuo-auditory feedback indicating the outcome of the trial (success or failure) and whether the corresponding reward had been earned. Trials in which muscle activity was detected prior to the onset of the "Go" signal were immediately aborted, and an "Anticipated" message was displayed on the screen. Regardless of the outcome, a fixed 2-second inter-trial interval followed each trial to allow for recovery.
If participants declined the offer, the trial terminated immediately and was followed by a 2-second blank screen before the onset of the next trial (i.e., corresponding to the inter-trial interval). Similarly, if no response was recorded within the initial 5-second decision window, the trial was considered a missed response, and the subsequent trial started after the standard inter-trial interval.
\textbf{Blocks of trials and MVC: }
Participants first completed an MVC assessment to calibrate the target force for the EBDM task. During this procedure, they were verbally encouraged to generate maximal elbow flexion for 5 s across three trials, each separated by 30 s of rest. MVC was defined as the highest force achieved in any trial, quantified as the mean force produced between the 2nd and 4th second of the contraction.

Participants then completed a 16-trial practice block to familiarise themselves with the task and to experience the force levels corresponding to the different percentages of their MVC. After practicing blocks, they performed five experimental blocks of 32 trials, each comprising two repetitions of the 16 effort–reward conditions. This yielded a total of 160 trials, with 10 trials per condition. 

New values of MVC per participant were obtained 30 seconds after the end of each block. Participants then rested for 90 s before beginning the next block. In total, MVC was assessed six times: once prior to each of the five blocks and once following the final block. These repeated assessments allowed us to adjust the required force levels in each block to account for potential fatigue. Critically, participants were unaware that the force requirements were recalibrated based on their most recent MVC.

\subsection{Data analyses}
\subsubsection{Computational modelling of acceptance rates}
To quantify individual differences in effort and reward sensitivity, we applied computational modelling to the acceptance rates in the effort-based decision-making task. As such, for each participant, the model estimated the influence of effort and reward on offer valuation, which was translated into a probability to accept the offer and engage in the action. 
First, we tested a set of candidate models of subjective value computation informed by prior literature (\citealp{morris2025decision, gilmour2023neuropsychiatric, le2018anatomy, bonnelle2016individual}), evaluating model fits using Bayesian information criterion (BIC) minimization and visual inspection (Supplementary Figure \ref{fig:chapter4-suppfigure1}).
In the best-fitting model, the subjective value V of each offer was computed as:

\begin{math}
    V = \beta_{\text{Effort}} \times \text{EFFORT}_{\text{LEVEL}}^2 \times (1 + \beta_{\text{Time}} \times t) + \beta_{\text{Reward}} \times \text{REWARD}_{\text{LEVEL}} + \beta_0
\end{math}

Here, the subjective value of an offer was modelled as a quadratic function of the proposed $\text{EFFORT}_{\text{LEVEL}}$, and a linear function of the $\text{REWARD}_{\text{LEVEL}}$, weighted by individual parameters $\beta_{\text{Effort}}$ and $\beta_{\text{Reward}}$, reflecting effort and reward sensitivity, respectively (see below for more details). To account for potential time-dependent effects on acceptance rates across repeated trials, the model also incorporated a $\beta_{\text{Time}}$ parameter, which linearly modulated the cost component as a function of trial number (\citealp{le2018anatomy}). This parameter thus captured time-on-task effects on acceptance rates, whether reflecting fatigue (progressive decline in acceptance across trials) or habituation and rising motivation (progressive increase). The intercept $\beta_0$ captured baseline choice bias, reflecting the propensity to accept an offer with zero reward. 

After model fitting, we extracted individual $\beta_{\text{Effort}}$ and $\beta_{\text{Reward}}$ parameters as proxies for effort and reward sensitivity, respectively. Parameter estimation was performed using constrained nonlinear optimization (fmincon function in MATLAB 2022b, MathWorks) to minimize the negative log-likelihood of observed choices. Model-derived choice probabilities (range: 0 – 1) were compared against actual behavior.

In this framework, more negative $\beta_{\text{Effort}}$ values indicated stronger effort sensitivity, i.e. a steeper relative decline in acceptance rates with increasing effort. Due to the quadratic nature of the cost function in the winning model (i.e., $\beta_{\text{Effort}} \times \text{EFFORT}_{\text{LEVEL}}^2$), this effect is particularly pronounced at high effort levels: the more a participant's acceptance rate drops for high efforts (e.g., 80\% MVC), the steeper the curvature of the effort-sensitivity function and the more negative the $\beta_{\text{Effort}}$. For ease of interpretation in subsequent analyses, $\beta_{\text{Effort}}$ values were multiplied by $-1$, such that higher (more positive) values consistently reflected stronger effort sensitivity. Conversely, more positive $\beta_{\text{Reward}}$ values corresponded to stronger reward sensitivity, as reflected by steeper relative increase in acceptance with higher rewards. $\beta_{\text{Reward}}$ values were transformed using a log function to normalize the distribution, in accordance with former studies (\citealp{le2018anatomy, derosiere2025fronto}). Subject-specific $\beta_{\text{Effort}}$ and $\beta_{\text{Reward}}$ values were used as dependent variables in voxel-wise, cluster-based permutation analyses (see Statistical Analysis section). 

\subsubsection{MRI data analyses}
Diffusion data were preprocessed using the Elikopy pipeline (\citealp{dessain2024fast}), including brain extraction (\citealp{hoopes2022synthstrip}), thermal noise removal (\citealp{veraart2016denoising}), and corrections for susceptibility-induced distortions, eddy currents, and head motion using FSL (v6.0.7.8). As reversed phase-encoding b0 images were not acquired, susceptibility distortion correction was performed using Synb0-DISCO, which generates a synthetic distortion-free b0 image from the T1-weighted anatomical scan. Post-processing was also performed using Elikopy, which included the mathematical reconstruction of the diffusion-weighted images (64 directions) to derive the diffusion tensor imaging (DTI) model. From this, volume-weighted microstructural maps of FA and MD were generated for each participant in MNI space (Figure \ref{fig:chapter4-method}.C), providing a voxel-wise measure of microstructure across the entire brain. 

\subsection{Statistical analyses}
As mentioned in the Introduction, we used a two-stage strategy to first determine white matter circuits underlying inter-individual differences in effort and reward sensitivity and to then test whether these sensitivities could be decoded from microstructure metrics. First, four whole-brain cluster-based analyses were performed to detect clusters where FA or MD covaried with $\beta_{\text{Effort}}$ or $\beta_{\text{Reward}}$. Second, the FA and MD values from these clusters were used as features in supervised machine learning classifiers to assess whether they could decode individual differences in sensitivity in out-of-sample predictions. These two stages are presented in detail below.

\subsubsection{Whole-brain cluster-based analysis}
Voxel-based correlation analyses were conducted using BrainVoyager™ (Brain Innovation, version 23.2), examining the associations between FA and MD maps and $\beta_{\text{Effort}}$ or $\beta_{\text{Reward}}$ (Figure \ref{fig:chapter4-method}.C). These analyses controlled for potential confounding factors by including them as covariates, specifically gender, age, intracranial volume, and other variables that could influence effort and reward sensitivity, such as self-reported depression and anhedonia (assessed using the EDAS and SHAPS scales; (\citealp{bonnelle2016individual})). Statistical significance was set at a voxel-wise p-value of .005, and correction for multiple comparisons was performed using cluster-size thresholding (threshold: 44 mm3; Forman et al., 1995). This approach revealed clusters where white matter microstructure significantly covaried with $\beta_{\text{Effort}}$ or $\beta_{\text{Reward}}$.

To interpret these findings, we use the term white matter integrity to describe FA and MD values in our healthy cohort. This term reflects microstructural organization and does not imply pathology. FA and MD provide complementary information: higher FA indicates greater directional coherence of water diffusion, often linked to axonal alignment, density, or myelination, whereas higher MD reflects greater overall diffusivity, potentially due to increased extracellular space or reduced barriers. Because associations may arise in one metric, the other, or both (Hayakawa et al., 2014) analyzing both metrics offers a more comprehensive view of white matter organization. Effects observed across both FA and MD (i.e., with FA and MD clusters overlapping), when replicating in direction, provide stronger evidence for a robust microstructural-behavioral relationship. To align their interpretability, we computed the additive inverse of MD (-MD), allowing FA and -MD to be interpreted in the same direction, with higher values indicating greater integrity. For visualization, clusters showing significant negative associations between integrity (FA or -MD) and $\beta_{\text{Effort}}$ or $\beta_{\text{Reward}}$ are highlighted in red, and positive associations in green.

To assign tract labels, we used the XTRACT HCP Probabilistic Tract Atlas (\citealp{warrington2020xtract}), which reliably identifies the major canonical white matter bundles, including deep, long-range pathways such as the cingulum, corticospinal tract, and corpus callosum. These tracts, typically located deep, form the principal structural connections within and between hemispheres and are thus the primary focus of our study. A minority of clusters did not overlap with these canonical bundles, instead lying within more superficial white matter (i.e., closer to the cortex) that supports short-range connections between adjacent cortical areas. Because such clusters fall outside the coverage of canonical atlases like XTRACT, we used the EBRAINS Human Connectome Project superficial white matter atlas (\citealp{avila2020inference}) to facilitate their labelling and describe them in the Supplementary Materials (Supplementary Figures \ref{fig:chapter4-suppfigure2} and \ref{fig:chapter4-suppfigure3}). All tract visualizations presented in this study were derived from white matter bundle definitions and population-averaged templates in MNI ICBM 2009a space, based on previously published work (\citealp{garyfallidis2018recognition, yeh2018population}).

Finally, to characterize the relationship between white matter microstructure and individual differences in effort and reward sensitivity, we conducted partial Pearson correlation analyses. For each significant cluster identified in the voxel-wise analyses, we extracted the FA and -MD values from all voxels within the cluster and computed a mean value per cluster for each participant. These averaged microstructural metrics served as independent variables in separate partial correlation analyses with $\beta_{\text{Effort}}$ and $\beta_{\text{Reward}}$, controlling for the potential confounding factors mentioned above, including age, gender, intra-cranial volume and self-reported measures of depression and anhedonia. The resulting correlation R coefficients, p-values, and corresponding regression plots are presented in the Results section for each cluster. In these plots, the y-axis depicts residualised $\beta_{\text{Effort}}$ or $\beta_{\text{Reward}}$ values (after controlling for covariates), while the x-axis represents the mean FA or -MD values for each cluster, allowing for direct visualization of the strength and direction of the adjusted associations. As before, significant negative and positive partial correlations are shown in red and green, respectively, to facilitate visualization throughout the Results section. 

\section{Results}
\subsection{Inter-individual variability in effort and reward sensitivity}

From computational modelling, behavioural data showed strong inter-individual differences in how participants changed their acceptance rates based on effort and reward magnitudes, reflecting differences in effort and reward sensitivity (Figure \ref{fig:chapter4-computational-results}.A). This variability was quantitatively captured by the density distribution of the $\beta_{\text{Effort}}$ and $\beta_{\text{Reward}}$ parameters, which showed broad dispersion across individuals (Figure \ref{fig:chapter4-computational-results}.B; $\beta_{\text{Effort}}$: mean = 16.92, SD = 11.10; $\beta_{\text{Reward}}$: mean = -0.30, SD = 0.98). The coefficient of variation reached 65.6\% for $\beta_{\text{Effort}}$ and 318.1\% for $\beta_{\text{Reward}}$, confirming substantial heterogeneity in sensitivity to effort and reward. To illustrate this diversity of profiles, Figure \ref{fig:chapter4-computational-results}.C presents individual examples of acceptance behavior in participants with low vs. high $\beta_{\text{Effort}}$ and $\beta_{\text{Reward}}$ values. These findings confirm the presence of heterogenous sensitivity profiles in the healthy population.

% First page
\begin{figure}
\centering
\includegraphics[width=1\linewidth]{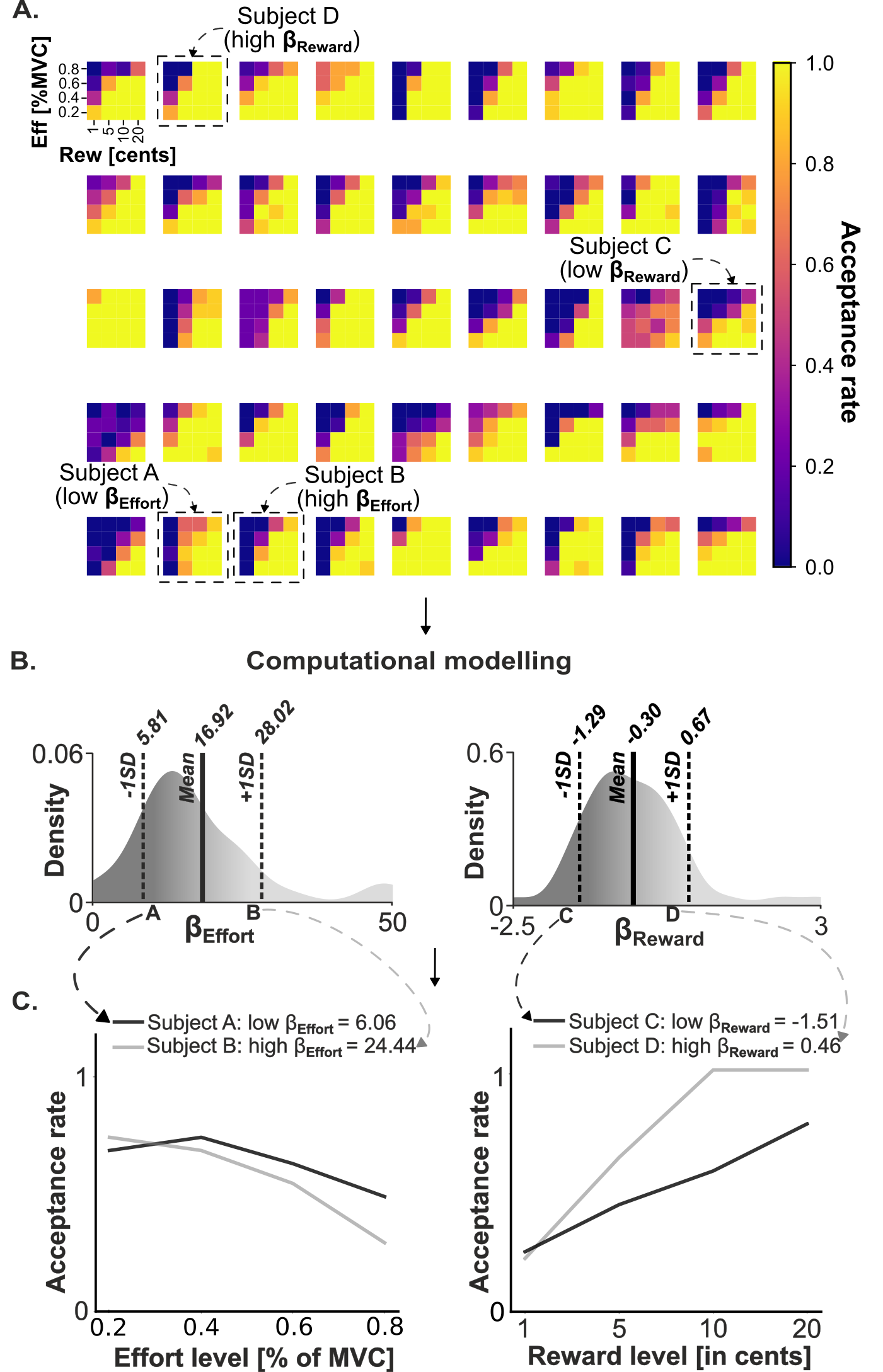}
\caption[Computational modelling of effort and reward sensitivity]{Effort and reward sensitivity show substantial inter-individual variability across healthy subjects. A. Individual acceptance rate maps. (Caption continued next page)}
\label{fig:chapter4-computational-results}
\end{figure}

% Second page
\begin{figure}
\ContinuedFloat
\caption[]{Each map represents the acceptance rate of a given subject as a function of effort (in \% MVC; y-axes) and reward (x-axes). These maps show the strong inter-individual differences in how participants modulated their acceptance rates based on effort and reward magnitudes. B. Distributions of $\beta_{\text{Effort}}$ and $\beta_{\text{Reward}}$. Parameters $\beta_{\text{Effort}}$ and $\beta_{\text{Reward}}$ were estimated from the modelling of acceptance rates presented in A using the equation presented in section Computational modelling of acceptance rates. Here again, the density distributions depict the high inter-individual variability in $\beta_{\text{Effort}}$ and $\beta_{\text{Reward}}$ values among our group of 45 subjects. As described in the section Computational modelling of acceptance rates, for ease of interpretation, $\beta_{\text{Effort}}$ values initially obtained following our modelling procedure were multiplied by $-1$, such that higher (more positive) values reflect stronger effort sensitivity in the left panel. Further, more positive $\beta_{\text{Reward}}$ values correspond to stronger reward sensitivity in the right panel. C. Representative acceptance curves from individuals with divergent $\beta_{\text{Effort}}$ and $\beta_{\text{Reward}}$ values. Left panel: Compared to subject A, subject B showed a stronger decrease in acceptance rates with increasing effort, reflected by a higher $\beta_{\text{Effort}}$ (24.44 vs. 6.06). Right panel: Compared to subject C, subject D showed a stronger increase in acceptance rates with increasing reward, reflected by a higher $\beta_{\text{Reward}}$ ($0.46$ vs. $-1.51$). These examples illustrate how modulation of decision behavior by effort and reward varies across healthy individuals.}
\end{figure}

\subsection{Key findings on the microstructural correlates of effort and reward sensitivity}
As detailed in the Methods, in the first stage of our statistical approach, we conducted four cluster-based analyses to examine the microstructural correlates of individual differences in $\beta_{\text{Effort}}$ and $\beta_{\text{Reward}}$, using FA and the additive inverse of MD (-MD, with higher values reflecting greater integrity) to capture complementary aspects of white matter integrity (see section Whole-brain cluster-based analysis, for more details). Across these analyses, we identified 12 significant clusters located within major, canonical white matter bundles: 5 covarying with effort sensitivity and 7 with reward sensitivity. All of the 5 effort-related clusters were situated within tracts connected to frontal valuation regions, such as SMA, dACC, and OFC (e.g., within the anterior cingulum tract). In contrast, the 7 reward-related clusters were more widely distributed: 3 were located within tracts linked to frontal valuation regions, while 4 lay outside this network, encompassing fronto-parietal and sensorimotor pathways. The most robust associations included one cluster shared across effort and reward sensitivity analyses and another consistent across FA and -MD metrics, both within SMA-connected tracts. Machine learning analyses further showed that microstructural metrics from these 5 effort-related and 7 reward-related clusters reliably predicted effort and reward sensitivity, respectively, whereas randomly positioned clusters did not. Table 1 summarizes all clusters identified across the four analyses (i.e., FA and -MD analyses for $\beta_{\text{Effort}}$ and $\beta_{\text{Reward}}$), their tract labels and associated partial correlation results, which are described individually in the following sections.

\begin{table}[!htbp]
\centering
\small
\caption{Clusters showing significant associations between white matter integrity and effort/reward}
 \resizebox{\textwidth}{!}{
\begin{tabular}{rrrcccc}
\hline \\[-2pt]
\multicolumn{3}{c}{\textbf{MNI coordinates}} & \textbf{Volume} & & \multicolumn{2}{c}{\textbf{Partial correlations}} \\
\textbf{x} & \textbf{y} & \textbf{z} & \textbf{(mm$^3$)} & \textbf{Tract label} & \textbf{R} & \textbf{p} \\ \\[-2pt]
\hline
\\[-4pt]
\multicolumn{7}{l}{\textbf{5 clusters: white matter integrity and $\boldsymbol{\beta}_{\textbf{Effort}}$}} \\
\multicolumn{7}{l}{\textit{FA analysis}} \\
-13.95 & -10.01 & 58.02 & 94 & SMA portion of the left corticospinal tract & -0.697 & $<$0.001 \\
-10.93 & 15.96 & 19.02 & 56 & Left mid-ant. segment of the corpus callosum & -0.556 & $<$0.001 \\
-9.95 & 34.7 & 11.22 & 209 & Left anterior cingulum bundle & 0.623 & $<$.001 \\
\multicolumn{7}{l}{\textit{MD analysis}} \\
-7.55 & 14.03 & 20.11 & 322 & Left mid-ant. segment of the corpus callosum & -0.641 & $<$0.001 \\
8.71 & 15.55 & 20.32 & 139 & Right mid-ant. segment of the corpus callosum & -0.575 & $<$0.001 \\
\\
\\
\multicolumn{7}{l}{\textbf{7 clusters: white matter integrity and $\boldsymbol{\beta}_{\textbf{Reward}}$}} \\
\multicolumn{7}{l}{\textit{FA analysis}} \\
-13.81 & -10.44 & 59.95 & 79 & SMA portion of the left corticospinal tract & -0.614 & $<$0.001 \\
-17.82 & 53.12 & -2.3 & 73 & Left forceps minor & 0.561 & $<$0.001 \\
17.9 & 53.81 & 4.19 & 67 & Right forceps minor & 0.572 & $<$0.001 \\
39.85 & 15.76 & 17.7 & 126 & Right superior lateral fasciculus & 0.579 & $<$0.001 \\
36.22 & -24.6 & 33.78 & 86 & Right superior lateral fasciculus & 0.561 & $<$0.001 \\
-9.61 & 5.02 & 26.19 & 137 & Left corpus callosum (motor portion) & 0.566 & $<$0.001 \\
\multicolumn{7}{l}{\textit{MD analysis}} \\
11.05 & -29.94 & -34.32 & 84 & Right mid. cerebellar peduncle / left CPC tract & -0.547 & $<$0.001 \\[2pt]
\hline
\end{tabular}}
\end{table}

\subsubsection{Microstructure in tracts connected to key frontal valuation regions is associated with effort sensitivity}
Among the 5 significant clusters associated with $\beta_{\text{Effort}}$, 4 showed negative associations, where reduced FA or -MD (i.e., lower microstructural integrity) was linked to higher $\beta_{\text{Effort}}$ (i.e., stronger effort sensitivity), while 1 showed a positive association. Below, we detail these clusters.
\begin{figure}
    \centering
    \includegraphics[width=.95\linewidth]{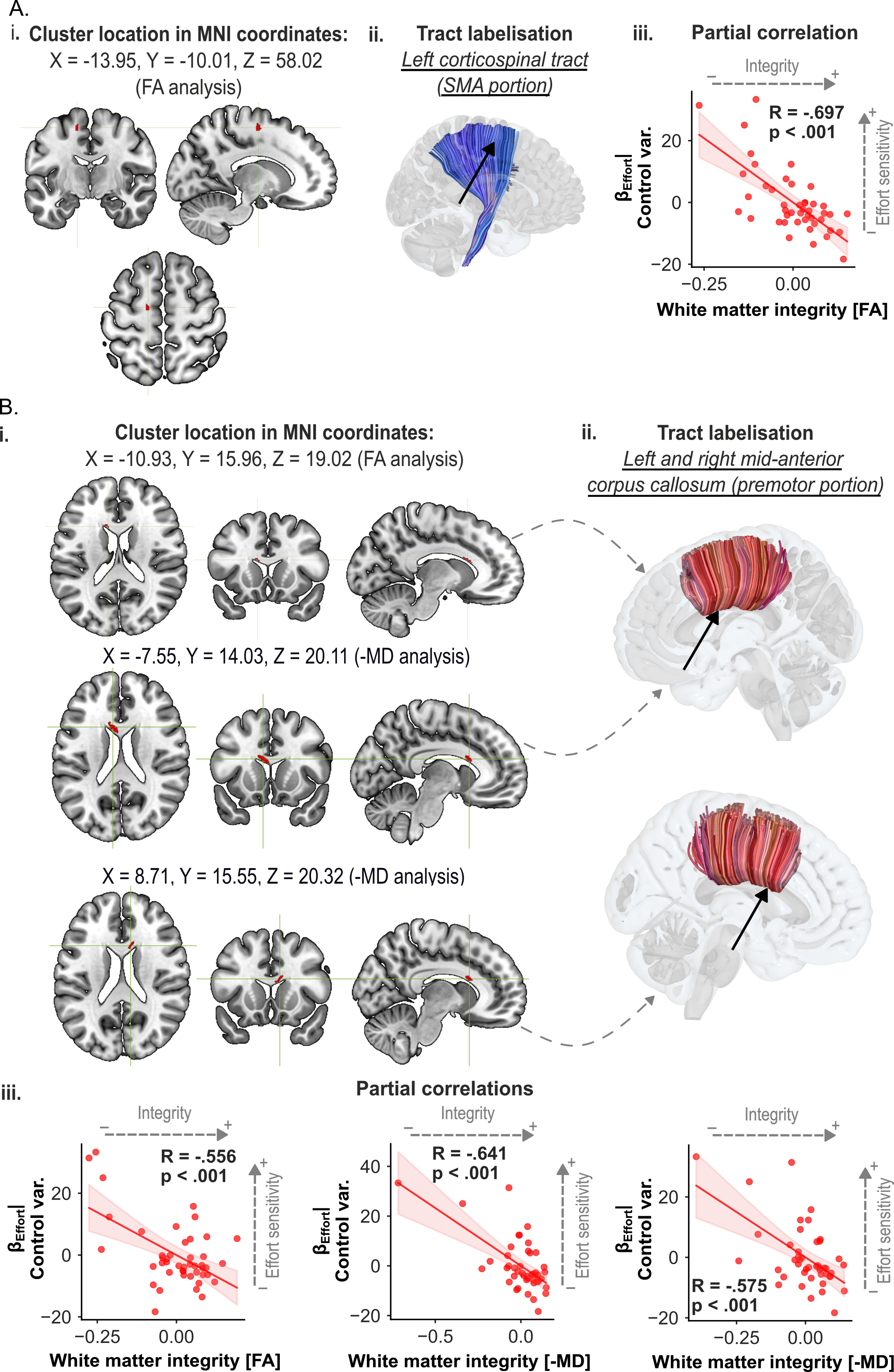}
    \caption[Microstructure in tracts connected to frontal valuation regions showing negative associations with effort sensitivity ($\beta_{\text{Effort}}$). ]{\textbf{Microstructure in tracts connected to frontal valuation regions showing negative associations with effort sensitivity ($\beta_{\text{Effort}}$).} (Caption continued on next page.)}
    \label{fig:chapter4-figure3}
\end{figure}

 \begin{figure}
\ContinuedFloat
\caption[]{\textbf{A. Cluster in the SMA portion of the left corticospinal tract.} (i) A significant cluster (red) showed a negative correlation between FA and $\beta_{\text{Effort}}$ (MNI: -13.95, -10.01, 58.02). (ii) Probabilistic tract labelling identified the corticospinal tract; anatomical masks (\citealp{bonnelle2016individual, beckmann2009connectivity}) confirmed the cluster’s location within SMA proper. The cluster extended beyond the tract (46\% of its voxels), likely encompassing SMA-related projections such as the SMA–NAcc pathway, previously linked to $\beta_{\text{Effort}}$ via macroscopic connectivity measures (\citealp{derosiere2025fronto}). (iii) Partial correlation confirmed reduced FA in this cluster was significantly correlated with higher effort sensitivity (R = -0.697, $p < 0.001$). Across all panels, residualized $\beta$ values (adjusted for age, gender, intracranial volume, depression, and anhedonia) are plotted against mean FA or -MD per cluster to visualize adjusted associations. \textbf{B. Clusters in the mid-anterior corpus callosum.} (i) Three significant clusters (red) exhibited consistent negative correlations between microstructural integrity and $\beta_{\text{Effort}}$. The FA cluster (top row, left) overlapped with one -MD cluster (middle row, left), while another -MD cluster was located symmetrically in the right hemisphere (bottom row). (ii) Probabilistic tract labelling localized all clusters to the mid-anterior corpus callosum, specifically in its premotor portion containing interhemispheric fibers connecting bilateral SMAs (\citealp{xiong2024cortical}). (iii) Partial correlation analyses confirmed strong negative associations for each cluster (R = -0.556, R = -0.641, and R = -0.575, all $p < 0.001$). The replication across hemispheres and both FA and -MD provides particularly strong evidence for a robust link between effort sensitivity and compromised integrity in SMA-related interhemispheric pathways.}
\end{figure}
The first negative association with $\beta_{\text{Effort}}$ was found for FA in a cluster located within the SMA portion of the left corticospinal tract (MNI: -13.95, -10.01, 58.02; 94 mm³; R = -0.697; $p < 0.001$; Figure \ref{fig:chapter4-figure3}.A). Probabilistic atlas labelling identified the corticospinal tract, and anatomical masks (\citealp{bonnelle2016individual, beckmann2009connectivity}) showed it was located within SMA proper. As typical in such labelling, the cluster extended beyond the tract (46\% of voxels in the tract), likely involving other SMA-related projections, including the SMA-NAcc pathway, which showed in prior work a negative association between effort sensitivity and macroscopic structural connectivity (i.e., streamline count; \citealp{derosiere2025fronto}). Importantly, the same cluster also emerged as a key correlate of $\beta_{\text{Reward}}$ (see below), underscoring its importance as a shared microstructural substrate of both effort and reward sensitivity.

A second robust pattern involved the mid-anterior corpus callosum. FA analyses revealed a cluster in its left portion (MNI: -10.93, 15.96, 19.02; 56 mm³; R = -0.556; p < 0.001; Figure \ref{fig:chapter4-figure3}.B), in a locus containing interhemispheric fibers predominantly connecting bilateral SMAs (\citealp{xiong2024cortical}). Supporting this, -MD analyses revealed two additional bilateral clusters with overlapping spatial distributions with the FA cluster (left: -7.55, 14.03, 20.11; 322 mm³; R = -0.641; $p < 0.001$; right: 8.71, 15.55, 20.32; 139 mm³; R = -0.575; $p < 0.001$; Figure \ref{fig:chapter4-figure3}.B). The fact that these effects replicate bilaterally and across both FA and -MD, consistently showing decreased integrity with higher $\beta_{\text{Effort}}$, provides particularly strong evidence for a robust relationship. These results suggest that higher effort sensitivity is linked to reduced directional integrity (lower FA, reflecting reduced axonal alignment or myelination) and increased water diffusion (lower -MD, suggesting greater extracellular space), which may disrupt interhemispheric communication between bilateral SMAs and other medial frontal regions critical for effort processing.

\begin{figure}
    \centering
    \includegraphics[width=.8\linewidth]{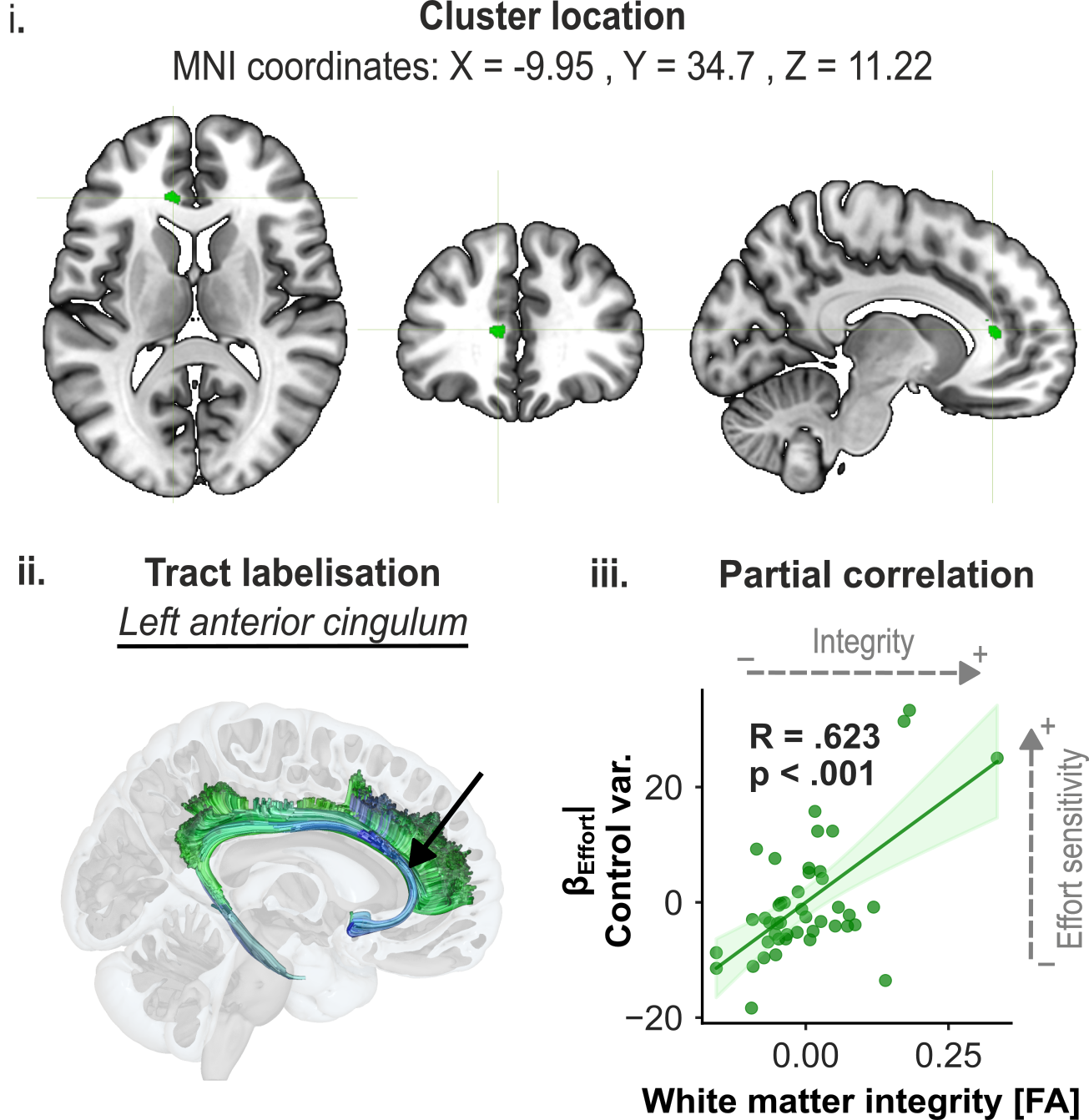}
    \caption[Microstructure in a tract connected to frontal valuation regions showing a positive association with effort sensitivity ]{\textbf{Microstructure in a tract connected to frontal valuation regions showing a positive association with effort sensitivity ($\beta_{\text{Effort}}$).} (i) A significant cluster (green) exhibited a positive correlation between FA and $\beta_{\text{Effort}}$ (MNI: -9.95, 34.7, 11.22). (ii) Probabilistic tract labelling identified the anterior cingulum. To clarify which medial frontal regions were specifically connected by this cluster, we projected it onto a high-resolution atlas, revealing fibers linking the dACC with SMA and medial OFC, core valuation hubs. (iii) Partial correlation confirmed that greater FA in this cluster was associated with higher effort sensitivity (R = 0.623, $p < 0.001$). No additional clusters were identified in major tracts, and the -MD analysis did not reveal any clusters with positive associations. }
    \label{fig:chapter4-figure4}
\end{figure}

In contrast, one cluster showed a positive association between FA and $\beta_{\text{Effort}}$. This cluster was located in the left anterior cingulum bundle (MNI: -9.95, 34.7, 11.22; 209 mm³; R = .623; $p < 0.001$; Figure \ref{fig:chapter4-figure4}), a tract that links medial frontal regions. To identify which medial frontal regions were specifically concerned by this cluster, we projected it onto a high-resolution atlas, which revealed it was located in fibers connecting the dACC with SMA and medial OFC, key frontal valuation hubs. No additional clusters emerged in major tracts, and the -MD analysis revealed none with positive associations.

Together, these findings indicate that individual differences in effort sensitivity are anchored in several white matter pathways connected to frontal valuation regions. Reduced microstructural integrity in SMA-connected clusters, including clusters identified in the corticospinal tract and interhemispheric fibers of the mid-anterior corpus callosum, is consistently associated with stronger effort sensitivity, representing some of the most robust effects in the present study (consistent across metrics and, for one cluster, across valuation dimensions [see below]). Conversely, increased integrity of the left anterior cingulum bundle, encompassing connections between SMA, dACC, and medial OFC, is associated with increased effort sensitivity, potentially reflecting a distinct mechanism whereby enhanced communication among these valuation hubs amplifies effort sensitivity (see Discussion section).

\subsubsection{Microstructure in tracts connected to frontal valuation regions, as well as fronto-parietal and sensorimotor structures, is associated with reward sensitivity}

As mentioned earlier, the FA and -MD analyses revealed 7 significant clusters, more widely distributed than the effort-related clusters: 3 were located within tracts linked to frontal valuation regions, while 4 lay outside this network, encompassing fronto-parietal and sensorimotor pathways. 2 clusters showed negative associations, where lower FA or -MD (i.e., lower microstructural integrity) was linked to higher $\beta_{\text{Reward}}$ (i.e., stronger reward sensitivity), while 5 clusters showed positive associations. Below, we detail these clusters, beginning with the negative associations.

\begin{figure}
    \centering
    \includegraphics[width=.75\linewidth]{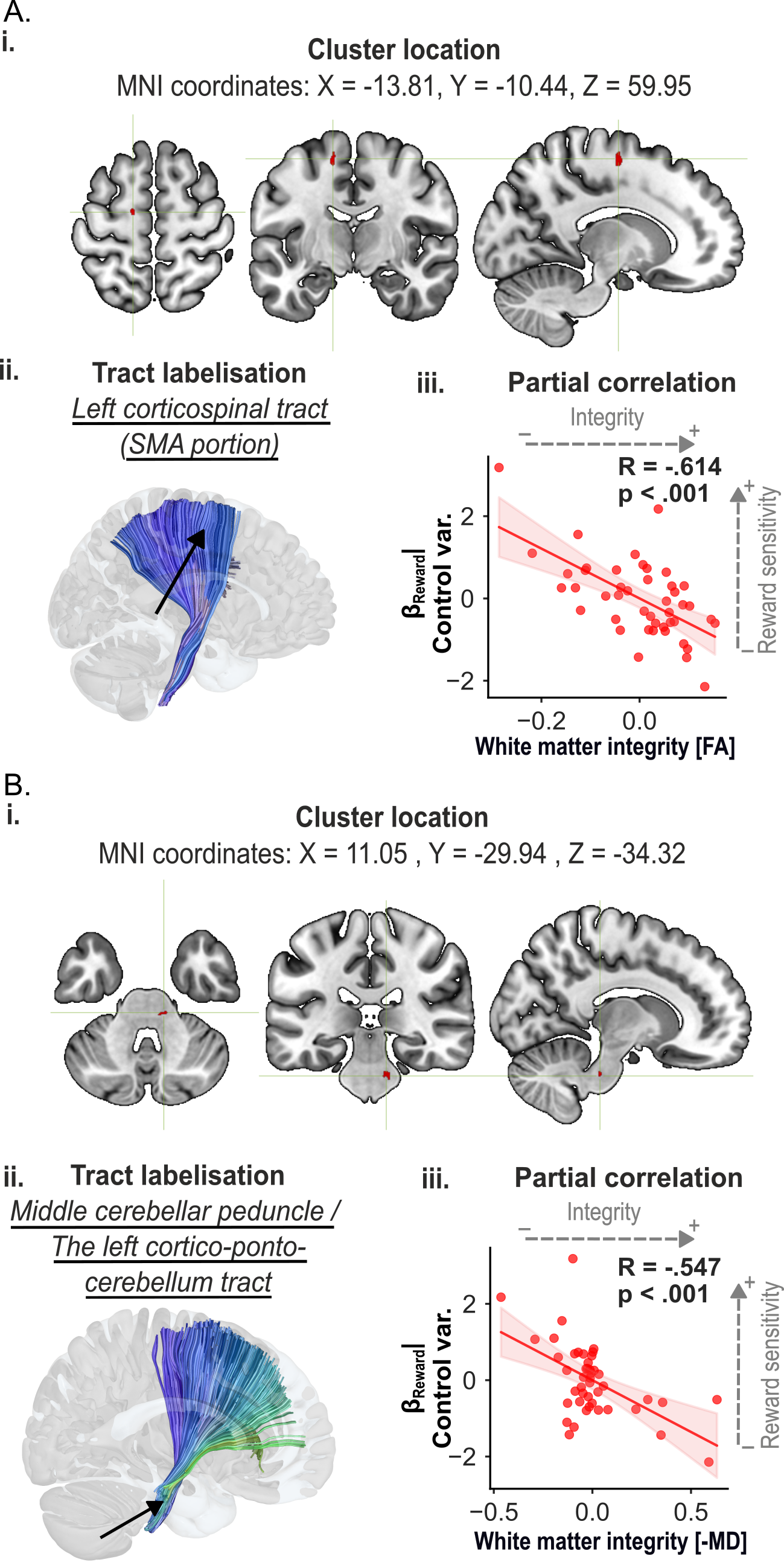}
    \caption[Microstructure in tracts connected to frontal valuation regions and sensorimotor structures and showing negative associations with reward sensitivity]{\textbf{Microstructure in tracts connected to frontal valuation regions and sensorimotor structures and showing negative associations with reward sensitivity ($\beta_{\text{Reward}}$)} (Figure caption continued on next page.)}
    \label{fig:chapter4-figure5}
\end{figure}

\begin{figure}
    \ContinuedFloat
    \caption[]{\textbf{A. Cluster in the SMA portion of the left corticospinal tract.} (i) A significant cluster (red) exhibited a negative correlation between FA and ($\beta_{\text{Reward}}$) (MNI: -13.81, -10.44, 59.95). (ii) Probabilistic tract labelling identified the corticospinal tract; anatomical masks (\citealp{bonnelle2016individual, beckmann2009connectivity}) confirmed the cluster’s location within SMA proper. The cluster overlapped substantially with the $\beta_{\text{Effort}}$ cluster (see Figure \ref{fig:chapter4-figure6}), suggesting a shared white matter substrate where reduced integrity may amplify both effort and reward sensitivity. (iii) Partial correlation confirmed that reduced FA in this region was significantly correlated with greater reward sensitivity (R = -0.614, $p < 0.001$). B. Cluster in the right middle cerebellar peduncle, extending to the left cerebello-ponto-cortical tract. (i) A significant cluster (red) exhibited a negative correlation between -MD and $\beta_{\text{Reward}}$ (MNI: 11.05, -29.94, -34.32). (ii) Probabilistic tract labelling localized the cluster right middle cerebellar peduncle, extending to the left cerebello-ponto-cortical tract. (iii) Partial correlation confirmed that reduced integrity in this region was significantly correlated with greater reward sensitivity (R = -0.547, p < 0.001). Across all panels, residualized $\beta$ values (adjusted for age, gender, intracranial volume, depression, and anhedonia) are plotted against mean FA or -MD per cluster to visualize adjusted associations.}
\end{figure}
\begin{figure}
    \centering
    \includegraphics[width=.6\linewidth]{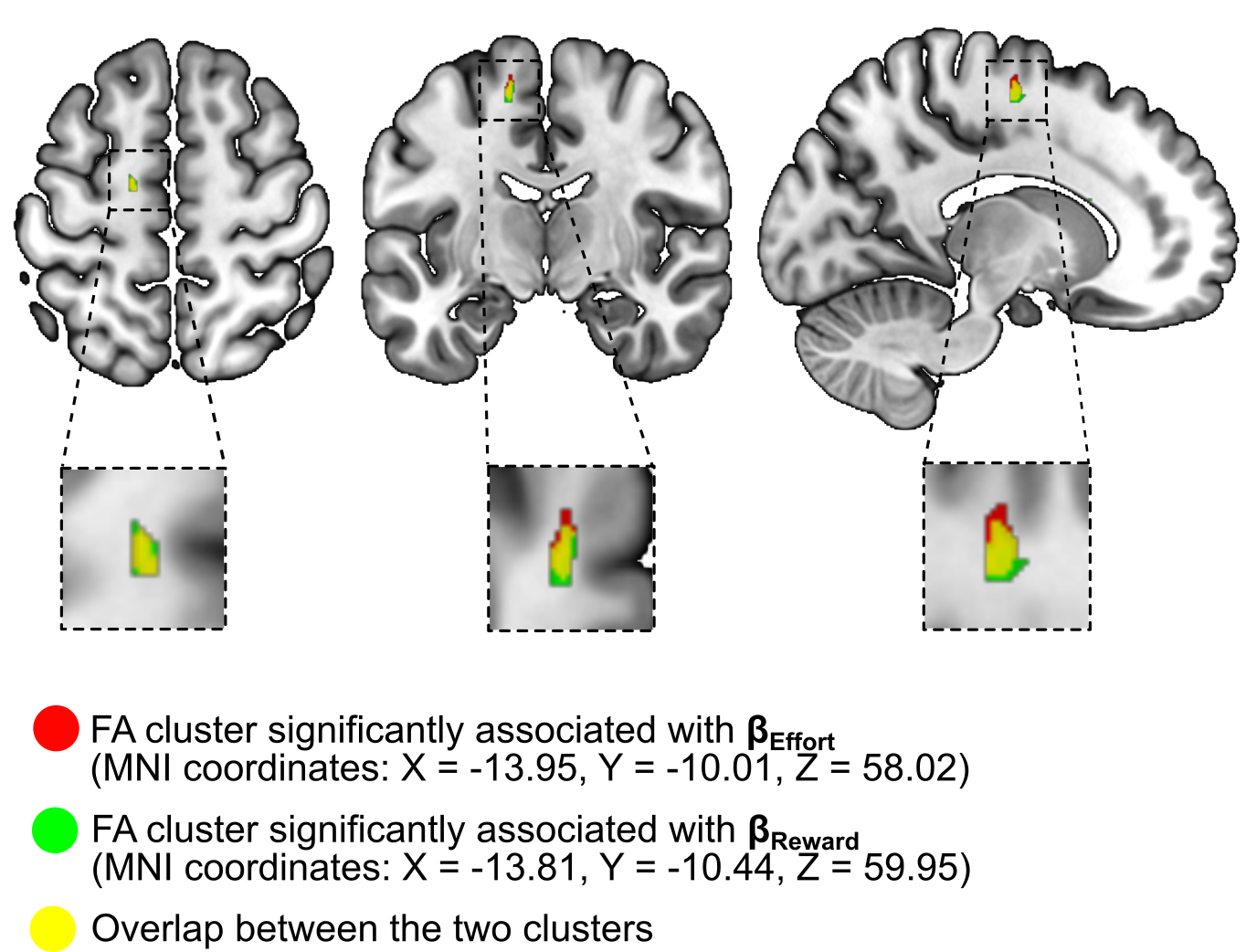}
    \caption[Exclusive overlap of white matter clusters associated with effort and reward sensitivity within the SMA portion of the corticospinal tract. ]{\textbf{Exclusive overlap of white matter clusters associated with effort and reward sensitivity within the SMA portion of the corticospinal tract.} FA analyses identified two significant clusters within the SMA portion of the left corticospinal tract. The cluster associated with $\beta_{\text{Effort}}$ (red) was centered at MNI: X = –13.95, Y = –10.01, Z = 58.02 (94 mm³), while the cluster associated with $\beta_{\text{Reward}}$ (green) was centered at MNI: X = –13.81, Y = –10.44, Z = 59.95. These clusters overlapped extensively (yellow), forming the only locus across the entire FA and -MD maps where microstructural integrity covaried with both $\beta_{\text{Effort}}$ and $\beta_{\text{Reward}}$. This convergence underscores the specificity of this SMA corticospinal segment as a shared structural substrate for individual differences in sensitivity to both effort and reward.}
    \label{fig:chapter4-figure6}
\end{figure}

The first negative association with $\beta_{\text{Reward}}$ was for FA in a cluster located within the SMA portion of the left corticospinal tract (MNI: -13.81, -10.44, 59.95; 79 mm³; R = -0.614; $p < 0.001$; Figure \ref{fig:chapter4-figure5}.A), as identified with probabilistic atlas labelling and anatomical masks (\citealp{bonnelle2016individual, beckmann2009connectivity} and overlapped extensively with the cluster identified for $\beta_{\text{Effort}}$ (MNI: -13.95, -10.01, 58.02; 94 mm³; Figure \ref{fig:chapter4-figure6}). Importantly, this shared cluster represents the sole locus where microstructural integrity covaries with both $\beta_{\text{Effort}}$ and $\beta_{\text{Reward}}$, suggesting a shared white matter substrate where reduced integrity may amplify both effort and reward sensitivity. To confirm that this dual association was not trivially driven by the fact that $\beta_{\text{Effort}}$ and $\beta_{\text{Reward}}$ were derived from the same acceptance rates, we tested whether FA values in these clusters correlated with the two other model-derived parameters: $\beta_{\text{Time}}$ and $\beta_{\text{0}}$. No significant correlations were observed either with the effort-related FA values ($\beta_{\text{Time}}$: R = -0.180, p = 0.237; $\beta_{\text{0}}$: R = -0.163, p = 0.284) or with the reward-related FA values ($\beta_{\text{Time}}$: R = -0.111, p = 0.466; $\beta_{\text{0}}$: R = -0.148, p = 0.332; Supplementary Figure \ref{fig:chapter4-suppfigure4}), underscoring the specificity of this SMA white matter locus for $\beta_{\text{Effort}}$ and $\beta_{\text{Reward}}$.

Five clusters showed positive associations between FA and $\beta_{\text{Reward}}$. Two bilateral clusters were located in the anterior portion of the forceps minor (left: MNI: -17.82, 53.12, -2.3; 73 mm³; R = 0.572; $p < 0.001$; right: 17.9, 53.81, 4.19; 67 mm³; R = 0.561; $p < 0.001$; Figure \ref{fig:chapter4-figure7}.A), a tract interconnecting bilateral OFCs (\citealp{filbey2014long}), key hubs for reward valuation. Two additional clusters were found in a fronto-parietal pathway, namely along the right superior longitudinal fasciculus, one in its frontal portion (MNI: 39.85, 15.76, 17.7; 126 mm³; R = 0.579; $p < .001$) and one in its parietal portion (MNI: 36.22, -24.6, 33.78; 86 mm³; R = 0.561; $p < 0.001$; Figure \ref{fig:chapter4-figure7}.B). Finally, one other cluster was located in a tract connected to motor structures, namely in the left mid-body of the corpus callosum (MNI: -9.61, 5.02, 26.19; 137 mm³; R = 0.566; p $<$ 0.001; Figure \ref{fig:chapter4-figure7}.C), a locus of fibers primarily connecting bilateral M1s (\citealp{hofer2006topography, tarumi2022microstructural, wahl2007human}). No additional clusters emerged in major tracts, and the -MD analysis revealed none with positive associations.

\begin{figure}[hp]
    \centering
    \includegraphics[width=.9\linewidth]{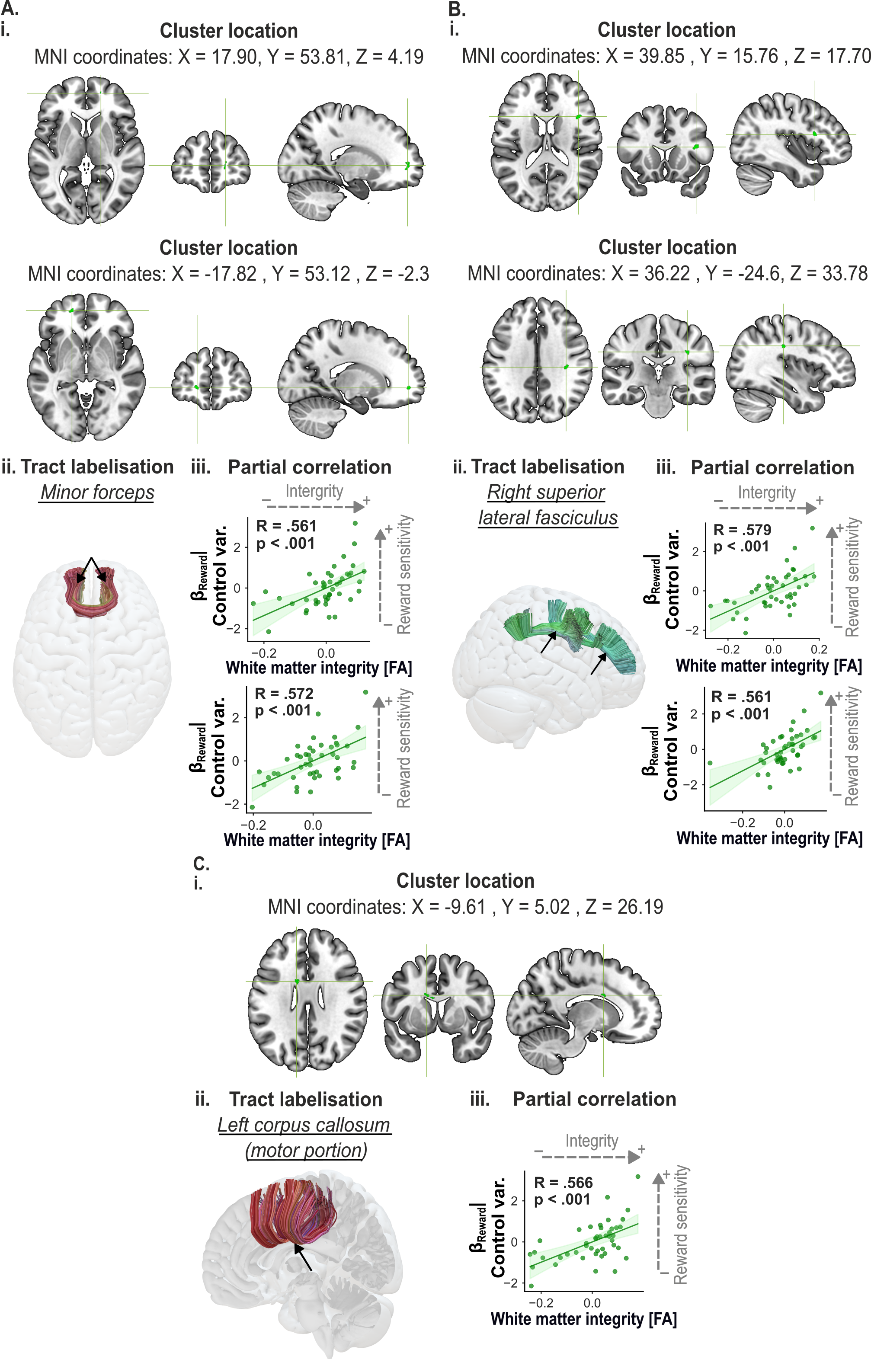}
    \caption[Microstructure in tracts connected to frontal valuation regions, fronto-parietal and sensorimotor structures showing positive associations with reward sensitivity ]{\textbf{Microstructure in tracts connected to frontal valuation regions, fronto-parietal and sensorimotor structures showing positive associations with reward sensitivity.} (Caption continued on next page.)}
    \label{fig:chapter4-figure7}
\end{figure}
% Second page
\begin{figure}
\ContinuedFloat
\caption[]{\textbf{A. Bilateral clusters in the forceps minor.} (i) Two significant clusters (green) exhibited a positive correlation between FA and $\beta_{\text{Reward}}$ (MNI: left: -17.82, 53.12, -2.3; right: 17.9, 53.81, 4.19). (ii) Probabilistic tract labelling localized both clusters to the forceps minor, interconnecting bilateral OFCs, key hubs for reward valuation. (iii) Partial correlation confirmed significant positive associations with reward sensitivity (R = 0.572 and R = 0.561, both $p < 0.001$). \textbf{B. Clusters in the right superior longitudinal fasciculus.} (i) Two significant clusters (green) exhibited a positive correlation between FA and $\beta_{\text{Reward}}$ (MNI: anterior: 39.85, 15.76, 17.7; posterior: 36.22, -24.6, 33.78). (ii) Probabilistic tract labelling indicated both clusters were located within the superior longitudinal fasciculus. (iii) Partial correlation confirmed significant associations with reward sensitivity (R = 0.579 and R = 0.561, both $p < 0.001$). \textbf{C. Cluster in the left mid-body of the corpus callosum (motor portion).} (i) One significant cluster (green) exhibited a positive correlation between FA and $\beta_{\text{Reward}}$ (MNI: -9.61, 5.02, 26.19). (ii) Probabilistic tract labelling indicated the cluster was located in callosal fibers connecting bilateral motor cortices. (iii) Partial correlation confirmed a significant association (R = 0.566, $p < 0.001$).}
\end{figure}
Collectively, these results show that differences in reward sensitivity are associated with white matter integrity across distributed systems, encompassing circuits directly connected to frontal valuation regions as well as fronto-parietal and sensorimotor pathways. Regarding frontal valuation regions, heightened sensitivity was linked to reduced integrity within an SMA cluster, comprising SMA-connected corticospinal fibers, and increased integrity within the OFC-connected anterior forceps minor. In addition, reduced integrity in cerebellum-connected tracts and greater integrity within a major fronto-parietal bundle (i.e., the superior longitudinal fasciculus) and motor callosal fibers were linked to stronger reward sensitivity, highlighting additional contributions from fronto-parietal and sensorimotor pathways. This pattern suggests that reward sensitivity reflects the integrated influence of frontal valuation circuits together with fronto-parietal and motor-related networks, rather than any single network.

\subsection{Decoding effort and reward sensitivity from DTI extracted microstructural measures using machine learning classifiers}
We next tested whether $\beta_{\text{Effort}}$ and $\beta_{\text{Reward}}$ could be predicted from the microstructural properties of the clusters identified above. Our machine learning models used as input the averaged FA and -MD values extracted from the significant clusters identified in voxel-wise analyses and were trained to classify participants into low versus high $\beta_{\text{Effort}}$ and $\beta_{\text{Reward}}$ groups. Performance metrics (accuracy and AUC) were estimated using nested cross-validation on true labels and compared against null distributions generated by 1000 label permutations.

For $\beta_{\text{Effort}}$, the best classifier achieved a mean accuracy of $0.648 \pm 0.053$ on the true-label data, well above the 0.5 chance level for binary classification. In contrast, accuracy on the permuted-label data averaged $0.506 \pm 0.086$, with only 37 out of 1{,}000 permutations reaching or exceeding the accuracy obtained with true-label data, yielding a Monte Carlo $p$-value of 0.037 (Figure \ref{fig:chapter4-figure8}.A). This difference in accuracy between true-label and permuted-label data was statistically significant ($t_{998} = 45.57$, $p < 0.001$, Cohen’s $d > 2.5$), and accuracy on true-label data was significantly above chance ($t_{999} = 88.19$, $p < 0.001$), unlike the permuted-label data ($t_{999} = -1.466$, $p = 0.93$). To complement these results, we also examined the AUC, which provides a threshold-independent measure of classification performance. The AUC obtained on true-label data was $0.719 \pm 0.050$, higher than the AUC from permuted-label data ($0.498 \pm 0.118$), with only 28 out of 1{,}000 permutations matching or exceeding it (Monte Carlo $p = 0.028$). The difference between true-label and permuted-label data was also statistically significant ($t_{998} = 54.40$, $p < 0.001$, Cohen’s $d > 2.5$), and only the AUC obtained on the true-label data was significantly greater than chance ($t_{999} = 137.3$, $p < 0.001$; permuted-label data: $t_{999} = -0.553$, $p = 0.580$).

\begin{figure}[hp]
    \centering
    \includegraphics[width=1\linewidth]{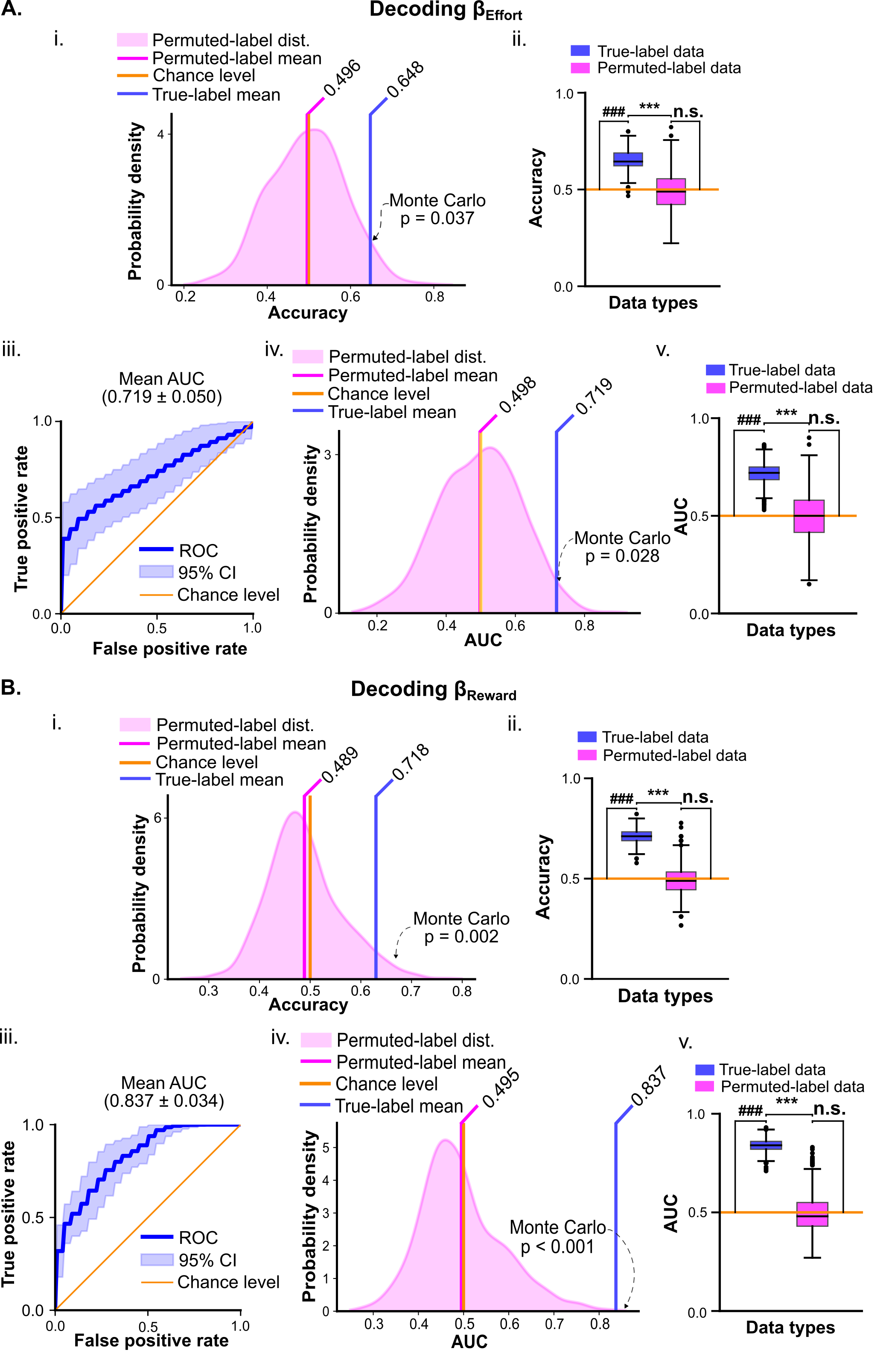}
    \caption[White matter microstructure predicts individual differences in effort and reward sensitivity.]{\textbf{White matter microstructure predicts individual differences in effort and reward sensitivity.} (Caption continued next page.)}
    \label{fig:chapter4-figure8}
\end{figure}

\begin{figure}
\ContinuedFloat
\caption[]{\textbf{(A)} $\beta_{\text{Effort}}$ decoding. (i) Classifier accuracy on true-label data (blue line) compared to the null distribution obtained from 1{,}000 label permutations (pink); only 37/1{,}000 permutations matched or exceeded true-label accuracy (Monte Carlo $p = 0.037$). (ii) Accuracy was significantly higher for true-label data compared to permuted-label and chance-level data (true vs.\ permuted: $t_{998} = 45.57$, $p < 0.001$; true vs.\ chance: $t_{999} = 88.19$, $p < 0.001$; permuted vs.\ chance: $t_{999} = -1.466$, $p = 0.93$). (iii) ROC curve showing classifier performance (mean AUC $= 0.720 \pm 0.039$; 95\% CI shown in shading). (iv) True-label AUC (blue line) compared to null distribution from permuted-label data (pink); 28/1{,}000 permutations matched or exceeded true-label AUC (Monte Carlo $p = 0.028$). (v) AUC was significantly higher for true-label data (true vs.\ permuted: $t_{998} = 54.40$, $p < 0.001$; true vs.\ chance: $t_{999} = 137.3$, $p < 0.001$; permuted vs.\ chance: $t_{999} = -0.553$, $p = 0.580$). \textbf{(B)} $\beta_{\text{Reward}}$ decoding. (i) True-label accuracy (blue line) exceeded that of 1{,}000 permutations (pink); only 2 permutations matched or exceeded it (Monte Carlo $p = 0.002$). (ii) Accuracy was significantly higher for true-label data (true vs.\ permuted: $t_{998} = 88.68$, $p < 0.001$; true vs.\ chance: $t_{999} = 183.2$, $p < 0.001$; permuted vs.\ chance: $t_{999} = -4.95$, $p = 0.99$). (iii) ROC curve (mean AUC $= 0.829 \pm 0.022$). (iv) AUC distribution from permuted-label data (pink) with true-label AUC (blue line); no permutations matched or exceeded the true-label AUC (Monte Carlo $p < 0.001$). (v) AUC was significantly higher for true-label data (true vs.\ permuted: $t_{998} = 112.7$, $p < 0.001$; true vs.\ chance: $t_{999} = 314.5$, $p < 0.001$; permuted vs.\ chance: $t_{999} = -1.698$, $p = 0.955$). Boxplots show median, interquartile range, and full range. Symbols including \# and * indicate statistical significance: \#\#\#$p < .001$, ***$p < .001$; n.s.\ = not significant.}
\end{figure}

Similarly, for $\beta_{\text{Reward}}$, the best classifier achieved a mean accuracy of $0.718 \pm 0.038$ on the true-label data. In contrast, accuracy on the permuted-label data averaged $0.489 \pm 0.073$, with only 2 out of 1{,}000 permutations reaching or exceeding the accuracy obtained with true-label data, yielding a Monte Carlo $p$-value of .002 (Figure \ref{fig:chapter4-figure8}.B). This difference between true-label and permuted-label data was statistically significant ($t_{998} = 88.68$, $p < 0.001$, Cohen’s $d > 2.5$), and accuracy on true-label data was significantly above chance ($t_{999} = 183.2$, $p < 0.001$), unlike the permuted-label data ($t_{999} = -4.95$, $p = 0.99$). The AUC from true-label data was $0.837 \pm 0.033$, while the one from permuted data-label was $0.495 \pm 0.090$, with 0 out of 1{,}000 permutations matching or exceeding it (Monte Carlo $p < .001$). This difference was also statistically significant ($t_{998} = 112.7$, $p < 0.001$, Cohen’s $d > 2.5$), and only the AUC from true-label data was significantly greater than chance ($t_{999} = 314.5$, $p < 0.001$; permuted-label data: $t_{999} = -1.698$, $p = 0.955$).

\begin{figure}[hp]
    \centering
    \includegraphics[width=1\linewidth]{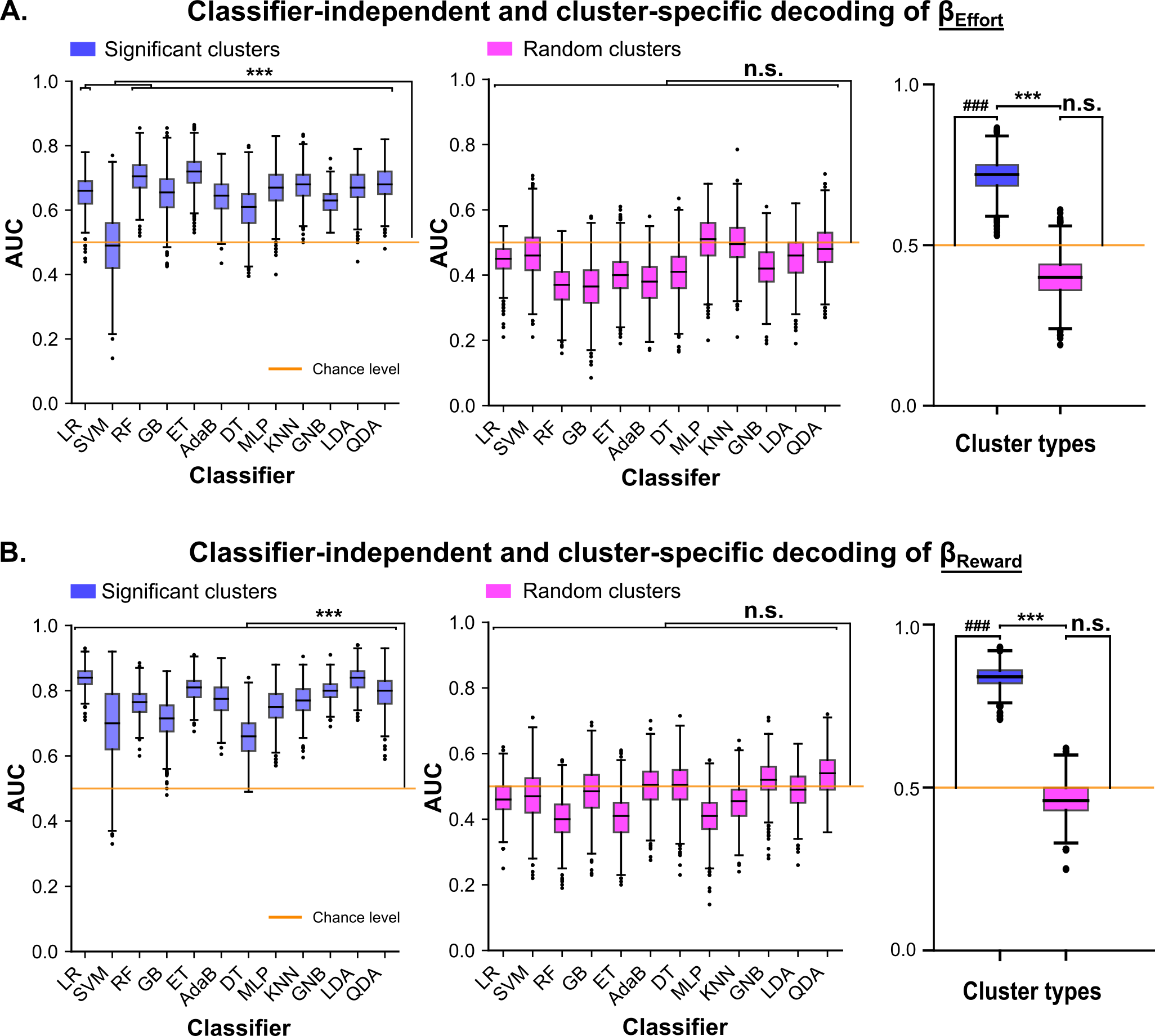}
    \caption[Decoding effort and reward sensitivity is classifier-independent and cluster-specific.]{\textbf{Decoding effort and reward sensitivity is classifier-independent and cluster-specific.} \textbf{A. Classifier-independent and cluster-specific decoding of $\beta_{\text{Effort}}$.} Left panel (blue): Area under the curve (AUC) values for 12 machine learning classifiers trained on microstructural measures from the 5 clusters significantly associated with $\beta_{\text{Effort}}$. Decoding performance was significantly above chance for nearly all classifiers (all FDR-corrected $p < 0.001$), except for SVM (FDR-corrected $p = 0.99$), demonstrating classifier-independent decoding. Classifier names: LR: logistic regression; SVM: support vector machine; RF: random forest; GB: gradient boosting; ET: ExtraTrees; AdaB: AdaBoost; DT: decision tree; MLP: multilayer perceptron; KNN: k-nearest neighbors; GNB: Gaussian naive Bayes; LDA: linear discriminant analysis; QDA: quadratic discriminant analysis. Middle panel (pink): AUC values for the same classifiers trained on microstructural measures from randomly sampled white matter clusters, matched in size and number to the original clusters, showing performance near chance. Right panel: Statistical comparison of AUC values from the best-performing classifier (ExtraTrees) trained on the 5 significant clusters (blue) versus 5 random clusters (pink). Performance with significant clusters was significantly higher than with random clusters (***FDR-corrected $p < 0.001$) and above chance (\#\#\#FDR-corrected $p < 0.001$), whereas performance with random clusters did not differ from chance (n.s.). Box plots indicate median and interquartile range; whiskers denote data within $1.5 \times \text{IQR}$. \textbf{B. Classifier-independent and cluster-specific decoding of $\beta_{\text{Reward}}$.} The panels follow the same organization as A. but for $\beta_{\text{Reward}}$.}
    \label{fig:chapter4-figure9}
\end{figure}

To further establish the robustness and specificity of our machine learning findings, we conducted two additional analyses. First, we verified that classification performance was not dependent on the choice of the best classifier by repeating the $t$-tests against chance level across the AUC values obtained for all 12 tested classifiers (i.e., with $p$ values FDR-corrected). AUC was significantly above chance level across nearly all classifiers for $\beta_{\text{Effort}}$ (all $p$-values $< .001$, except SVM: $p = .999$) and across all classifiers for $\beta_{\text{Reward}}$ (all $p$-values $< .001$; see Figure \ref{fig:chapter4-figure9} \& Supplementary Tables \ref{table:chap4_effort_prediction} and \ref{table:chap4_reward_prediction}), underscoring that predictive performance was classifier-independent and generalizable. Second, we tested whether decoding performance was specific to the white matter clusters identified in our voxel-wise analyses. For each target variable, we randomly sampled clusters from the whole-brain white matter mask, matched in number and size to the original clusters, and re-ran the full classification pipeline, including evaluation across all classifiers and nested cross-validation. In contrast to the primary analyses, AUC for these randomly located clusters was consistently near chance ($\beta_{\text{Effort}}$ prediction: AUC values = [.364 to .506], all $p$-values = [.360 to 1.00]; $\beta_{\text{Reward}}$ prediction: AUC values = [.401 to .534], all $p$-values = [.368 to 1.00]; Figure~9, pink). In fact, the AUC obtained from the original, significant clusters with the best-performing classifier was significantly higher than the AUC obtained from the randomly located clusters ($t_{999} = 192.4$, $p < 0.001$). This second analysis demonstrates that the decoding of $\beta_{\text{Effort}}$ and $\beta_{\text{Reward}}$ is driven by microstructural features of the specific clusters identified in the voxel-wise analyses, rather than reflecting generic inter-individual variability in white matter.

To identify which white-matter clusters contributed most strongly to the decoding of $\beta_{\text{Effort}}$ and $\beta_{\text{Reward}}$, we performed feature importance and recursive feature elimination analyses (see Methods section). In this context, the term ``feature'' refers specifically to the individual white matter clusters included as input variables in the classifier, a standard term in machine learning used to denote each parameter or predictor considered during the analysis. 

\begin{figure}[hp]
    \centering
    \includegraphics[width=1\linewidth]{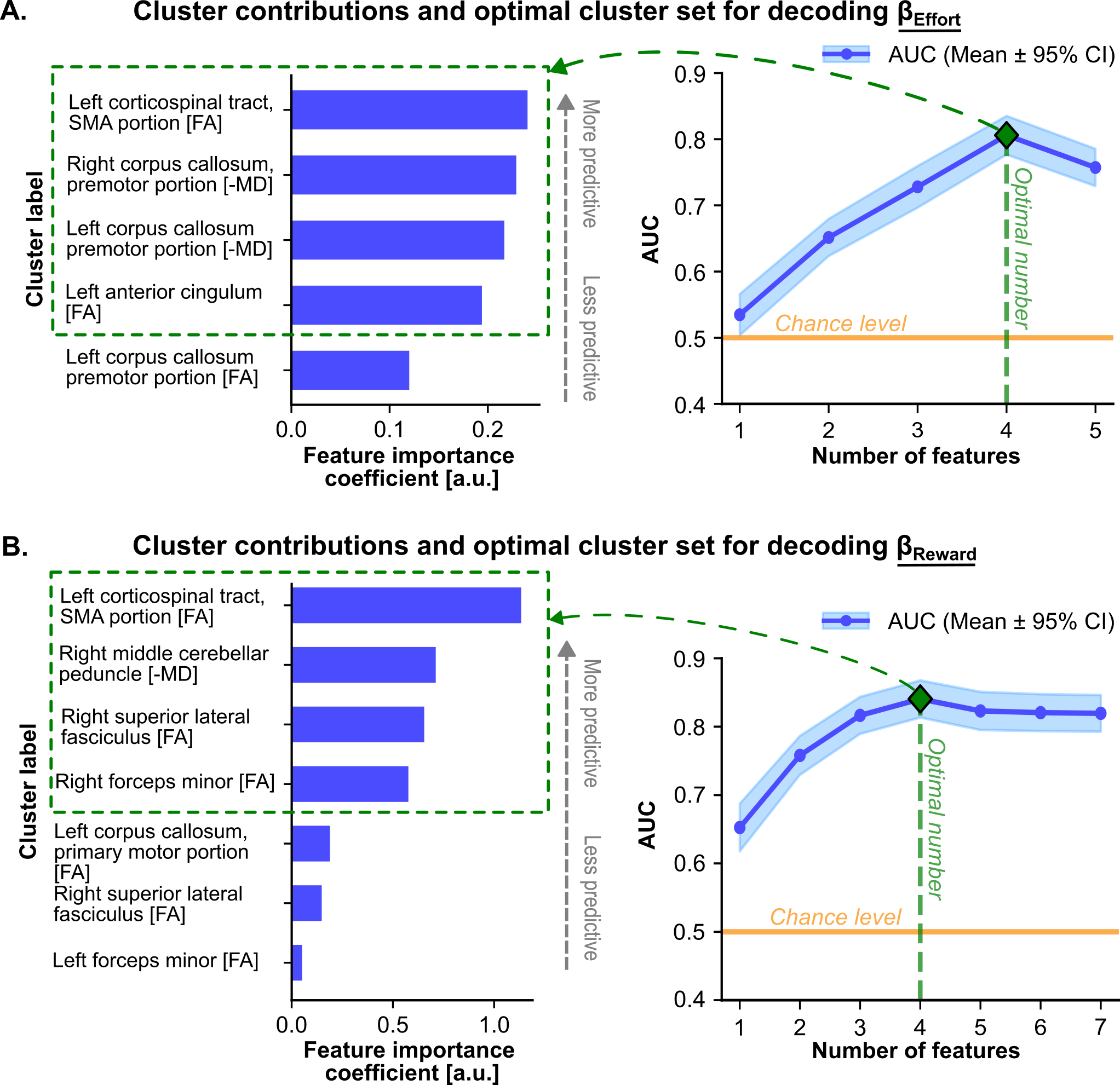}
    \caption[SMA-connected clusters dominate in terms of predictive power, with distributed circuits contributing to reward sensitivity decoding]{\textbf{SMA-connected clusters dominate in terms of predictive power, with distributed circuits contributing to reward sensitivity decoding.} \textbf{A. Cluster contributions and optimal cluster set for decoding $\beta_{\text{Effort}}$.} Left panel: Feature importance analysis, where ``features'' denote the individual white matter clusters used as input variables in the model (a standard machine learning term for predictors). The most predictive cluster was the left corticospinal tract, SMA portion [FA], followed by the right and left premotor portions of the corpus callosum [-MD and FA], and the left anterior cingulum [FA]. Right panel: Recursive feature elimination analysis. Classification performance (AUC) peaked when restricted to the four most informative clusters (green diamond), which are highlighted By the green dashed box in the left panel. \textbf{B. Cluster contributions and optimal cluster set for decoding $\beta_{\text{Reward}}$.} The panels follow the same organization as A. but for $\beta_{\text{Reward}}$. Here, the left corticospinal tract, SMA portion [FA], was again the most predictive feature, followed by clusters in the cortico-ponto-cerebellar tract [-MD], superior lateral fasciculus [FA], and forceps minor [FA]. Maximal classification performance (AUC) was also reached with four features, indicating that reward sensitivity is best predicted by a combination of SMA-connected, fronto-parietal, and cerebellar pathways.
}
    \label{fig:chapter4-figure10}
\end{figure}

Feature importance rankings (derived from the best-performing classifiers), identified the cluster located within the SMA portion of corticospinal tract as the most predictive feature for both $\beta_{\text{Effort}}$ and $\beta_{\text{Reward}}$ (Figure \ref{fig:chapter4-figure10}.A and .B). This finding aligns with the overlap of this cluster across the $\beta_{\text{Effort}}$ and $\beta_{\text{Reward}}$ cluster-based analyses and its lack of correlation with other computational parameters (i.e., both $\beta_{\text{Time}}$ and $\beta_{0}$), underscoring its specificity to effort and reward sensitivities. Recursive feature elimination indicated that high decoding accuracy could be achieved without the full cluster set. For $\beta_{\text{Effort}}$, classification performance peaked when restricted to the four most informative clusters (AUC $= 0.806 \pm 0.147$; Figure \ref{fig:chapter4-figure10}), three of which were connected to SMA and one with the anterior cingulum. For $\beta_{\text{Reward}}$, classification performance also reached maximal performance with the four most predictive clusters (AUC $= 0.840 \pm 0.136$). Of these, two clusters belonged to tracts directly connected to frontal valuation regions (SMA corticospinal tract and OFC-connected forceps minor), while the remaining two involved fronto-parietal and sensorimotor pathways (right superior longitudinal fasciculus and right middle cerebellar peduncle/left cortico-ponto-cerebellar tract). Hence, these analyses indicate that while SMA-connected pathways are the most consistent predictors of both $\beta_{\text{Effort}}$ and $\beta_{\text{Reward}}$, optimal decoding of reward sensitivity additionally relies on contributions from fronto-parietal and motor pathways, reinforcing the idea that inter-individual differences in reward sensitivity emerge from distributed and functionally diverse white matter circuits.

Altogether, these results demonstrate that white matter microstructure within the clusters identified through voxel-wise analyses carries sufficient information to reliably decode inter-individual differences in both effort and reward sensitivity. Predictions were consistently above chance across complementary metrics (accuracy and AUC) and robust across all tested classifiers. Importantly, classification performance did not exceed chance when using randomly sampled white matter clusters of matched size and number, confirming that predictive information is specific to the loci uncovered by the voxel-wise analyses rather than reflecting generic microstructural variability. Feature importance and recursive elimination further revealed that decoding performance is driven by a subset of clusters, most prominently those connected to SMA, while additional contributions from fronto-parietal and motor pathways are required for optimal prediction of reward sensitivity.

%%% Discussion
\section{Discussion}
\subsubsection{Effort sensitivity relates to SMA-, dACC- and OFC-connected white matter clusters while reward sensitivity involves additional fronto-parietal and sensorimotor clusters}

White matter pathways are increasingly recognised as active modulators of neural function, capable of amplifying or attenuating electrophysiological signals (\citealp{innocenti2022functional, thiebaut2022emergent}), thereby shaping brain activity and behaviour (\citealp{lopez2013word}. Variability in white matter microstructure is therefore thought to drive individual differences in cognition and behaviour (\citealp{forkel2022white}), and mapping these structure-behavior relationships can help identify biomarkers for neurological and psychiatric conditions (\citealp{thiebaut2022emergent}) and predict the behavioural impact of structural lesions (\citealp{koch2021structural}). Recent work highlights the clinical potential of such predictive modelling, from forecasting disease progression (e.g., Parkinson’s, catatonia; \citealp{huang2024structural, peretzke2025deciphering}) to anticipating behavioral outcomes after neurosurgery (\citealp{aylmore2025use, essayed2017white, ordonez2023preoperative}), suggesting that extending these approaches to motivational constructs could improve risk assessment for apathy and impulsivity and guide personalized interventions. Here, using a whole-brain, data-driven approach combining DWI with computational modelling of decision-making, we show that white matter integrity covaries with individual differences in effort and reward sensitivity, two key determinants of goal-directed behavior. Twelve clusters emerged: 5 linked to effort sensitivity, all within tracts connected to frontal valuation hubs (SMA, dACC, OFC), and 7 linked to reward sensitivity, spanning frontal valuation, fronto-parietal, and sensorimotor pathways. The most robust effects localized to two SMA-connected clusters, one common to effort and reward sensitivity and another converging across FA and -MD metrics and related to effort sensitivity. FA and -MD metrics from the 5 effort-related and 7 reward-related clusters reliably predicted effort and reward sensitivity in out-of-sample machine learning analyses, respectively, whereas randomly sampled clusters did not. SMA-connected tracts dominated decoding in these analyses, but fronto-parietal and sensorimotor pathways also strongly contributed to the decoding of reward sensitivity.

We found both negative and positive associations between white matter integrity and effort/reward sensitivity, supporting the view that white matter modulates neural signals (\citealp{innocenti2022functional}) and that changes in its integrity, whether reductions or increases, alter signal modulation between gray matter regions and influence behavior. Similar bidirectional patterns have been reported across networks in disorders affecting effort and reward sensitivity, including apathy (\citealp{baggio2015resting, tay2019apathy}), depression (\citealp{leaver2016desynchronization, lynch2024frontostriatal, oestreich2022hyperconnectivity}), and addictions (\citealp{tolomeo2022brain}). These observations suggest that both hypo- and hyperconnectivity can heighten sensitivity to effort and reward, depending on the circuit and the stage of dysfunction.

As mentioned above, one of the strongest effects of our data-driven analyses localized to SMA-connected pathways, with a shared cluster in the SMA segment of the corticospinal tract covarying with both effort and reward sensitivity, suggesting a generic role in action valuation. This dual association was construct-specific, as FA values in this cluster did not correlate with $\beta_{\text{Time}}$ and $\beta_0$, which index sensitivity to time-on-task and overall baseline acceptance, respectively. Machine learning feature importance analyses independently ranked these SMA clusters as dominant predictors, reinforcing their central role in predicting both effort and reward sensitivity. Although SMA activity has been consistently linked to value computation, particularly in relation to effort processing (\citealp{zenon2015disrupting, bonnelle2016individual, husain2018neuroscience, le2018anatomy}), the prevailing view attributes these changes primarily to interactions with other fronto-striatal valuation hubs, such as the dACC and NAcc (see \citealp{husain2018neuroscience}, for review). Our findings suggest a potentially complementary mechanism: changes in SMA activity may not only reflect interactions with other fronto-striatal valuation regions, but also indicate SMA’s direct influence on behavioural engagement via its corticospinal projections, a pathway largely overlooked in current models. Future studies assessing corticospinal excitability via transcranial magnetic stimulation of the SMA (\citealp{entakli2014tms, spieser2013involvement, neige2023connecting}) during effort-reward decisions could further characterize the contribution of SMA corticospinal projections to effort and reward sensitivity. 

Probabilistic tract labelling further showed that, while the shared cluster indeed mainly overlapped with the SMA corticospinal tract, it also extended beyond this tract, likely engaging additional, more classical valuation-related pathways, such as the SMA-NAcc tract, where reduced macroscopic connectivity has likewise been linked to heightened effort sensitivity (\citealp{derosiere2025fronto}). Altogether, reduced integrity in these different SMA-originating circuits may disrupt SMA’s dual roles: that is, invigorating action initiation when movement execution is costly (\citealp{fried1991functional, potgieser2014insights, zimnik2019perturbation}) and exerting inhibitory control to suppress action initiation when reward incentives are high (\citealp{chen2010supplementary}), thereby amplifying sensitivity to both effort and reward.

A second major effect involved reduced integrity in the mid-anterior corpus callosum, a key pathway supporting interhemispheric communication, in part between bilateral SMAs, which was robustly associated with heightened effort sensitivity. This effect was strong and spatially consistent, spanning both FA and -MD metrics and forming symmetrical clusters across hemispheres. Extending the view that SMA contributions are not limited to interactions with other frontal valuation nodes (such as the OFC, dACC or NAcc), these findings point to interhemispheric communication as another critical substrate for individual differences in effort computation. Notably, our forceps minor results suggest a parallel role for commissural fibers linking bilateral OFCs in reward sensitivity (see below), underscoring the broader importance of interhemispheric integration in value computation.

In addition to these SMA-connected pathways, additional clusters were identified in classical frontal valuation circuits. As described above, reward sensitivity was positively associated with integrity in the forceps minor, which interconnects bilateral OFCs, regions central to reward processing. Increased connectivity in this tract has also been linked to impulsivity (\citealp{jeong2016white}), suggesting that stronger OFC communication may enhance reward signal integration and drive greater behavioral engagement when incentives are high. Similarly, higher integrity in the anterior cingulum bundle, connecting SMA and dACC, both key to effort processing, was associated with heightened effort sensitivity. While this might seem at odds with prior findings of reduced cingulum integrity in apathy (\citealp{bonnelle2016individual}), it is consistent with clinical evidence from anterior cingulotomy in obsessive-compulsive disorder, where lesions to this tract reduce action initiation (\citealp{bubb2018cingulum}). This may suggest that increased integrity in the anterior cingulum might enhance interactions between effort processing regions such as the SMA and dACC (\citealp{innocenti2022functional, thiebaut2022emergent}), amplifying the influence of high effort costs on behavioral disengagement.
Notably, beyond the canonical frontal valuation regions, our data-driven analysis uncovered less expected white matter pathways linked to reward sensitivity, including the right superior longitudinal fasciculus, cerebellar connections, and the motor segment of the corpus callosum. These associations, unlikely to emerge from hypothesis-driven tractography restricted to predefined valuation circuits, highlight the capacity of whole-brain approaches to reveal non-canonical circuits. While the functional implications remain uncertain, these findings raise the possibility that reward-related signals are transmitted and modulated not only through core valuation hubs but also via broader fronto-parietal and motor networks, potentially influencing how individuals adjust behavioral engagement as rewards increase. Such a role would align with emerging evidence implicating fronto-parietal regions (\citealp{etzel2016reward}), the cerebellum (\citealp{kostadinov2022reward}) and M1 (\citealp{derosiere2017primary, derosiere2017learning, derosiere2025reward, pessiglione2018not, pessiglione2007brain}) in reward-related computations during decision-making.

A key contribution of the present study is the demonstration that individual differences in effort and reward sensitivity can be accurately predicted from white matter microstructure using machine learning. By moving beyond correlational analyses and using a decoding framework, we show that DWI metrics within specific clusters carry sufficient information to decode effort and reward sensitivity across multiple classifiers, highlighting the robustness of these structure-behavior associations. This decoding framework builds on recent advances applying machine learning on DWI metrics in clinical settings (e.g., to predict Parkinson’s progression, \citealp{huang2024structural} or catatonia, \citealp{peretzke2025deciphering}) and extends them to dimensional constructs relevant to motivation. Crucially, our findings suggest that microstructure metrics could be leveraged to anticipate the impact of white matter alterations, whether pathological (e.g., following stroke, multiple sclerosis, tumor infiltration, etc) or iatrogenic (e.g., following surgical tumor resection, callosotomy, etc), on goal-directed behavior. Such predictive modelling may support personalized risk assessment and pre-operative planning in the context of surgery (\citealp{aylmore2025use, essayed2017white, ordonez2023preoperative}), as well as inform targeted rehabilitation strategies (e.g., reward-based interventions, \citealp{vassiliadis2021reward, vassiliadis2022reward}), and guide the development of neurotechnological interventions targeting key hubs of the valuation network (\citealp{vassiliadis2024non}) particularly in cases of motivational impairments such as apathy or impulsivity, which remain challenging to anticipate.
\subsubsection{Machine learning classifiers reliably predict individual sensitivities from microstructural features.}
Decoding performance was systematically higher for reward sensitivity than for effort sensitivity across both accuracy and AUC metrics. This asymmetry likely reflects fundamental differences in the neurobiological organization of these two motivational constructs. Reward valuation engages a well‐defined mesocorticolimbic network, and white‐matter microstructure in key tracts (such as the forceps minor and anterior cingulum) has been shown to relate to individual differences in reward-related processing. (\citealp{koch2014association}). In contrast, effort sensitivity may depend more strongly on dynamic integration between valuation, control, and motor systems, yielding weaker or more distributed microstructural correlates. Consistent with this, white matter correlates of effort sensitivity are more sparsely localized (\citealp{derosiere2025fronto}). Methodologically, higher classification performance for $\beta_{\text{Reward}}$ may also arise from greater behavioural reliability and variance in reward sensitivity estimates, or from lower anatomical variability in the corresponding tracts. Together, these findings suggest that reward sensitivity is more tightly anchored to stable anatomical features of the white matter, whereas effort sensitivity likely relies on flexible, state-dependent network dynamics that extend beyond what can be captured by diffusion metrics alone.

These findings provide a robust anatomical mapping of individual differences in effort and reward sensitivity. By integrating data-driven tract identification with predictive modelling, we isolate specific white matter pathways, particularly SMA-connected tracts, that reliably account for variability in motivational processes. This framework offers a concrete basis for anticipating how structural disruptions may affect goal-directed behaviour, with potential relevance for clinical applications in neurology and psychiatry.

\subsubsection{Methodological considerations and future work}
Several methodological considerations should be further examined. First, although the decoding framework incorporated nested cross-validation and permutation testing to minimize overfitting, the sample size (n = 45) remains moderate for machine learning experiments. Future studies should aim to replicate these findings in larger and more diverse cohorts, ideally using standardized imaging methods to ensure results can be compared across different scanners and research sites. Additionally, we restricted our analyses to structural predictors. Given the inherently dynamic nature of motivation, future work should integrate multimodal data—combining diffusion metrics with resting-state or task-based functional connectivity to test whether effort and reward sensitivity are better explained by structural–functional coupling rather than structure alone. Finally, testing this finding in clinical or long term studies will be an important next step. Future work should examine whether measures of white matter microstructure can predict changes in motivation after brain surgery, during disease progression, or following brain stimulation. Showing that these structural features remain stable over time and are causally linked to motivational outcomes would provide strong evidence for their clinical usefulness and help build more predictive models of motivation in the human brain.

\section{Conclusion}
Using a whole-brain, cluster-based approach to white matter integrity measures extracted from DTI combined with machine learning classifiers, we showed that inter-individual variability in effort and reward sensitivity could be reliably predicted with white matter microstructural measures including FA and MD. Twelve clusters emerged in which five of them linked to effort sensitivity, all within pathways connected to frontal valuation hubs (SMA, dACC, medial OFC) and the remaining seven associated with reward sensitivity including fronto-parietal, cerebellar, and sensorimotor circuits. We found a shared cluster in the SMA segment of the corticospinal tract associated with both effort and reward sensitivity. More broadly, reduced integrity within SMA originating circuits including both corticospinal and interhemispheric callosal pathways may disrupt SMA-driven coordination of action initiation and inhibition, thereby amplifying individual sensitivity to effort and reward through altered intra- and interhemispheric communication within the valuation network. Microstructural metrics from these clusters decoded high and low sensitivity with accuracies and AUCs consistently above chance, were classifier-independent, and were specific to the clusters uncovered by the voxel-wise analyses. 

Feature importance and recursive feature elimination converged on SMA-connected tracts as the most important predictors in both predicting effort and reward sensitivity, with additional contributions from fronto-parietal and motor pathways for reward sensitivity. Together, these results show that differences in motivation can be linked to specific white matter pathways and that the integrity of these pathways contains enough information to predict how people differ in effort- and reward-based decisions.

Building on this chapter about associations between white matter integrity and effort and reward sensitivity, the next chapter turns to grey matter volumes. Specifically, we will examine whether regional gray matter volumes relate to variability in subclinical apathy, as well as in individual differences in effort and reward sensitivity. By incorporating a whole-brain data-driven analysis with gray matter volume and the behavioural parameters established here including measures of subclinical apathy and effort–reward sensitivities, we aim to further investigate how gray matter morphology contributes to the neural architecture underlying subclinical apathy among healthy participants.

\chapter{Grey matter volume and computational modelling of effort sensitivity, reward sensitivity and subclinical apathy}
In the previous chapter, we examined the relationships between reward and effort sensitivity and white matter integrity measures include FA and -MD, providing crucial insights into the structural connectivity underlying motivational processes. An essential complementary question arises: what are the associations between regional grey matter and these sensitivities? Understanding the grey matter correlates of individual differences in effort and reward sensitivity is fundamental to developing a comprehensive neural correlates of apathy and EBDM. While recent work emphasises that white matter does not simply act as passive pathways but actively contributes to modulating neural activity, grey matter remains crucial because it contains the local computational units where synaptic integration and neuronal processing occur. This chapter aims to investigate the associations between grey matter morphometry, specifically regional grey matter volumes, and individual differences in effort and reward sensitivity and then, extends the analysis to subclinical apathy.

\section{Introduction}
Apathy, defined as a diminished motivation to initiate and sustain goal-directed behaviour, represents one of the most prevalent and debilitating symptoms observed across various neuropsychiatric and neurodegenerative disorders (\citealp{levy2006apathy, marin1991apathy}). Clinically significant apathy is highly prevalent among individuals with behavioural variant frontotemporal dementia, Alzheimer's Disease (\citealp{kumfor2018apathy}), Parkinson’s Disease (\citealp{PEDERSEN2009295}) and major depressive disorder (\citealp{morin2025chronic}). Beyond its high prevalence, apathy is associated with poorer functional outcomes, increased caregiver burden, reduced treatment adherence, and diminished quality of life across diagnostic categories. Despite its clinical significance, the neurobiological mechanisms underlying apathy remain incompletely understood (\citealp{steffens2022neurobiology}). Even within healthy populations, subclinical variations in apathy are present and represent meaningful individual differences (\citealp{ang2017distinct, sockeel2006lille}). These variations can affect everyday functioning, occupational performance, social engagement, and overall well-being, but their neural underpinnings remain unclear (\citealp{bonnelle2016individual, husain2018neuroscience}). Understanding the structural brain correlates of subclinical apathy in healthy individuals offers several advantages: it allows investigation of apathy-related neural mechanisms without the confounding effects of disease processes, medication effects, or comorbid symptoms (\citealp{levy2006apathy, radakovic2014developing}); it may identify neural markers that predict vulnerability to developing clinical apathy (\citealp{groeneweg2014quality}); and it can reveal whether the neural substrates of subclinical apathy resemble those observed in clinical populations. Although prior work has identified associations between apathy and structural brain changes, particularly in fronto-temporal regions, the anterior cingulate cortex, and basal ganglia structures (\citealp{le2018anatomy, nair2022imbalanced, caga2021apathy, morris2023altered}), these findings largely derive from patient populations. In healthy individuals, a study indicated that subclinical apathy reflects different neural mechanisms between men and women and suggested that microstructural brain alterations observed in healthy females may represent an early marker that could help predict progression toward clinical apathetic symptoms. Despite this work, structural correlates of subclinical apathy in healthy individuals remains poorly understood.

One key cognitive process implicated in apathy is effort-based decision-making, which involves evaluating the trade-off between rewards and the physical or mental effort required to obtain them (\citealp{hartmann2018effort, bonnelle2015characterization}). This process is fundamental to motivated behaviour, as individuals must constantly decide whether potential rewards justify the expenditure of effort. Dysfunction in this cost-benefit computation has been linked to impairments in goal-directed behaviour, a core symptom of apathy (\citealp{le2018apathy}). Grey matter volume (GMV), measured using structural MRI, provides a valuable index of individual differences in brain structure. Variations in GMV have been associated with differences in cognitive abilities, personality traits, and psychiatric symptoms (\citealp{takeuchi2017global, hilger2020predicting}). Examining the GMV correlates of apathy and effort-based decision-making may therefore reveal the structural brain regions supporting motivation.

In chapter 4, we used a computational modeling approach to model effort-based decision making with separable parameters: effort sensitivity (how much an individual values the effort required to obtain rewards) and reward sensitivity (how much an individual values potential rewards). We then attempted to find correlations between white matter integrity measures and these parameters. In this chapter, we aim to investigate the grey matter correlates of effort and reward sensitivity in a healthy population. In addition to these analyses, we also examine the relationship between apathy scores and grey matter volumes, as apathy has been consistently associated with structural alterations in specific cortical and subcortical regions involved in motivation, such as the anterior cingulate cortex, orbitofrontal cortex and basal ganglia. Specifically, we examine how GMV relates to (1) effort sensitivity, derived from computational models, (2) reward sensitivity, similarly quantified through computational modeling of behavioural choices; and (3) subclinical apathy scores, as measured by validated self-report questionnaires. By employing data-driven approaches, including LASSO regression, our goal is to identify the specific brain regions whose GMVs predict individual differences in these motivational aspects.

%We hypothesized that subclinical apathy would be associated with reduced GMV in frontal-subcortical regions previously implicated in clinical apathy, including the prefrontal cortex, anterior cingulate cortex, and basal ganglia. We further hypothesized that reward and effort valuation would show partially dissociable structural correlates, with reward valuation associated with striatal and orbitofrontal regions involved in reward processing, and effort valuation associated with motor and insular regions involved in effort representation and interoception. Finally, we predicted that the structural correlates of apathy would overlap partially, but not completely, with those of reward and effort valuation, consistent with the hypothesis that apathy reflects dysfunction in multiple component processes of motivated behavior.

\section{Methods}
\subsection{Participants and MRI data acquisition}
The same cohort of forty five participants described in Chapter~4 was included in all analyses presented in this chapter.
High-resolution structural images were acquired for each participant using a 3D T1-weighted sequence on a 3T \textit{GE SIGNA\texttrademark{} Premier} scanner (GE Healthcare, Chicago, IL). MRI acquisition parameters were as follows: echo time (\textit{TE}) = 2.96~ms, repetition time (\textit{TR}) = 2238.93~ms, inversion time (\textit{TI}) = 900~ms, 170 slices, slice thickness = 1~mm, in-plane field of view (FOV) = 256~$\times$~256~mm$^2$, and matrix size = 256~$\times$~256, yielding 1~mm isotropic voxels.

\subsection{Image preprocessing and parcellation}
Structural MRI data were processed using the \texttt{FreeSurfer} software package (\url{https://surfer.nmr.mgh.harvard.edu/}) for automated volumetric segmentation (see Figure \ref{fig:chapter-5-method}). Cortical and subcortical structures were parcellated using the Brainnetome Atlas, yielding volumetric estimates for each brain region of interest (ROI). We employed an atlas-based parcellation approach because it provides reliable, anatomically interpretable volumetric estimates for both cortical and subcortical regions, including key structures such as the basal ganglia that are central to effort-based decision-making. All segmented volumes were visually inspected for accuracy and corrected when necessary. Each participant's regional GMV was normalized by total intracranial volume to control for inter-individual differences in brain size.

\subsection{Questionnaires and processing}
In this study, we examined the relationship between regional GMV and apathy. Given the frequent co-occurrence of apathy, depression, and anhedonia (\citealp{husain2018neuroscience}), apathy was considered as the dependent variable, while depression and anhedonia were included as covariates to control for potential confounding effects. Participants were instructed to complete three computerized self-report questionnaires assessing apathy, depression, and anhedonia, respectively, implemented using the \textit{PsyToolkit} software.

Apathy was assessed using the extended version of the Lille Apathy Rating Scale (LARS-e), adapted for use in healthy individuals by \citealp{bonnelle2015characterization}. The LARS-e comprises 51 items, each rated on a 5-point scale (1–5) reflecting motivational level, with higher values indicating greater motivation. The apathy score was computed by subtracting each item’s score from the maximum (5), summing across items, and dividing by the total number of items, yielding a mean score between 1 and 5, where higher scores indicate greater apathy (\citealp{bonnelle2016individual}).

Because apathy frequently co-occurs with depression and anhedonia (\citealp{husain2018neuroscience}), considered these two measures as covariates in our analysis. Depression and anhedonia were assessed using the Depression Anxiety Stress Scales (DASS-21; \citealp{bonnelle2016individual}) and the French version of the Snaith–Hamilton Pleasure Scale (SHAPS; \citealp{snaith1995scale}), respectively.  

\subsection{Statistical data analysis}
\begin{figure}
    \centering
    \includegraphics[width=1\linewidth]{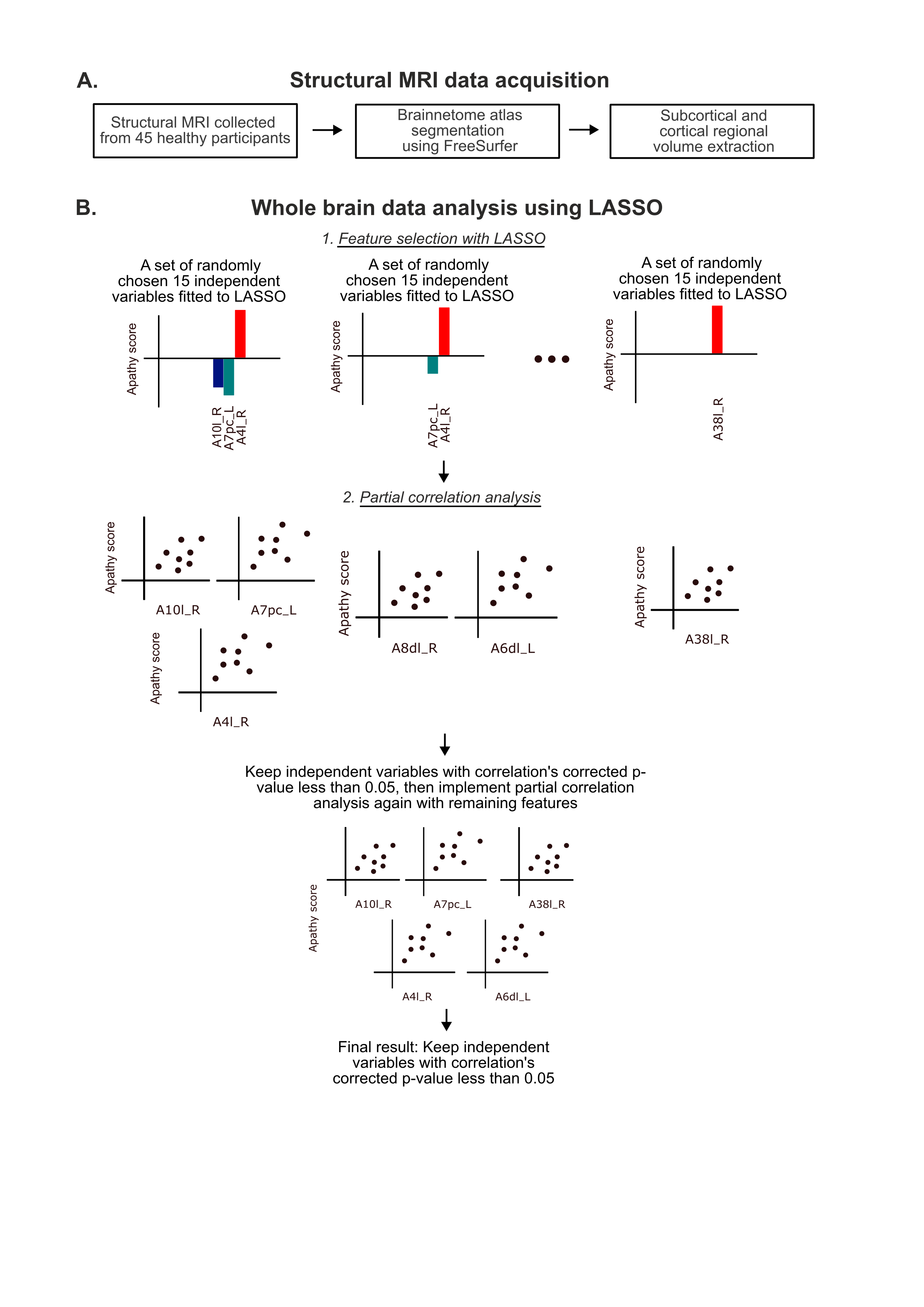}
    \caption[GMVs from structural MRI and correlational analysis method]{\textbf{GMV from structural MRI and correlational analysis method. }\textbf{A.} Structural MRI data were acquired from 45 healthy participants and processed using FreeSurfer for volumetric segmentation and the Brainnetome Atlas for parcellation
(Caption continued next page).}   
    \label{fig:chapter-5-method}
\end{figure}
\begin{figure}
    \ContinuedFloat
    \caption[]{\textbf{B.} Whole-brain feature selection and statistical testing pipeline were conducted to find significantly correlational regional GMVs. In each iteration, 15 randomly selected regional brain volumes were entered into a LASSO regression model with a target variable as the dependent variable. The regularization parameter was optimized using 10-fold cross-validation. Regions with non-zero coefficients across repetitions were retained for further analysis. Surviving features were then tested for association with target variables using partial correlation analyses controlling for age, gender, depression, and anhedonia. Regions showing significant associations after false discovery rate (FDR) correction (p $<$ 0.05) were retained.}
\end{figure}

To identify brain regions that showed significant associations with the target variables (apathy scores from the apathy questionnaires and effort sensitivity ($\beta_{\text{Effort}}$) and reward sensitivity ($\beta_{Reward}$) from the computational modelling of the effort-based decision making task as in Chapter 4), we used a similar two-stage procedure combining feature selection with elastic absolute shrinkage and selection operator (LASSO) and partial correlation analysis proposed by \citealp{derosiere2025fronto}, but expanded for a whole-brain, data-driven analysis (see Figure \ref{fig:chapter-5-method}.B):

\begin{enumerate}
    \item Feature selection: In each iteration, a random subset of approximately 10--15 regional GMVs was sampled from the full feature set, and a LASSO was fitted to predict the target behavioral measure. Features with non-zero regression coefficients were considered provisionally associated with the behavioural measure and retained for further testing. The procedure was repeated over many iterations to minimize sampling bias, and the total number of unique features selected at least once was recorded.
    \item Partial correlation: each regional GMV identified during the feature selection stage was then tested for its association with the target variables using partial correlation analyses that controlled for demographic and clinical covariates (age, gender, depression score, and anhedonia score). Partial correlations were computed using two-tailed tests, and the corresponding effect sizes and P values were extracted.
\end{enumerate}
To ensure robustness and minimize the influence of random sampling, the full analysis was repeated across multiple cross-validation configurations and random initializations. Specifically, $K$-fold cross-validation was performed with $K \in \{3, \ldots, 9\}$, and for each $K$, the procedure was repeated across ten different random resamples. Within each configuration, the analysis was iterated 100 times, yielding a total of 100 iterations that jointly varied both the number of cross-validation folds and the random state.
False discovery rate (FDR) correction for multiple comparisons was applied at two hierarchical levels. First, within each iteration, P values were adjusted across all LASSO-based selected features after the first step. Second, multiple comparison correction was applied again when a final round of partial correlation analyses was performed across this aggregated set. Statistical significance was defined as corrected P $<$ 0.05.

Regarding implementation, all analyses were conducted in Python using standard machine-learning and statistical libraries. The LASSO feature selection stage were implemented with scikit-learn (\citealp{pedregosa2011scikit}), partial correlations were computed using Pingouin (\citealp{Vallat2018}), and multiple comparison corrections were performed with the statsmodels library (\citealp{seabold2010statsmodels}). All statistical tests were two-tailed with a threshold of 0.05 after correction.

\subsection{Machine learning analysis}
We applied the same supervised machine learning analysis method described in Chapter 4 to test whether individual differences in  \(\beta_{\text{Effort}}\), \(\beta_{\text{Reward}}\) and apathy scores could be predicted from regional GMVs. This approach provides a predictive rather than purely correlative framework, allowing us to estimate how structural variations in cortical grey matter may contribute to inter-individual variability in motivational sensitivity.

For each analysis, GMVs extracted from significant clusters identified in the previous LASSO feature selection analysis and used as input features. Participants were classified into high and low groups for apathy scores, \(\beta_{\text{Effort}}\) and \(\beta_{\text{Reward}}\) based on a median split of individual parameter estimates, yielding a binary classification task. We evaluated 12 popular standard supervised classifiers: logistic regression, support vector machines, random forests, gradient boosting, extra trees, adaptive boosting, decision trees, multilayer perceptron, k-nearest neighbors, Gaussian naive Bayes, linear discriminant analysis, and quadratic discriminant analysis—using the same nested cross-validation strategy as in Chapter~4. The outer loop consisted of stratified 5-fold cross-validation to estimate generalization performance, while the inner loop optimized hyperparameters via grid search. This procedure was repeated 1{,}000 times with random resampling to ensure robust and unbiased performance estimates.

For each iteration, classification accuracy and the area under the receiver operating characteristic curve (AUC) were computed as performance metrics. Accuracy provides an intuitive measure of overall performance, reflecting the proportion of correctly predictions. However, accuracy alone can be biased in the presence of class imbalance or unequal decision thresholds. To complement this, the AUC summarizes classifier sensitivity and specificity across all possible thresholds, providing a threshold-independent measure of discriminability between classes. The best-performing classifier was identified as the one achieving the highest mean AUC across iterations. Statistical significance of classification performance was assessed using permutation testing (1{,}000 iterations with randomly shuffled class labels), computing Monte Carlo \(p\)-values, one-sided \(t\)-tests against permuted distributions, and comparisons to chance-level performance (AUC = 0.5). Significance was set at \(p < 0.05\).

Together, this analysis tested whether regional GMVs encode sufficient information to predict motivational parameters, providing a structural substrate complementary to the white matter analyses presented in Chapter~4.

\section{Results}
\subsection{Associations between GMVs and effort and reward sensitivities and apathy scores}

To examine the neural correlates of apathy and motivational sensitivity, we conducted whole-brain data-driven analyses with LASSO correlating GMVs to individual differences in effort sensitivity, reward sensitivity and apathy scores. Across these analyses, distinct yet partially overlapping fronto-temporo-parietal and striatal regions showed significant associations between regional GMV and behavioural measurements (see Table \ref{tab:chapter-5-gmv-correlations} for a summary of the region name, MNI coordinates, correlation coefficients, and significance values for each identified region).

\begin{table}
\centering
\caption{Brain regions showing significant correlations between GMV and apathy scores, effort, and reward sensitivity}
\resizebox{\textwidth}{!}{
\begin{tabular}{lccc}
\hline
\\[-4pt]
\textbf{Region} & \textbf{MNI coordinates (x, y, z)} & \textbf{r} & \textbf{p} \\ \\[-4pt]
\hline \\[-2pt]

%\hline \\[-2pt]
\multicolumn{4}{l}{\textbf{Effort sensitivity}} \\
Right dorsal insula & 39, $-7$, 8 & 0.315 & 0.0400 \\
Left postcentral superior parietal lobule & $-22$, $-47$, 65 & 0.397 & 0.00832 \\
Left primary motor cortex  & 5, $-21$, 61 & 0.372 & 0.0142 \\
Right rostroventral inferior parietal lobule & 55, $-26$, 26 & 0.331 & 0.0303 \\ \\[-4pt]
%\hline \\[-2pt]
\multicolumn{4}{l}{\textbf{Reward sensitivity}} \\
Left ventromedial putamen & $-23$, 7, $-4$ & 0.356 & 0.0193 \\
Right postcentral area of superior parietal lobule & 23, $-43$, 67 & 0.428 & 0.00422 \\ \\[-4pt]
\multicolumn{4}{l}{\textbf{Apathy}} \\
Right supplementary eye field & 7, 16, 54 & $-0.378$ & 0.0125 \\
Right medial prefrontal cortex & 13, 48, 40 & $-0.390$ & 0.00975 \\
Right supplementary motor area & 39, $-7$, 8 & $-0.413$ & 0.00587 \\
Left ventrolateral fusiform gyrus & $-55$, $-60$, $-6$ & $-0.460$ & 0.00189 \\
Right dorsolateral fusiform gyrus & 51, 6, $-32$ & $-0.325$ & 0.0333 \\
Right caudal superior temporal gyrus & 39, $-7$, 8 & $-0.313$ & 0.0410 \\ \\[-4pt]
\hline 
\end{tabular}}
\label{tab:chapter-5-gmv-correlations}
\end{table}

\subsubsection{Effort sensitivity and grey matter volume}
Partial correlation analyses revealed significant positive associations between effort sensitivity ($\beta_{\text{Effort}}$) and regional GMV across several cortical regions. Higher effort sensitivity was associated with greater GMV in the right dorsal insula (MNI coordinates: 39, –7, 8; R = 0.315, p = 0.040), the left postcentral superior parietal lobule (MNI coordinates: –22, –47, 65; R = 0.397, p = 0.00832), the left primary motor cortex (leg area) (MNI coordinates: 5, –21, 61; R = 0.372, p = 0.0142), and the right rostroventral inferior parietal lobule (MNI coordinates: 55, –26, 26; R = 0.331, p = 0.0303). These associations suggests that individuals with greater sensitivity to effort demands increased cortical volume within insula, parietal and motor regions.
\begin{figure}
    \centering
    \includegraphics[width=.85\linewidth]{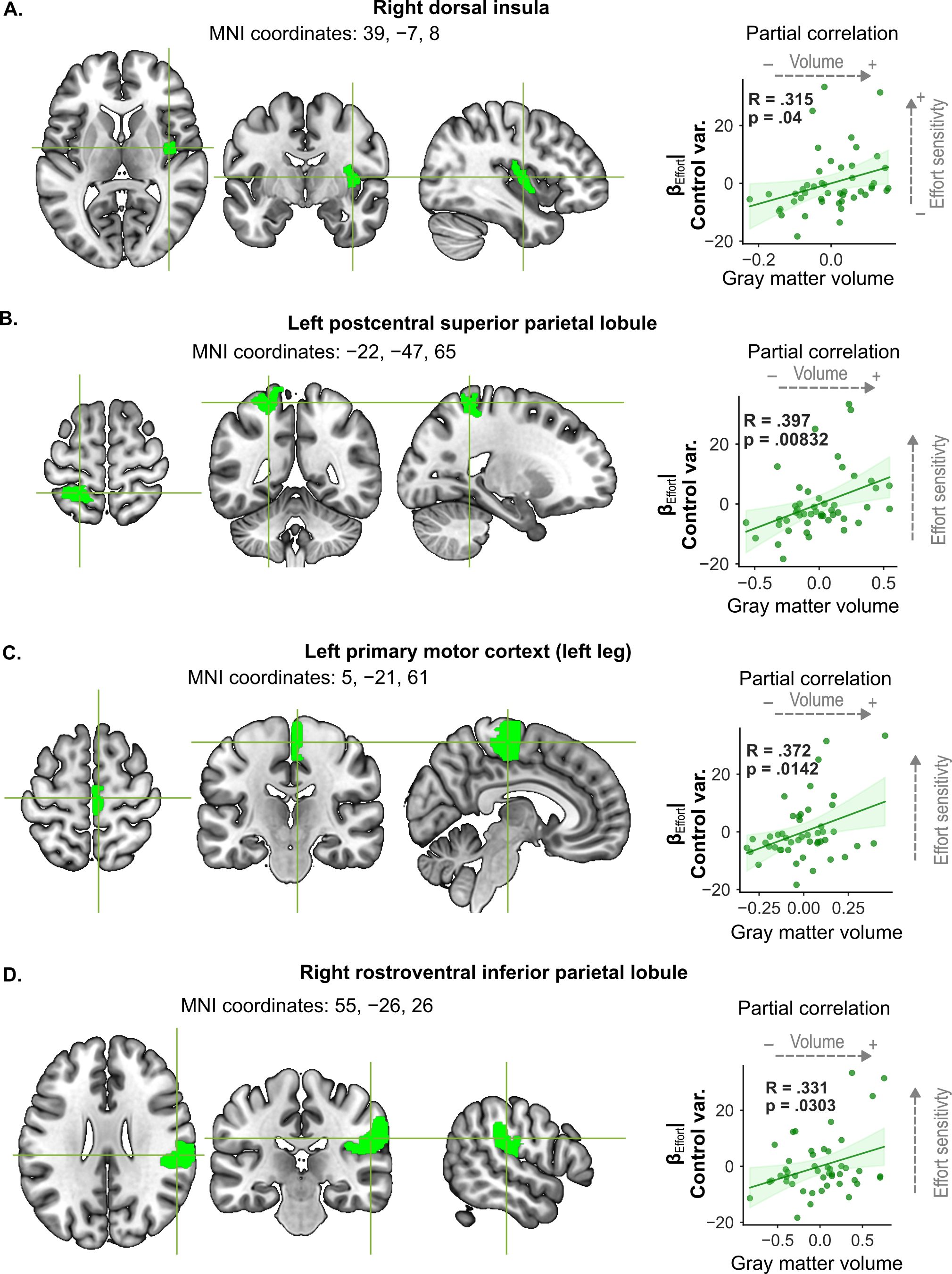}
    \caption[Associations between GMVs and effort sensitivity]{
Higher effort sensitivity was associated with greater GMV in 
\textbf{A.} the right dorsal insula (MNI coordinates: 39, $-7$, 8; $R = 0.315$, $p = 0.040$), 
\textbf{B.} the left postcentral superior parietal lobule (MNI coordinates: $-22$, $-47$, 65; $R = 0.397$, $p = 0.00832$), 
\textbf{C.} the left primary motor cortex (leg area) (MNI coordinates: 5, $-21$, 61; $R = 0.372$, $p = 0.0142$), 
and \textbf{D.} the right rostroventral inferior parietal lobule (MNI coordinates: 55, $-26$, 26; $R = 0.331$, $p = 0.0303$).
Scatter plots on the right depict partial correlations between regional GMV and effort sensitivity after controlling for age, gender, depression, and anhedonia.}
    \label{fig:chapter-5-beff-gmv-correlation}
\end{figure}

\subsubsection{Reward sensitivity and grey matter volume}
\begin{figure}
    \centering
    \includegraphics[width=.85\linewidth]{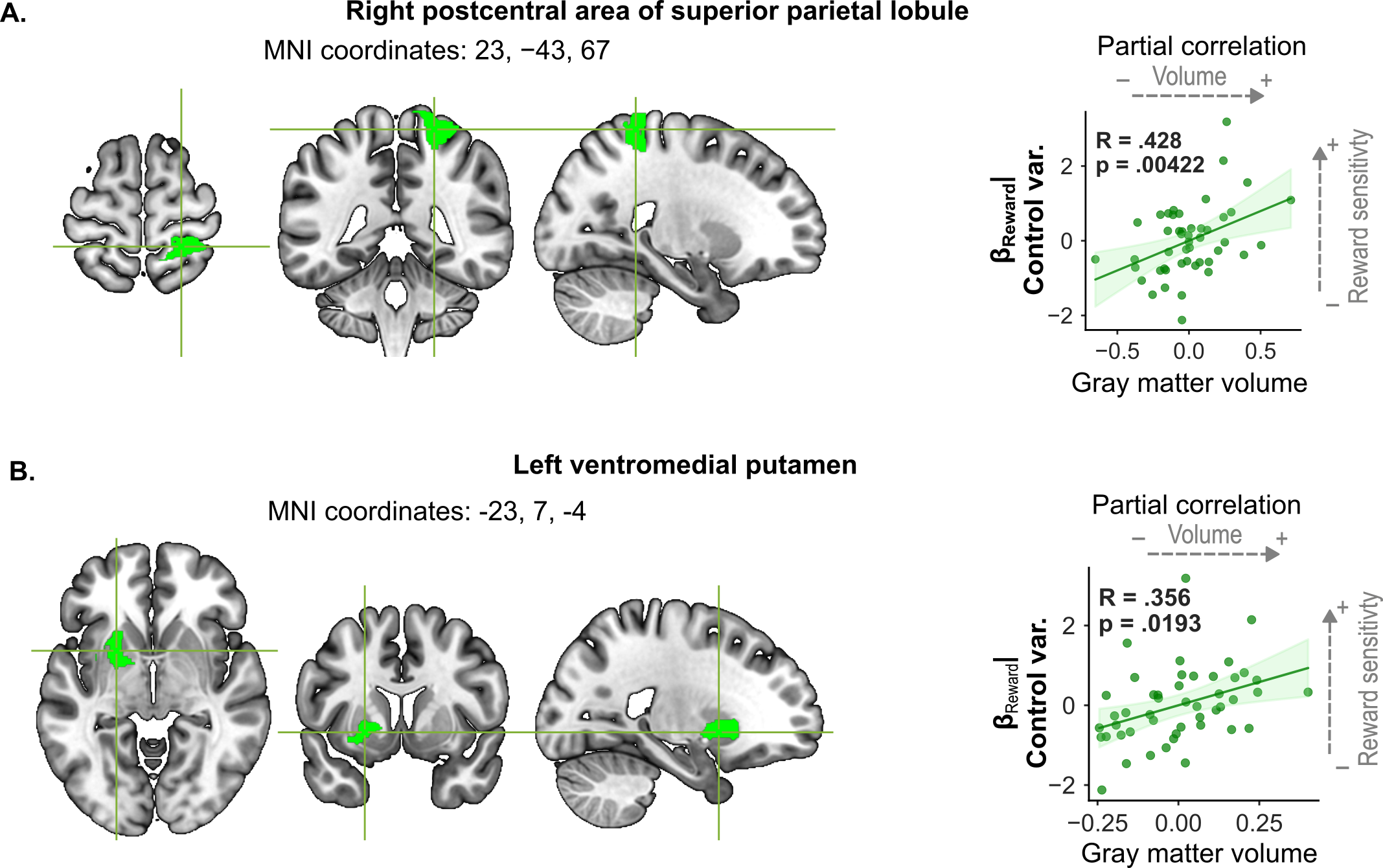}
    \caption[Associations between GMVs and reward sensitivity]{
Reward sensitivity ($\beta_{\text{Reward}}$) showed positive correlations with GMVs in 
\textbf{A.} the right postcentral area of the superior parietal lobule (MNI coordinates: 23, $-43$, 67; $R = 0.428$, $p = 0.00422$) 
and \textbf{B.} the left ventromedial putamen (MNI coordinates: $-23$, 7, $-4$; $R = 0.356$, $p = 0.0193$). 
Individuals with lower regional volume in these areas exhibited reduced reward sensitivity. 
Scatter plots show partial correlations after controlling for age, gender, depression, and anhedonia.}
    \label{fig:chapter-5-brew-gmv-correlation}
\end{figure}

Lower volumes in the putamen and superior parietal lobule were associated with lower reward valuation, indicating decreased sensitivity to reward cues (see Figure \ref{fig:chapter-5-brew-gmv-correlation}). Specifically, we found positive correlations between reward sensitivity ($\beta_\text{Reward}$) and GMVs of the left ventromedial putamen (MNI coordinates: –23, 7, –4; R = 0.356, p = 0.0193) and the right postcentral area of the superior parietal lobule (MNI coordinates: 23, –43, 67; R = 0.428, p = 0.00422).

\subsubsection{Apathy scores and grey matter volume}
Higher apathy scores were significantly associated with lower GMVs in fronto-temporal regions (see Figure \ref{fig:chapter-5-lars-gmv-correlation}). In particular, we found negative correlations between subclinical apathy scores and GMVs in the supplementary eye field (MNI coordinates: 7, 16, 54, R = –0.378, p = 0.0125), the medial prefrontal cortex (MNI coordinates: 13, 48, 40, R = –0.390, p = 0.00975), the supplementary motor area (MNI coordinates: 39, –7, 8, R = –0.413, p = 0.00587), the left ventrolateral fusiform gyrus (MNI coordinates: –55, –60, –6, R = –0.460, p = 0.00189), the right dorsolateral fusiform gyrus (MNI coordinates: 51, 6, –32, R = –0.325, p = 0.0333), and the right caudal superior temporal gyrus (MNI coordinates: 39, –7, 8, R = –0.313, p = 0.041).

\begin{figure}
    \centering
    \includegraphics[width=.9\linewidth]{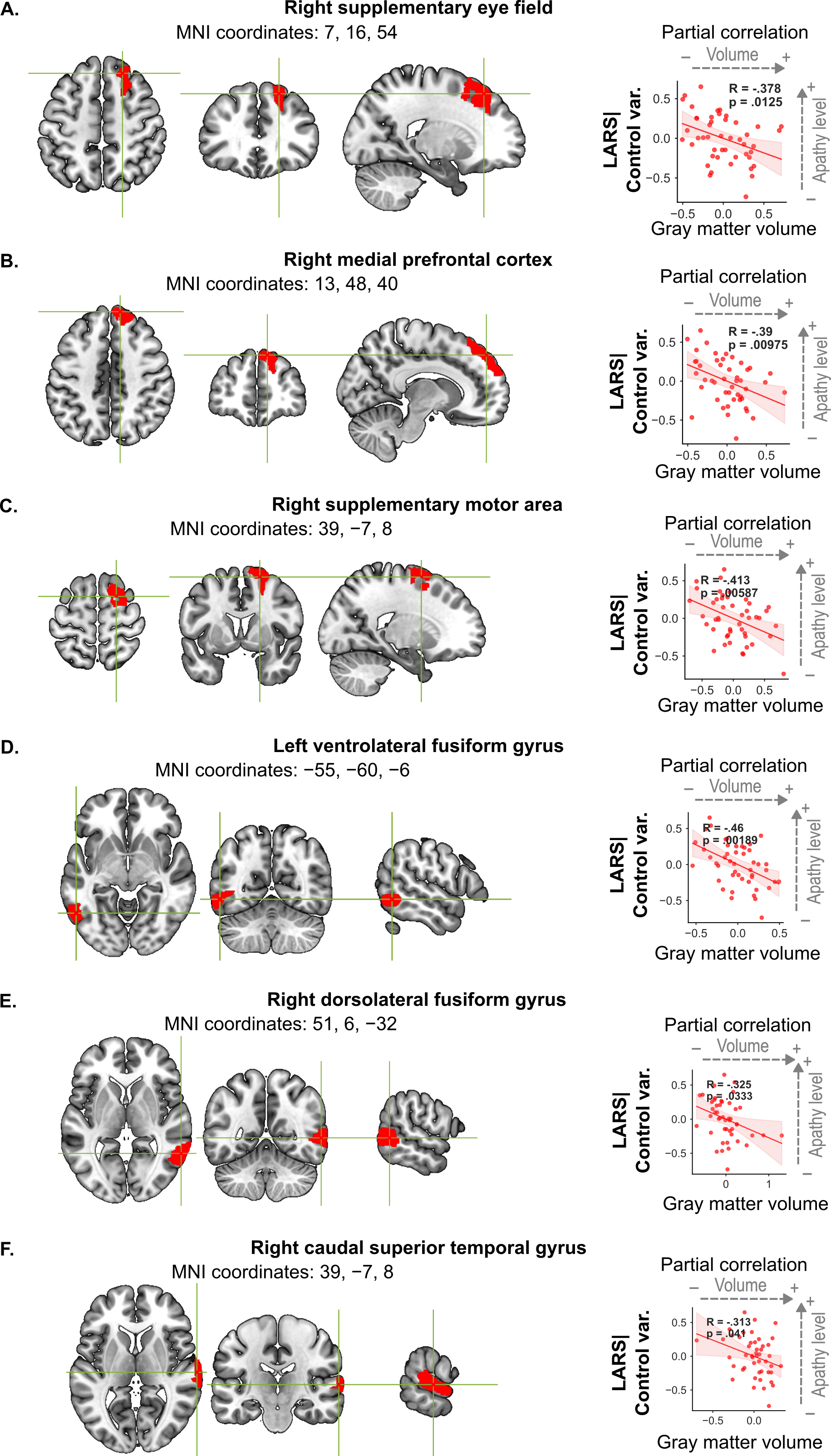}
    \caption[Associations between GMVs and apathy scores]{Apathy scores and GMV correlation (Caption continued on next page)}
    \label{fig:chapter-5-lars-gmv-correlation}
\end{figure}

\begin{figure}
    \ContinuedFloat
    \caption[]{Whole brain data-driven analyses revealed that higher apathy scores were significantly associated with lower GMVs across fronto-temporal regions. Axial, coronal, and sagittal slices show clusters identified in 
\textbf{A.} the right supplementary eye field (MNI coordinates: 7, 16, 54; $R = -0.378$, $p = 0.0125$), 
\textbf{B.} the right medial prefrontal cortex (MNI coordinates: 13, 48, 40; $R = -0.390$, $p = 0.00975$), 
\textbf{C.} the right supplementary motor area (MNI coordinates: 39, $-7$, 8; $R = -0.413$, $p = 0.00587$), 
\textbf{D.} the left ventrolateral fusiform gyrus (MNI coordinates: $-55$, $-60$, $-6$; $R = -0.460$, $p = 0.00189$), 
\textbf{E.} the right dorsolateral fusiform gyrus (MNI coordinates: 51, 6, $-32$; $R = -0.325$, $p = 0.0333$), 
and \textbf{F.} the right caudal superior temporal gyrus (MNI coordinates: 39, $-7$, 8; $R = -0.313$, $p = 0.041$). 
Scatter plots on the right depict partial correlations between regional GMV and apathy scores after controlling for age, gender, depression, and anhedonia.}

\end{figure}

\subsection{Decoding effort sensitivity, reward sensitivity and apathy levels with machine learning}
\subsubsection{Classification performance}
\begin{figure}
    \centering
    \includegraphics[width=1\linewidth]{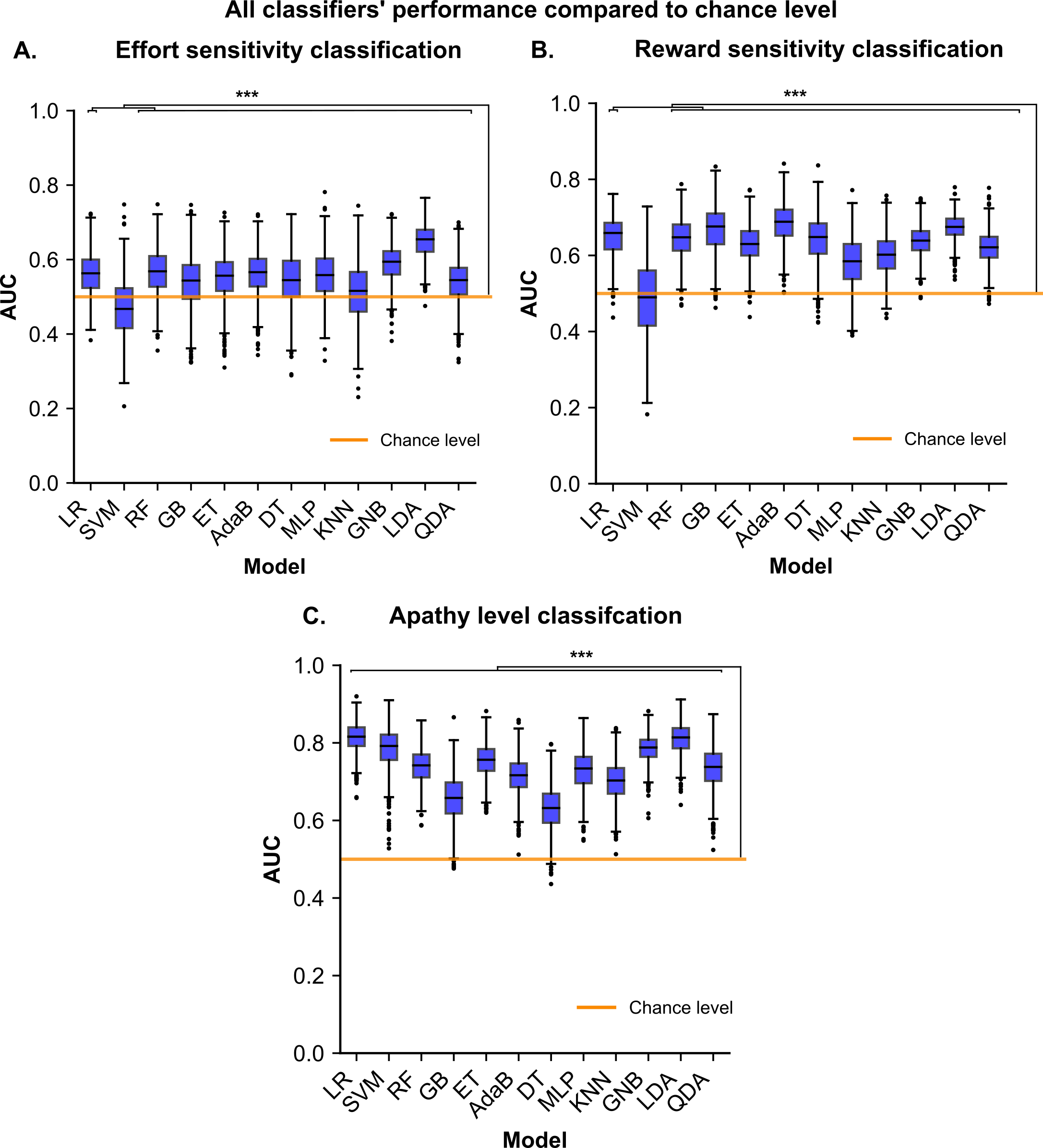}
    \caption[Classifier performance compared to chance level across classification tasks]{\textbf{Classifier performance compared to chance level across classification tasks.}
\textbf{A, B.} For effort sensitivity and reward sensitivity classifications, all classifiers except the support vector machine (SVM) achieved performance significantly higher than chance. \textbf{C.} In the apathy level classification, all models performed significantly above chance. Boxplots show the distribution of the area under the receiver operating characteristic curve (AUC) across cross-validation folds for each model. The orange line indicates the theoretical chance level (AUC = 0.5). ***p $<$ 0.001.}
    \label{fig:chapter-5-model-performance-gmv}
\end{figure}

In this study we applied supervised learning approach with input features as GMVs that we found from our correlation analysis and with binary target variable as high and low levels of effort sensitivity, reward sensitivity and apathy scores. With apathy score level classification, all classifiers performed significantly higher than chance (see Figure \ref{fig:chapter-5-model-performance-gmv}.A). With effort and reward sensitivities, all classifiers (except SVM) performed significantly higher than chance (see Figure \ref{fig:chapter-5-model-performance-gmv}.B and C). 

\subsubsection{Permutation test}
To further examine the statistical significance of our classification, we performed permutation tests using our observed performance of the three best machine learning models across three classification tasks (logistic regression for apathy score decoding, linear discriminant analysis for effort sensitivity decoding and adaptive boosting for reward sensitivity decoding) by permuting class labels and predicting these permuted labels over 1000 iterations (see Figure \ref{fig:chapter-5-ml-stat}). These iterations then yielded a classification performance of 1000, forming a null distribution. 

\begin{figure}
    \centering
    \includegraphics[width=1\linewidth]{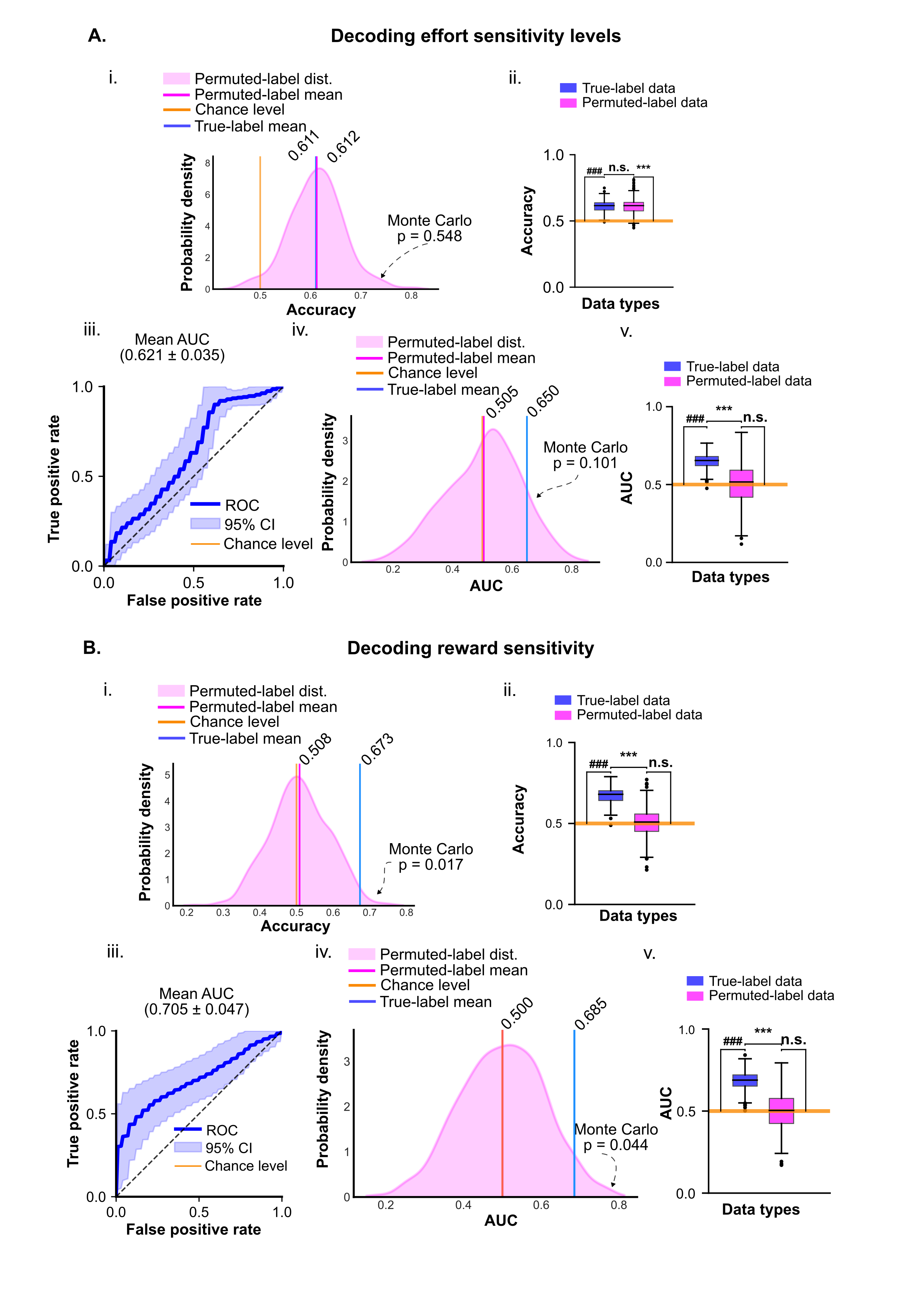}
    \caption[Permutation test on apathy score and effort sensitivity classification performance]{\textbf{Permutation test on apathy score, effort sensitivity and reward sensitivity classification performance.} Caption continued on next page. }
    \label{fig:chapter-5-ml-stat}
\end{figure}

\begin{figure}
    \ContinuedFloat
    \centering
    \includegraphics[width=1\linewidth]{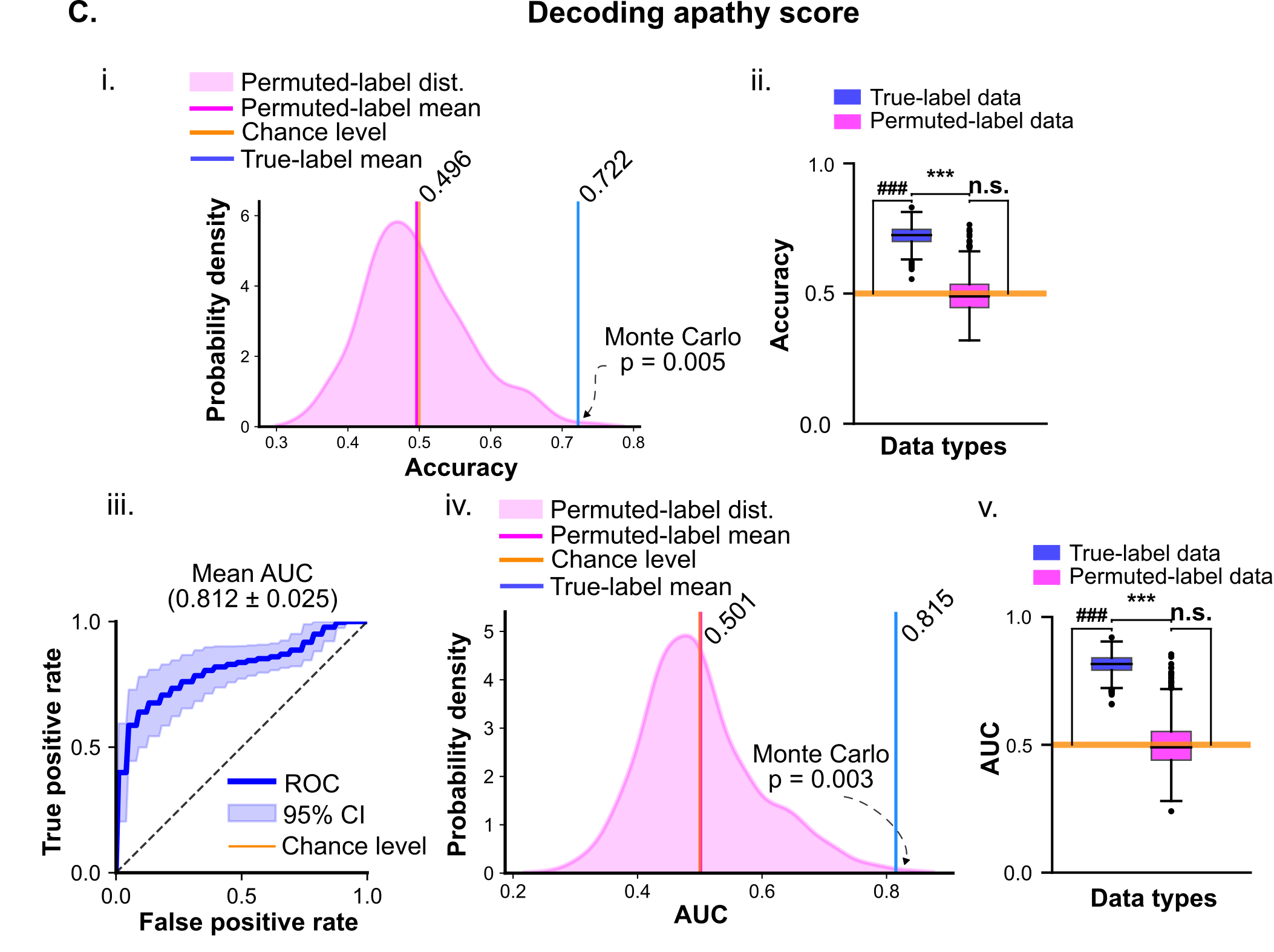}
    \caption[]{\textbf{A. Effort sensitivity classification.} (i) Classifier accuracy for true-label data (blue line) did not differ significantly from the permuted-label distribution (pink; Monte Carlo $p = 0.548$). (ii) Accuracy for true-label data was higher than chance but not significantly different from permuted-label data (true vs. permuted: $t_{998} = 1.13$, $p = 0.26$; true vs. chance: $t_{999} = 48.19$, $p < 0.001$). (iii) ROC curve showing mean AUC = $0.621 \pm 0.035$ (95\% CI). (iv) True-label AUC did not differ significantly from the permuted-label distribution (Monte Carlo $p = 0.101$). (v) AUC was significantly higher than chance but not from permuted data (true vs. permuted: $t_{998} = 1.71$, $p = 0.087$; true vs. chance: $t_{999} = 54.62$, $p < 0.001$). 
\textbf{B. Reward sensitivity classification.} (i) Classifier accuracy on true-label data (blue line) was significantly higher than the null distribution from 1,000 permutations (pink); only 17 permutations matched or exceeded true-label accuracy (Monte Carlo $p = 0.017$). (ii) Accuracy was significantly higher for true-label data (true vs. permuted: $t_{998} = 44.83$, $p < 0.001$; true vs. chance: $t_{999} = 89.72$, $p < 0.001$). (iii) ROC curve showing classifier performance (mean AUC = $0.705 \pm 0.047$). (iv) True-label AUC exceeded the permuted-label distribution (Monte Carlo $p = 0.044$). (v) AUC was significantly higher for true-label data compared to permuted and chance-level data (true vs. permuted: $t_{998} = 39.62$, $p < 0.001$; true vs. chance: $t_{999} = 96.84$, $p < 0.001$). Boxplots show median, interquartile range, and full range. Symbols denote statistical significance: \#\#\#$p < .001$, ***$p < .001$, n.s. = not significant.
Caption continued on next page.}
\end{figure}

\begin{figure}
    \ContinuedFloat
    \caption[]{
    \textbf{C. Apathy classification.} (i) Classifier accuracy on true-label data (blue line) was significantly higher than the null distribution obtained from 1,000 label permutations (pink); only 5 permutations matched or exceeded true-label accuracy (Monte Carlo $p = 0.005$). (ii) Accuracy was significantly higher for true-label data compared to permuted-label and chance-level data (true vs. permuted: $t_{998} = 47.83$, $p < 0.001$; true vs. chance: $t_{999} = 93.26$, $p < 0.001$). (iii) Receiver operating characteristic (ROC) curve showing classifier performance (mean AUC = $0.812 \pm 0.025$, 95\% CI shown in shading). (iv) True-label AUC (blue line) exceeded the null distribution from permuted-label data; only 3 permutations matched or exceeded the true-label AUC (Monte Carlo $p = 0.003$). (v) AUC was significantly higher for true-label data compared to permuted-label and chance-level data (true vs. permuted: $t_{998} = 52.41$, $p < 0.001$; true vs. chance: $t_{999} = 121.5$, $p < 0.001$).    
}
\end{figure}

Effort sensitivity classification did not generalize beyond chance-level patterns in GMV. True-label accuracy did not differ from the permuted-label distribution (Monte Carlo $p = 0.548$; $t_{998} = 1.13$, $p = 0.26$), although accuracy remained above chance ($t_{999} = 48.19$, $p < 0.001$). The classifier achieved a mean AUC of $0.621 \pm 0.035$ (95\% CI), which was higher than chance ($t_{999} = 54.62$, $p < 0.001$) but not significantly different from the permuted-label distribution (Monte Carlo $p = 0.101$; $t_{998} = 1.71$, $p = 0.087$).

For reward sensitivity classification, true-label accuracy was significantly higher than the permuted-label distribution (Monte Carlo $p = 0.017$), with only 17 permutations matching or exceeding true-label accuracy. Accuracy for true-label data exceeded both permuted and chance-level data ($t_{998} = 44.83$, $p < 0.001$; $t_{999} = 89.72$, $p < 0.001$). The classifier yielded a mean AUC of $0.705 \pm 0.047$, with true-label AUC exceeding the permuted-label distribution (Monte Carlo $p = 0.044$; $t_{998} = 39.62$, $p < 0.001$; $t_{999} = 96.84$, $p < 0.001$). 

For apathy classification, model accuracy for true-label data significantly exceeded the null distribution obtained from 1,000 label permutations (Monte Carlo $p = 0.005$), with only 5 permutations matching or exceeding true-label performance. Accuracy was significantly higher for true label versus permuted label ($t_{998} = 47.83$, $p < 0.001$) and versus chance ($t_{999} = 93.26$, $p < 0.001$). The receiver operating characteristic (ROC) curve revealed a robust mean AUC of $0.812 \pm 0.025$ (95\% CI), with true-label AUC exceeding the permuted-label distribution (Monte Carlo $p = 0.003$; true vs. permuted: $t_{998} = 52.41$, $p < 0.001$; true vs. chance: $t_{999} = 121.5$, $p < 0.001$).
\subsubsection{Feature importance analysis}
\begin{figure}
    \centering
    \includegraphics[width=1\linewidth]{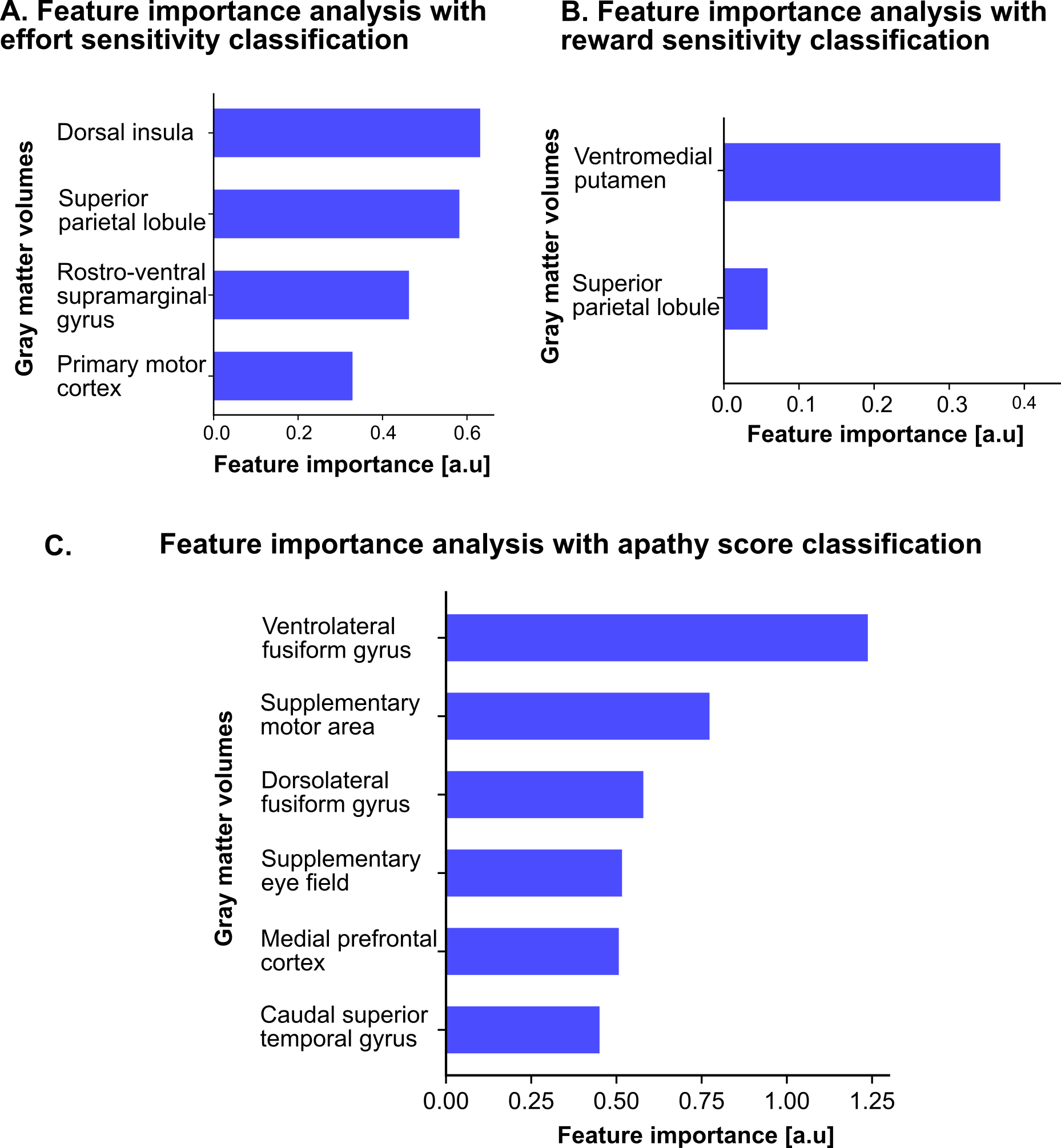}
    \caption[Feature importance analysis of the best classifiers among three classification experiments.]{\textbf{Feature importance analysis of the best classifiers among three classification experiments.} Feature importance values were derived from the best-performing classifiers predicting individual differences in apathy, effort sensitivity, and reward sensitivity from regional GMVs. \textbf{A.} For apathy classification, the ventrolateral fusiform gyrus and supplementary motor area showed the strongest contributions. \textbf{B.} For effort sensitivity classification, the dorsal insula and superior parietal lobule were the most predictive regions. \textbf{C.} For reward sensitivity classification, the ventromedial putamen and superior parietal lobule contributed most strongly to model performance.}
    \label{fig:chapter-5-feature-importance}
\end{figure}

To investigate the extent which regional GMVs contribute to the classifiers, we conducted feature importance analyses for three best classifiers (based on the AUC) from three classification experiments (see Figure \ref{fig:chapter-5-feature-importance}). Our analysis identified the regional GMVs most predictive of individual differences in apathy, effort sensitivity, and reward sensitivity. For effort sensitivity classification, the dorsal insula and superior parietal lobule showed the highest predictive value. For reward sensitivity classification, the ventromedial putamen and superior parietal lobule emerged as the strongest predictors. For apathy classification, the ventrolateral fusiform gyrus and supplementary motor area contributed most strongly to model performance.

\section{Discussion}
%% Draft

\subsubsection{Structural correlates of effort sensitivity in the insula, the superior parietal lobule, the primary motor cortex and the inferior parietal lobule}

In our study, effort sensitivity was derived from computational modeling of participants’ behaviour in a task requiring them to exert varying levels of force to obtain rewards. We found significantly positive correlations between effort sensitivity and GMVs in the dorsal insula, the caudal superior temporal lobule (STL), the primary motor cortex (M1) and the inferior parietal lobule (IPL). 

The insula has emerged as a critical region in the perception and processing of effort-related information (\citealp{treadway2012dopaminergic}). The anterior insula (AI) has been reported as being consistently activated during effortful tasks and is thought to play an essential role in interoceptive awareness (the perception of internal body states) (\citealp{craig2009you, werner2013interoceptive}). \citealp{paulus2006insular} has demonstrated that the insula plays a central role in anticipating aversive bodily states and generating avoidance behaviours. \citealp{segerdahl2015dorsal} reported that dorsal posterior insula served an important role in human experience of pain. In the context of effort, a more developed dorsal insula may amplify the salience of effort-related interoceptive signals, such as increased heart rate, muscle fatigue or pain, or metabolic demand  (\citealp{critchley2004neural}). This heightened awareness could translate into a stronger subjective experience of effort as costly or unpleasant, even when objective task demands remain constant. An alternative interpretation would be, given that the insula is a central node of the salience network, which acts to detect and integrate relevant internal and external stimuli to guide behaviour (\citealp{seeley2007dissociable}), individuals with a larger dorsal insula might be more sensitive to effort-related signals, meaning that such signals could carry greater weight in their cost-benefit decision computations. This interpretation is consistent with evidence that activity in the insula (and the anterior cingulate cortex) during decision making about effort-costly options tracks how much the reward is discounted by effort (\citealp{prevost2010separate}).

%The SPL has been implicated in a diverse array of functions, including auditory processing, language comprehension, and social cognition (1). An fMRI study combined with effort-based decision making reported a result similar to our finding, where there was a reduced response in the STG between high- and low probability tasks in patients with major depressive disorder (2).
%1 The cortical organization of speech processing
%2 Diminished caudate and superior temporal gyrus responses to effort-based decision making in patients with first-episode major depressive disorder

The parietal cortex, including regions such as the superior parietal lobule (SPL), inferior parietal lobule (IPL), and posterior parietal cortex (PPC), plays multiple roles in sensorimotor integration, attention, and decision-making (\citealp{park2014encoding}). These regions play a critical role in transforming sensory information into motor plans and in monitoring the outcomes of actions (\citealp{andersen2009intention, park2014encoding}). The correlation between larger parietal volumes and higher effort sensitivity likely reflects the parietal cortex's role in integrating information such as task demands, bodily states, and action outcomes. 
Additionally, the parietal cortex is involved in metacognitive monitoring (the ability to evaluate one's own cognitive processes and performance; \citealp{fleming2012neural}). Enhanced metacognitive sensitivity, potentially associated with larger parietal volumes, might lead to more accurate or amplified awareness of the effort being exerted during tasks. This heightened awareness could contribute to higher perceived effort costs, even when objective performance remains unchanged.

The primary motor cortex (M1) is traditionally understood as the final common pathway for voluntary movement execution, containing neurons that directly control muscle activity (\citealp{bhattacharjee2021role, kakei1999muscle}). However, contemporary neuroscience recognises that M1 is not merely a passive relay station but also actively participates in motor planning, effort scaling, and the representation of movement costs (\citealp{svoboda2018neural, li2015motor}). Structural MRI research has demonstrated that motor cortical thickness, including regions encompassing M1, correlates with individual differences in effort-based decision-making: people with thicker motor cortices tend to show greater willingness to exert effort for reward (\citealp{umesh2020motor}). Moreover, fronto–motor–basal ganglia circuits have recently been linked to effort-based valuation, suggesting that M1 interacts with prefrontal and striatal regions involved in cost–benefit computations (\citealp{derosiere2025fronto}). The correlation between larger M1 volumes and higher effort sensitivity suggests that this region may encode not just motor commands but also the anticipated costs of executing those commands. Although empirical evidence linking structural measures of M1 (e.g., volume) with effort sensitivity in behaviour is still limited, one possible interpretation is that individuals with larger motor cortices might have more elaborate representations of motor costs, making anticipated efforts more salient and potentially more aversive. 

Together, these findings suggest that effort sensitivity arises from a distributed network encompassing interoceptive, metacognitive, and motor systems.
\subsubsection{Structural correlates of reward sensitivity in the ventromedial putamen and the superior parietal lobule}
The present findings reveal significant positive correlations between reward sensitivity and GMVs in two distinct neural regions: the right superior parietal lobule (SPL) and the left ventromedial putamen (vmPu). 

The positive correlation between left ventromedial putamen volume and reward sensitivity aligns directly with the literature that the striatum has been recognised as a central hub for reward-related learning and motivation (\citealp{haber2010reward, schultz2015neuronal}). The ventromedial part of the putamen occupies a transitional region between the dorsal and ventral striatum, receiving inputs from both limbic and cortical regions, and appears to play a particularly important role in reward processing (\citealp{mawlawi2001imaging, draganski2008evidence}). The finding that individuals with greater reward sensitivity possess larger vmPu volumes suggests that individual differences in reward processing may be grounded in the anatomical properties of striatal reward circuits. Several mechanisms could account for this relationship. First, larger GMV in the vmPu may reflect a greater number of neurons potentially providing enhanced computational capacity for processing reward-related information. This increased neural substrate could support more sensitive detection of reward cues or more precise encoding of reward values. Second, the relationship between the vmPu's volume and reward sensitivity may reflect experience-dependent structural plasticity. Animal studies have demonstrated that exposure to rewarding stimuli and engagement in reward-seeking behaviours can induce structural changes in the striatum, including alterations in dendritic spine density and neuronal morphology (\citealp{colom2024conditioning, kolb2003experience, robinson2004structural}). In humans, longitudinal studies have shown that learning and experience can drive changes in GMV in task-relevant brain regions (\citealp{olivo2022estimated, draganski2004changes, maguire2000navigation}). It is therefore possible that individuals who are more reward-sensitive engage more frequently in reward-seeking behaviours, and this repeated engagement drives structural expansion of the vmPu through use-dependent plasticity. This interpretation also suggests a bidirectional relationship: structural properties of the putamen may drive individuals toward reward sensitivity, and this repeated engagement drives plasticity within reward-related striatal regions, potentially contributing to the structural expansion of the vmPu.

The posterior parietal cortex (PPC), including the SPL, has been shown to encode value-related signals and support evidence accumulation and value comparison during choice (\citealp{massar2015separate, xie2024graded}), with causal studies reporting PPC associated in reward valuation with risky decision making (\citealp{panidi2024posterior}) and with visual categorial decision making (\citealp{zhou2023posterior}). Our finding that greater GMV in the SPL correlates with higher reward sensitivity aligns with this literature, suggesting that the SPL might support an individual's ability to integrate value information and weigh potential outcomes when making decisions. Furthermore, the SPL's role in action selection and motor planning (\citealp{randerath2017contributions}) suggests that individual differences in this region's volume may relate to the translation of reward motivation into action. Reward sensitivity, as conceptualized in Reinforcement Sensitivity Theory, includes not only the subjective experience of reward but also the motivational tendency to approach and pursue rewarding stimuli (\citealp{gray1990brain, firth2024neural}). Individuals with larger SPL volumes may exhibit more efficient coupling between reward detection and behavioural response, manifesting as higher reward sensitivity.

Together, our findings extend our understanding of the neural substrates of reward processing by highlighting the importance of both traditional reward circuitry in the striatum and attentional/sensorimotor integration systems in the posterior parietal cortex. The involvement of the SPL suggests that reward sensitivity depends not only on the capacity to encode reward values but also on the ability to detect reward-relevant stimuli and translate motivational information into action. The vmPu's involvement aligns with established models emphasizing the central role of the striatum in reward processing and reinforcement learning.

\subsubsection{Structural correlates of subclinical apathy in fronto-temporal regions}
Our results revealed a significant association between subclinical apathy and reduced GMVs in a distributed network of fronto-temporal regions, specifically, the medial prefrontal cortex (mPFC), the supplementary eye field (SEF), the supplementary motor area (SMA), the fusiform gyrus and the caudal superior temporal gyrus. 

Structural MRI studies, including voxel-based morphometry, have linked apathy to grey matter reductions within frontal–striatal circuits, encompassing the anterior cingulate, medial prefrontal, and orbitofrontal cortices, as well as subcortical structures such as the ventral striatum (\citealp{reijnders2010neuroanatomical, zamboni2008apathy}). The medial prefrontal and cingulate regions have been found to support goal-directed behaviour, effort-based decision-making, and reward valuation, providing a framework that structural alterations in these areas may lead to motivational deficits (\citealp{le2018anatomy, husain2018neuroscience}). Our finding that reduced GMV in the mPFC was associated with higher levels of apathy is in line with this framework. This result supports the view that structural alterations within the prefrontal – cingulate network contributes to diminished motivation, possibly through impaired evaluation of effort and reward or disruption of goal maintenance processes (\citealp{le2018anatomy, steffens2022neurobiology}). Importantly, the association between mPFC volume and apathy remained significant after controlling for depression and anhedonia in our analysis, suggesting this brain–behaviour association is specific to motivational deficits rather than general emotional or hedonic symptoms.

The observed associations involving the SMA and SEF further underscore the contribution of medial frontal regions to motivated behaviour. The SMA, a critical node in motor preparation and self-initiated movement (\citealp{nachev2008functional, fried1991functional}), is also increasingly recognised as an integration hub integrating motivational signals with motor planning (\citealp{bonini2014action, gordon2023somato}). Lesion studies have shown that damage to the SMA produces akinetic mutism and profound apathy despite preserved motor capacity (\citealp{krainik2001role}), highlighting its critical role in behavioural initiation. Our findings suggest that structural variations in this region may manifest as subclinical motivational deficits, supporting a dimensional view of apathy that exists along a continuum from normal variation to clinical impairment. Interestingly, in Chapter 4, we also found that the FA of an overlapped cluster in the SMA was significantly associated with both effort and reward sensitivity and was the strongest predictors in our machine learning analysis (see Figure \ref{fig:chapter4-figure10}). Together, these results point to the SMA as a hub that directly supports behavioural engagement through corticospinal projections, in addition to receiving modulatory input from valuation networks. 

The SEF, located in the dorsomedial frontal cortex anterior to the SMA, plays a specialized role in the voluntary control of eye movements and visual attention (\citealp{schall1999neural}). While traditionally conceptualized as part of the oculomotor system, recent evidence indicates that the SEF also supports higher‐order control of behaviour, for example, by facilitating the exploration of visual space and selecting behaviourally relevant stimuli prior to action initiation (\citealp{purcell2010neurally, herbet2022contribution}). Reduced SEF volume in individuals with higher apathy may therefore reflect a diminished ability to actively explore or focus on important, motivational information in their surroundings, resulting in lower engagement with the environment.(\citealp{manohar2016human}).

Finally, the involvement of the fusiform gyrus, particularly the ventrolateral and dorsolateral subregions, extends these findings beyond purely motor and valuation frameworks. This region, situated in the temporal cortex, is critically involved in high-level visual processing, including object recognition and the integration of visual features into coherent representations (\citealp{weiner2012improbable}). Recent studies have increasingly recognized the importance of ventral temporal regions in value-based decision-making and reward processing, with the fusiform gyrus showing sensitivity to motivationally salient or reward-relevant visual stimuli (\citealp{schiffer2014reward}). Reduced GMV in this region may thus indicate a weakened linkage between perceptual salience and motivational valuation, contributing to diminished motivational drive.

\subsubsection{Machine learning analysis}
In this study, we applied supervised machine learning to predict individual differences in apathy, effort sensitivity, and reward sensitivity from regional GMVs. All classifiers across three classification tasks performed significantly above chance level, as confirmed by permutation testing over 1000 label shuffles with apathy and reward sensitivity, except for effort sensitivity classification, where the observed performance did not exceed the null distribution. These results indicate that GMVs contains sufficient information to predict apathy and reward sensitivity but that effort sensitivity differences between individuals may be less robustly captured by structural volumetric features alone.

To better understand which regions contributed to classification performance, we examined feature importance values derived from the best-performing classifiers across the three classification experiments (see Figure \ref{fig:chapter-5-feature-importance}). For apathy, the ventrolateral fusiform gyrus and supplementary motor area (SMA) contributed most strongly to the classifier's performance, highlighting the integration of perceptual salience and volitional control processes in sustaining motivation. For effort sensitivity, the dorsal insula and superior parietal lobule (SPL) showed the highest feature importance, further supporting the role of interoceptive and sensorimotor integration in encoding perceived effort costs. For reward sensitivity, the strong contributions of the vmPu and SPL emphasize a coupling between striatal reward valuation and parietal attention/action systems in shaping individual differences in reward responsiveness.

Machine learning classification of GMVs revealed structural predictors for apathy and reward sensitivity, but not for effort sensitivity, emphasizing both the promise and the limitations of morphological decoding of motivation. Together with the white matter microstructure findings presented earlier, these results support a distributed architecture of motivation spanning cortical, subcortical, and network systems. The absence of predictive accuracy for effort sensitivity could reflect weaker structure–behaviour coupling, higher behavioural noise, or the need for additional modalities such as functional connectivity.
\subsubsection{Methodological considerations and future directions}
While the correlation between brain structure and effort/reward sensitivity is intriguing, several methodological considerations must be acknowledged. First, structural neuroimaging measures such as GMV reflect a complex mixture of factors including neuron number, dendritic arborization, glial cells, and vasculature (\citealp{zatorre2012plasticity}). The specific cellular and molecular mechanisms underlying volume differences remain unclear.

Second, correlational findings cannot establish causality. Longitudinal studies examining how changes in brain structure relate to changes in effort/reward sensitivity and subclinical apathy over time would provide stronger evidence for causal relationships. Intervention studies examining whether training or pharmacological manipulations also change brain structure would be particularly valuable.

Future research should also examine how brain structure relates to different types of effort. Physical effort and cognitive effort may have partially distinct neural substrates (\citealp{schmidt2012neural}), and the relationship between brain structure and effort sensitivity might differ across these domains. 
\section{Conclusion}
In this chapter, our findings revealed a distributed structural architecture underpinning individual variation in motivated behaviour. Subclinical apathy was associated with reduced GMV across a fronto–temporal network encompassing the medial prefrontal and supplementary motor cortices, the supplementary eye field, and ventral temporal regions. These associations converge on a framework in which diminished integrity of prefrontal–cingulate–motor circuits impairs the generation and maintenance of goal-directed behaviour, while alterations in ventral temporal regions may disrupt the representation of motivational salience. Complementary effects were observed for effort and reward sensitivity, implicating distinct yet overlapping structural substrates. Greater effort sensitivity was associated with larger volumes in the superior parietal lobule, primary motor cortex, basal ganglia, and dorsal insula. Conversely, greater reward sensitivity correlated with increased GMV in the superior parietal lobule and ventromedial putamen, linking striatal reward circuits and parietal systems involved in value-guided attention and sensorimotor transformation. 

By combining computational measures of motivational components with structural neuroimaging, this work demonstrates that individual differences in apathy, effort sensitivity, and reward sensitivity are reflected in partially separable anatomical networks including prefrontal, striatal, insular, and parietal cortices. These results support a dimensional view of motivation that individual differences in the anatomy of fronto–striatal–parietal systems modulate how effort and reward are internally represented and translated into action. More broadly, the findings bridge the gap between subclinical motivational variation and the neural substrates of clinically significant apathy, highlighting continuous structure–function relationships within the human motivation network.

\chapter{Conclusion}
\section{Summary}
Motivated behaviour arises from the valuations of reward and effort supported by distributed cortical and subcortical networks. Dysregulation within these systems manifests across a spectrum of conditions, from motivational deficits in ADHD to reduced motivated behaviour in apathy. Both syndromes may reflect disruptions within shared motivation-related circuits. This thesis therefore integrates electrophysiological signal (EEG), structural MRI, diffusion MRI and computational approaches to investigate the neural substrates of motivation across clinical and subclinical populations.
We proposed three research questions as follows:
\begin{itemize}
    \item Research Question 1: Can task-based EEG data outperform resting-state EEG data in classifying adults with ADHD and healthy controls, and what does this reveal about the neural dynamics of inhibitory control and motivation?
    
    \item Research Question 2: Which white matter pathways underlie individual differences in effort and reward sensitivity during effort-based decision-making?
    
    \item Research Question 3: How do regional differences in grey matter volume relate to effort sensitivity, reward sensitivity and subclinical apathy and what do these associations reveal about the structural basis of motivational regulation in healthy individuals?
\end{itemize}

For Research Question 1, Chapter 3 has shown that classification models trained on stop-signal task EEG consistently outperformed those based on resting-state data, indicating that ADHD-related neural abnormalities are most evident when executive control and inhibitory processes are actively recruited. Feature importance analyses highlighted reduced gamma-band power over fronto-central, temporal and parietal regions as the primary contributors to this enhanced performance, pointing to disruptions in high-frequency oscillatory dynamics that support attention and motivation. These findings suggests that task-evoked gamma activity reflects deficits in coordinated cortical and reward-related circuits more sensitively than resting-state measures such as TBR. The chapter therefore supports the value of engaging cognitive systems to develop robust neural markers, and motivates future work integrating ERPs, spectral features, behavioural measures and computational parameters with machine learning frameworks to improve diagnostic precision and clinical utility.

For Research Question 2, Chapter 4 demonstrated that individual differences in effort and reward sensitivity can be reliably predicted from white matter microstructure, revealing separate yet partially overlapping neural mechanisms of motivational processes. Using a whole-brain, data-driven approach, the chapter identified five clusters linked to effort sensitivity, primarily within SMA-, dACC-, and OFC-connected tracts, and seven clusters linked to reward sensitivity, which additionally involved fronto-parietal, sensorimotor, and cerebellar pathways. The strongest and most consistent effects centered on SMA-connected clusters suggesting a shared role for SMA pathways in effort valuation and reward processing. Machine learning classification further confirmed that diffusion metrics (FA and MD) within these clusters could reliably predict motivational parameters. Together, these findings advance the understanding of neural mechanisms in motivation by demonstrating that white matter integrity within these clusters is robustly associated with individual differences in effort and reward sensitivity. 

For Research Question 3, Chapter 5 extended the investigation of structural correlates of EBDM to grey matter, expanding the analysis beyond effort and reward sensitivity to also include apathy scores, given the central role of apathy in disrupted motivated behaviour. This chapter provided the evidence that individual differences in effort sensitivity, reward sensitivity, and subclinical apathy were reflected in distributed grey matter regions spanning interoceptive, sensorimotor, valuation, and frontal control circuits. Effort sensitivity was associated with larger volumes in the dorsal insula, parietal cortex, and primary motor cortex, supporting an interpretation in which subjective effort costs arise from increased interoceptive awareness, heightened metacognitive monitoring, and higher motor cost representations. Reward sensitivity was linked to greater GMV in the ventromedial putamen and superior parietal lobule, highlighting the joint contributions of striatal circuits in reward valuation and posterior parietal cortex in sensorimotor information integration. Subclinical apathy was predicted by reduced GMV across fronto-temporal regions, including the mPFC, SMA, SEF, and fusiform gyrus, implicating impairments in goal maintenance, initiation, attentional exploration, and perceptual salience processing. Machine learning analyses further revealed that GMV features reliably decoded apathy and reward sensitivity but not effort sensitivity. 

Both neural correlates found in Chapters 4 and 5 highlight the potential translational value of microstructural biomarkers for diagnosing motivational impairments and guiding personalised clinical decision-making in contexts such as neurosurgery, neurological disorders, and rehabilitative interventions. Taken together, Chapters 3, 4 and 5 in this thesis demonstrate that multiple levels of neural mechanisms including task-based electrophysiological activity, white matter integrity and grey matter morphometry could be linked to individual differences in motivated behaviour.

\section{Methodological considerations and future work}
In Chapter 3, task-based EEG emerged as a more sensitive approach than resting-state EEG for identifying ADHD-related neural signatures. While this supports the value of dynamic, context-dependent measures, it also highlights key methodological challenges. EEG signals are highly state-dependent, and individual differences in attention, fatigue, or compliance can influence spectral power estimates. The stop-signal task effectively engaged inhibitory control processes, but its reliance on a single cognitive paradigm limits generalization to other cognitive aspects of ADHD. Future studies could integrate multiple cognitive tasks or combine EEG with behavioural and eye-tracking measures to capture a broader picture of attentional and motivational control. The relatively small sample size may restrict generalisability and increase the risk of overfitting in the classification models. In addition, the absence of event-related potential (ERP) analyses limits insight into the temporal dynamics of inhibitory control, particularly in relation to N200 and P300 components known to differentiate ADHD populations. Incorporating ERP measures and expanding analyses to additional cognitive paradigms would strengthen the interpretability and robustness of these findings. Future studies combining EEG with complementary behavioural and imaging data could further clarify how neural oscillations underpin motivational and inhibitory processes in ADHD.

In Chapter 4, diffusion MRI and computational modelling were combined to map how white-matter microstructure relates to effort and reward sensitivity. While the data-driven approach minimised bias in region selection, white matter integrity metrics such as fractional anisotropy and mean diffusivity are inherently indirect measures of microstructural integrity. Factors like crossing fibers or partial volume effects may reduce the interpretation of tract-specific associations. Additionally, the cross-sectional design excludes causal inference about whether white matter differences shape or result from individual differences in motivational traits. Longitudinal imaging or interventional approaches, such as non-invasive stimulation or behavioural training, could help clarify these directional relationships.

In Chapter 5, grey matter volume extracted from structural MRI was used to link structural variability to subclinical apathy and motivational traits in healthy individuals. The findings offer strong evidence for grey matter correlates of motivational traits, yet volumetric analyses remain sensitive to sample size, parcellation schemes, and normalisation methods. Self-reported measures of apathy may also underestimate subtle behavioural manifestations, introducing variance unrelated to brain structure. Extending analyses to functional MRI data would strengthen the link between structure and behaviour. 

We only conducted analysis between apathy scores and grey matter volumes in Chapter 5 but not between apathy scores and white matter integrity in Chapter 4, it would be interesting to expand Chapter 4 with this analysis (apathy and white matter). A multimodal analysis combining both grey matter morphometry and white matter integrity could also be carried out to analyse the neural correlates of effort sensitivity, reward sensitivity and subclinical apathy, providing a more integrated understanding of the neural basis of individual differences in effort and reward sensitivity.
% Final thoughts
\section{My contributions}
This work is a collaborative effort between researchers in Trinity College Dublin, UC Louvain, Dublin City University and the Lyon Neuroscience Research Centre (CNRL):
\begin{itemize}
    \item In Chapter 3, the EEG dataset and ADHD scores were collected by the Whelan Lab in Trinity College Dublin. My contributions included preprocessing the EEG data (including cleaning noise and removing artifacts), developing machine learning models and feature importance analysis, interpreting the results and writing a conference paper.
    \item In Chapters 4 and 5, the MRI data and behavioural questionnaire data were collected and processed by researchers at UC Louvain. Dr. Laurence Dricot and Dr. Quentin Dessain (UC Louvain) extracted grey matter volumes following segmentation and parcellation, as well as white matter integrity measures, including fractional anisotropy and mean diffusivity. Dr. Pierre Vassiliadis (University College London and Imperial College London) provided the computational modelling method for the EBDM task. My contributions included implementing statistical analyses, developing and evaluating machine learning models, conducting feature importance analysis, interpreting and visualising the results, and writing a research article.
\end{itemize}
\section{Final remarks}
By integrating multiple modalities of data (including EEG, T1-weighted MRI, and diffusion MRI) along with computational modeling and machine learning techniques, the thesis enhances the understanding of neural mechanisms in ADHD and individual differences in effort and reward sensitivity. Beyond its theoretical contributions, the findings in this thesis potentially provide neural biomarkers in clinical diagnosis of ADHD and motivational impairments. In clinical contexts, such neural biomarkers might potentially \textit{complement} (rather than replace) traditional interview-based subjective assessments as a diagnostic tool, providing more objective diagnosis of diminished motivation. The EEG and MRI measures linked to motivational traits could also contribute to the development of individualised and targeted interventions, from neuromodulation strategies to behavioural or cognitive training. 

This research also highlights the promising application of machine learning techniques as a research tool in neuroscience. The integration of machine learning analysis potentially enables a new way of discovering multivariate patterns that may complement traditional univariate statistical analyses. In this thesis, machine learning models were used not only to predict computational modelling parameters and behavioural measures but also to identify important features that most strongly contributed to the classification performance. This feature importance analysis approach enabled a further investigation into the neural mechanisms of underlying predictions. These machine learning approaches could further support reliable predictive diagnostics, predict therapeutic outcomes, and personalise treatments for individuals with psychiatric disorders. 

%\printbibliography
\bibliographystyle{abbrvnat}

\bibliography{references}
\appendix
%\appendixpage
\addappheadtotoc

\renewcommand\thefigure{\thesection.\arabic{figure}} 
\renewcommand\thetable{\thesection.\arabic{table}}
\counterwithin{figure}{section}
\counterwithin{table}{section}
\chapter*{Appendices}
%\clearpage
\section{Appendix materials for chapter 3}
\begin{figure}[h]
    \centering
    \includegraphics[width=1\linewidth]{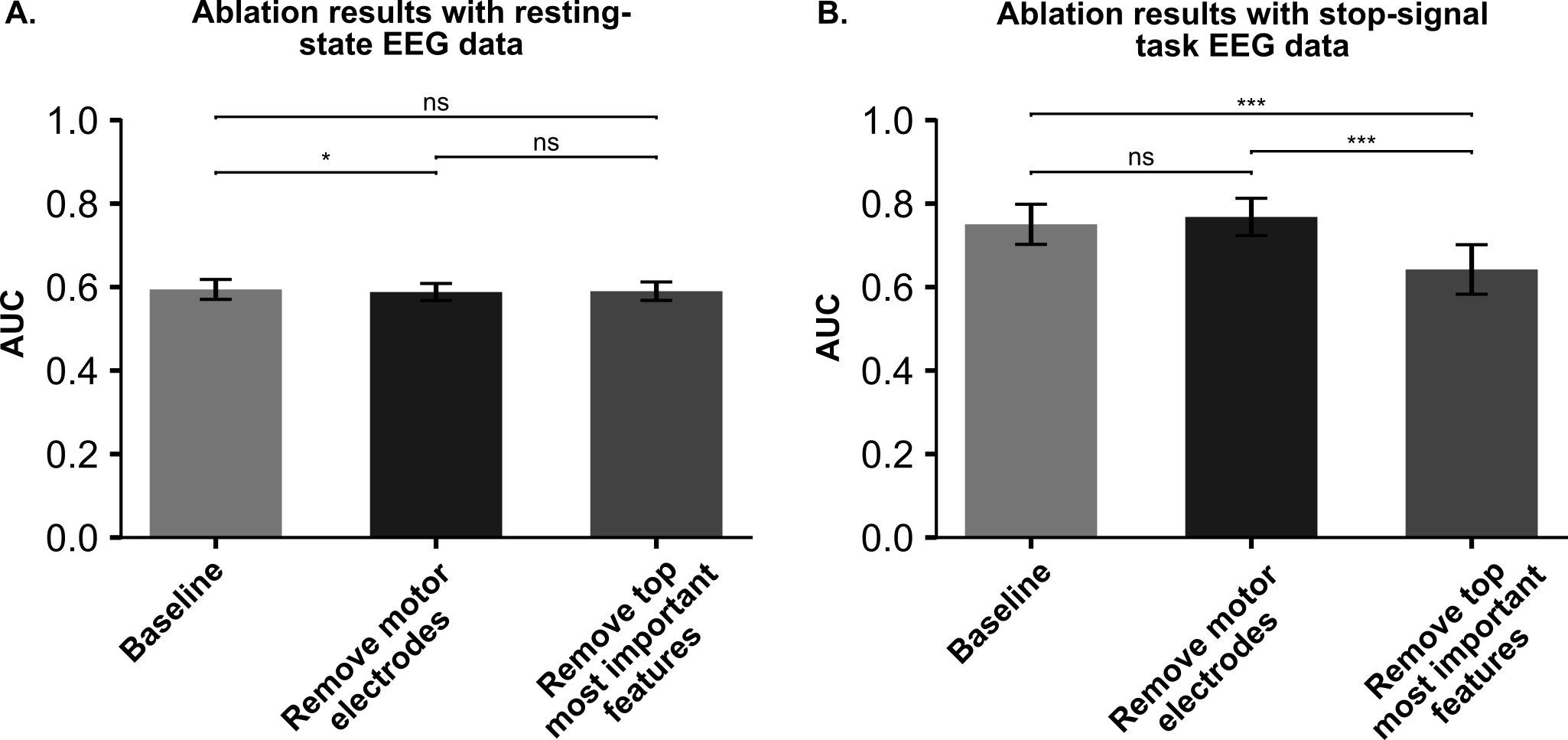}
    \caption[Ablation study of EEG feature contributions.]{\textbf{Ablation study of EEG feature contributions.}
(A) Resting-state EEG. Removing motor electrode features led to a significant reduction in AUC compared with the baseline model (*p $<$ 0.05), whereas removing the top-ranked features identified by feature importance analysis did not significantly affect performance.
(B) Stop-signal task EEG. Removing motor electrode features did not alter performance, but removing top-ranked features significantly reduced AUC compared with both baseline and motor-electrode ablation (***p $<$ 0.001). Error bars indicate ±1 SEM.}
    \label{fig:chapter-3-ablation-study}
\end{figure}

\textbf{Ablation study method: }In addition to the main analysis, we examined the impact of targeted feature removal on classification performance for resting-state and task-related EEG data (see Figure \ref{fig:chapter-3-ablation-study}). Performance was quantified using AUC, and statistical comparisons were corrected for multiple comparisons using the false discovery rate (FDR).

\textbf{Abaltion study results:}

\textbf{Resting-state EEG classification.}
Removing spectral power features from motor electrodes significantly reduced classification accuracy relative to the baseline model (t = 2.50, FDR-corrected p = 0.0186). By contrast, eliminating the top-ranked features identified through feature importance analysis did not yield a significant change in performance compared with the baseline (t = 1.69, FDR-corrected p = 0.0687). Direct comparison between the motor electrode removal and top-feature removal conditions also revealed no significant difference (t = –0.66, FDR-corrected p = 0.7446).

\textbf{Stop-signal task EEG classification.}
A different pattern emerged for task-related EEG. Removing motor electrode features had no measurable effect on performance relative to the baseline model (t = –3.52, FDR-corrected p = 0.9998). In contrast, eliminating the top-ranked features caused a marked reduction in ROC–AUC compared with both the baseline (t = 20.93, FDR-corrected $p < 0.001$) and the motor electrode removal condition (t = 16.86, FDR-corrected $p < 0.001$).

\textbf{Ablation study discussion:}

Ablation study showed two different patterns between resting-state EEG data and stop signal EEG data (see Tables \ref{tab:ablation_rsec_eeg_5_most_important_electrodes}, \ref{tab:ablation_rsec_eeg_motor_electrodes}, \ref{tab:ablation_sst_eeg_5_most_important_electrodes} and\ref{tab:ablation_sst_eeg_motor_electrodes} for more details). With resting-state, removing motor electrode features caused a significant drop in AUC relative to the baseline model (*$p < 0.05$). In contrast, removing the highest-ranked features identified through feature importance analysis did not significantly change performance. This finding supports the increasing recognition of motor system involvement in ADHD, as evidenced by studies showing altered motor cortex excitability and motor timing deficits in individuals with ADHD (\cite{gilbert2011motor, noreika2013timing}). Interestingly, the removal of the most informative features, as determined by feature-importance analysis, did not appreciably alter performance, suggesting a degree of redundancy among TBR features across electrodes (\cite{saad2018theta, kiiski2020eeg}). This redundancy may explain why, despite TBR features contributing disproportionately to classification, resting-state EEG ultimately yielded limited predictive power. With stop-signal task, removing motor electrode features had no impact on performance, but excluding the top-ranked features resulted in a significant decrease in AUC compared to both the baseline and the motor-electrode ablation conditions (***$p < 0.001$). This indicates that task-based classification relies heavily on a specific subset of gamma oscillation features involved with cognitive control and inhibitory processing and with little redundancy. Such dependence is consistent with prior evidence that task-evoked EEG signatures, particularly in the gamma range, reflect active recruitment of executive control and fronto-striatal networks (\cite{jensen2007human, ulloa2022control, dor2021high}).

\begin{table}[!h]
    \centering
      \resizebox{\textwidth}{!}{
    \begin{tabular}{cccccc}
\toprule
Model & AUC & Accuracy & Recall & Precision & F1 \\
\midrule
LR & 0.535 ± 0.030 & 0.628 ± 0.015 & 0.011 ± 0.024 & 0.027 ± 0.055 & 0.013 ± 0.026 \\
SVM & 0.464 ± 0.042 & 0.623 ± 0.020 & 0.018 ± 0.027 & 0.033 ± 0.048 & 0.021 ± 0.029 \\
RF & 0.519 ± 0.033 & 0.588 ± 0.025 & 0.171 ± 0.046 & 0.352 ± 0.108 & 0.216 ± 0.055 \\
GB & 0.495 ± 0.047 & 0.585 ± 0.033 & 0.188 ± 0.064 & 0.353 ± 0.123 & 0.225 ± 0.068 \\
ET & 0.533 ± 0.035 & 0.602 ± 0.025 & 0.130 ± 0.045 & 0.305 ± 0.121 & 0.169 ± 0.056 \\
AdaB & 0.533 ± 0.043 & 0.593 ± 0.035 & 0.271 ± 0.060 & 0.416 ± 0.096 & 0.307 ± 0.060 \\
DT & 0.502 ± 0.047 & 0.546 ± 0.041 & 0.347 ± 0.071 & 0.363 ± 0.060 & 0.342 ± 0.059 \\
KNN & 0.549 ± 0.035 & 0.597 ± 0.029 & 0.304 ± 0.048 & 0.422 ± 0.067 & 0.340 ± 0.050 \\
GNB & 0.590 ± 0.022 & 0.554 ± 0.022 & 0.689 ± 0.032 & 0.432 ± 0.022 & 0.526 ± 0.022 \\
\bottomrule
\end{tabular}}
    \caption{Appendix table: ADHD classification performance with resting-state EEG data when removing top 5 most important electrodes from feature importance analysis}
    \label{tab:ablation_rsec_eeg_5_most_important_electrodes}
\end{table}

\begin{table}[!h]
    \centering
      \resizebox{\textwidth}{!}{
    \begin{tabular}{cccccc}
\toprule
Model & AUC & Accuracy & Recall & Precision & F1 \\
\midrule
LR & 0.538 ± 0.032 & 0.628 ± 0.016 & 0.013 ± 0.024 & 0.026 ± 0.052 & 0.014 ± 0.025 \\
SVM & 0.471 ± 0.042 & 0.622 ± 0.021 & 0.013 ± 0.023 & 0.026 ± 0.046 & 0.015 ± 0.024 \\
RF & 0.517 ± 0.032 & 0.591 ± 0.025 & 0.163 ± 0.049 & 0.357 ± 0.112 & 0.208 ± 0.058 \\
GB & 0.505 ± 0.043 & 0.589 ± 0.037 & 0.194 ± 0.065 & 0.372 ± 0.130 & 0.234 ± 0.070 \\
ET & 0.527 ± 0.038 & 0.600 ± 0.025 & 0.117 ± 0.044 & 0.288 ± 0.128 & 0.155 ± 0.057 \\
AdaB & 0.513 ± 0.036 & 0.582 ± 0.035 & 0.254 ± 0.052 & 0.390 ± 0.093 & 0.287 ± 0.050 \\
DT & 0.501 ± 0.047 & 0.551 ± 0.042 & 0.339 ± 0.079 & 0.363 ± 0.068 & 0.338 ± 0.068 \\
KNN & 0.547 ± 0.035 & 0.599 ± 0.031 & 0.278 ± 0.048 & 0.431 ± 0.078 & 0.321 ± 0.050 \\
GNB & 0.588 ± 0.020 & 0.547 ± 0.021 & 0.679 ± 0.035 & 0.426 ± 0.021 & 0.518 ± 0.022 \\
\bottomrule
\end{tabular}}
    \caption{Appendix table: ADHD classification performance with resting-state EEG data when removing motor electrodes }
    \label{tab:ablation_rsec_eeg_motor_electrodes}
\end{table}

\begin{table}[!h]
    \centering
      \resizebox{\textwidth}{!}{
    \begin{tabular}{cccccc}
\toprule
Model & AUC & Accuracy & Recall & Precision & F1 \\
\midrule
LR & 0.639 ± 0.062 & 0.834 ± 0.015 & 0.977 ± 0.019 & 0.850 ± 0.009 & 0.908 ± 0.009 \\
SVM & 0.456 ± 0.095 & 0.834 ± 0.015 & 0.981 ± 0.022 & 0.847 ± 0.007 & 0.908 ± 0.010 \\
RF & 0.613 ± 0.047 & 0.829 ± 0.015 & 0.966 ± 0.013 & 0.852 ± 0.008 & 0.905 ± 0.009 \\
GB & 0.564 ± 0.058 & 0.837 ± 0.013 & 0.991 ± 0.016 & 0.843 ± 0.003 & 0.911 ± 0.008 \\
ET & 0.637 ± 0.041 & 0.828 ± 0.011 & 0.975 ± 0.012 & 0.845 ± 0.007 & 0.905 ± 0.007 \\
AdaB & 0.642 ± 0.059 & 0.815 ± 0.024 & 0.952 ± 0.024 & 0.848 ± 0.011 & 0.896 ± 0.014 \\
DT & 0.517 ± 0.053 & 0.731 ± 0.035 & 0.830 ± 0.043 & 0.848 ± 0.017 & 0.835 ± 0.025 \\
KNN & 0.572 ± 0.052 & 0.833 ± 0.013 & 0.986 ± 0.016 & 0.843 ± 0.004 & 0.909 ± 0.009 \\
GNB & 0.675 ± 0.026 & 0.630 ± 0.020 & 0.631 ± 0.021 & 0.904 ± 0.013 & 0.735 ± 0.018 \\
\bottomrule
\end{tabular}}
    \caption{Appendix table: ADHD classification performance with stop signal task EEG data when removing top 5 most important electrodes from feature importance analysis}
    \label{tab:ablation_sst_eeg_5_most_important_electrodes}
\end{table}

\begin{table}[!h]
    \centering
      \resizebox{\textwidth}{!}{
    \begin{tabular}{cccccc}
\toprule
LR & 0.698 ± 0.059 & 0.833 ± 0.019 & 0.967 ± 0.025 & 0.856 ± 0.013 & 0.906 ± 0.012 \\
SVM & 0.474 ± 0.099 & 0.831 ± 0.014 & 0.974 ± 0.021 & 0.849 ± 0.010 & 0.906 ± 0.009 \\
RF & 0.635 ± 0.049 & 0.835 ± 0.014 & 0.970 ± 0.011 & 0.855 ± 0.009 & 0.908 ± 0.008 \\
GB & 0.684 ± 0.054 & 0.836 ± 0.016 & 0.980 ± 0.021 & 0.849 ± 0.009 & 0.909 ± 0.010 \\
ET & 0.645 ± 0.047 & 0.831 ± 0.011 & 0.978 ± 0.009 & 0.846 ± 0.007 & 0.907 ± 0.006 \\
AdaB & 0.768 ± 0.045 & 0.841 ± 0.018 & 0.958 ± 0.018 & 0.869 ± 0.012 & 0.910 ± 0.011 \\
DT & 0.629 ± 0.056 & 0.794 ± 0.039 & 0.876 ± 0.043 & 0.881 ± 0.018 & 0.876 ± 0.027 \\
KNN & 0.567 ± 0.051 & 0.834 ± 0.014 & 0.987 ± 0.016 & 0.844 ± 0.005 & 0.909 ± 0.009 \\
GNB & 0.675 ± 0.025 & 0.629 ± 0.020 & 0.632 ± 0.023 & 0.903 ± 0.013 & 0.735 ± 0.020 \\
\bottomrule
\end{tabular}}
    \caption{Appendix table: ADHD classification performance with stop signal task EEG data when removing motor electrodes}
    \label{tab:ablation_sst_eeg_motor_electrodes}
\end{table}

\begin{table}[ht]
\centering
\small % or \footnotesize, \scriptsize
\clearpage
\section{Appendix materials for chapter 4}

  \resizebox{\textwidth}{!}{

\begin{tabular}{cccccc}
\toprule
Model & AUC & Accuracy & Recall & Precision & F1 \\
\midrule
LR   & 0.652 $\pm$ 0.053 & 0.613 $\pm$ 0.042 & 0.580 $\pm$ 0.063 & 0.644 $\pm$ 0.075 & 0.586 $\pm$ 0.058 \\
SVM  & 0.490 $\pm$ 0.099 & 0.561 $\pm$ 0.055 & 0.484 $\pm$ 0.096 & 0.590 $\pm$ 0.109 & 0.497 $\pm$ 0.085 \\
RF   & 0.705 $\pm$ 0.052 & 0.630 $\pm$ 0.054 & 0.612 $\pm$ 0.060 & 0.665 $\pm$ 0.080 & 0.613 $\pm$ 0.056 \\
GB   & 0.651 $\pm$ 0.068 & 0.601 $\pm$ 0.060 & 0.603 $\pm$ 0.080 & 0.627 $\pm$ 0.081 & 0.592 $\pm$ 0.067 \\
ET   & 0.719 $\pm$ 0.050 & 0.648 $\pm$ 0.053 & 0.602 $\pm$ 0.064 & 0.698 $\pm$ 0.084 & 0.619 $\pm$ 0.059 \\
AdaB & 0.644 $\pm$ 0.055 & 0.568 $\pm$ 0.051 & 0.601 $\pm$ 0.079 & 0.588 $\pm$ 0.072 & 0.568 $\pm$ 0.061 \\
DT   & 0.606 $\pm$ 0.067 & 0.592 $\pm$ 0.064 & 0.588 $\pm$ 0.088 & 0.620 $\pm$ 0.090 & 0.579 $\pm$ 0.073 \\
MLP  & 0.670 $\pm$ 0.061 & 0.605 $\pm$ 0.053 & 0.597 $\pm$ 0.075 & 0.633 $\pm$ 0.072 & 0.592 $\pm$ 0.062 \\
KNN  & 0.677 $\pm$ 0.049 & 0.604 $\pm$ 0.045 & 0.664 $\pm$ 0.054 & 0.618 $\pm$ 0.055 & 0.624 $\pm$ 0.044 \\
GNB  & 0.625 $\pm$ 0.034 & 0.572 $\pm$ 0.035 & 0.386 $\pm$ 0.044 & 0.627 $\pm$ 0.103 & 0.449 $\pm$ 0.048 \\
LDA  & 0.671 $\pm$ 0.048 & 0.634 $\pm$ 0.037 & 0.568 $\pm$ 0.049 & 0.679 $\pm$ 0.065 & 0.596 $\pm$ 0.048 \\
QDA  & 0.682 $\pm$ 0.055 & 0.612 $\pm$ 0.046 & 0.430 $\pm$ 0.062 & 0.698 $\pm$ 0.115 & 0.500 $\pm$ 0.065 \\
\bottomrule
\end{tabular}}
\caption[Appendix table: Effort sensitivity ML classification performance with white matter integrity]{Decoding effort sensitivity is classifier-independent: further evidence from multiple machine learning metrics. Performance of 12 machine learning classifiers in predicting effort sensitivity from white matter microstructural features, including fractional anisotropy and mean diffusivity. Reported values reflect the mean ± standard deviation across 1,000 iterations of stratified 5-fold cross-validation with randomized data partitioning. Classifier abbreviations: LR, logistic regression; SVM, support vector machine; RF, random forest; GB, gradient boosting; ET, extra trees; AdaB, AdaBoost; DT, decision tree; MLP, multi-layer perceptron; KNN, k-nearest neighbors; GNB, Gaussian naïve Bayes; LDA, linear discriminant analysis; QDA, quadratic discriminant analysis. Performance metrics include area under the receiver operating characteristic curve (AUC), accuracy, recall (sensitivity), precision (positive predictive value), and F1 score (harmonic mean of precision and recall). Among classifiers, the ET model achieved the highest AUC (0.719 ± 0.050), along with a balance between precision (0.698 ± 0.084) and F1 score (0.619 ± 0.059), indicating robust discriminative performance. RF and QDA also performed well (AUCs of 0.705 ± 0.052 and 0.682 ± 0.055, respectively), though QDA exhibited reduced recall (0.430 ± 0.062), suggesting a trade-off favoring specificity. Linear models such as LR and LDA demonstrated moderate AUCs (0.652 ± 0.053 and 0.671 ± 0.048, respectively), with LDA yielding relatively high precision (0.679 ± 0.065). In contrast, SVM showed the lowest AUC (0.490 ± 0.099), indicating performance at or below chance levels. GNB and QDA showed the largest discrepancies between precision and recall, highlighting imbalances in classification. Hence, despite variations in individual metrics, most classifiers achieved above-chance performance, reinforcing the robustness of the decoding. These results collectively indicate that effort sensitivity can be reliably predicted from white matter features across a wide range of machine learning approaches.}
  \label{table:chap4_effort_prediction}
\end{table}

\begin{table}[ht]
\centering
\small % or \footnotesize, \scriptsize

  \resizebox{\textwidth}{!}{

\begin{tabular}{cccccc}
\toprule
Model & AUC & Accuracy & Recall & Precision & F1 \\
\midrule
LR   & 0.837 $\pm$ 0.034 & 0.718 $\pm$ 0.038 & 0.749 $\pm$ 0.052 & 0.736 $\pm$ 0.047 & 0.722 $\pm$ 0.043 \\
SVM  & 0.695 $\pm$ 0.117 & 0.704 $\pm$ 0.041 & 0.787 $\pm$ 0.064 & 0.707 $\pm$ 0.049 & 0.721 $\pm$ 0.046 \\
RF   & 0.763 $\pm$ 0.042 & 0.678 $\pm$ 0.039 & 0.689 $\pm$ 0.059 & 0.702 $\pm$ 0.053 & 0.673 $\pm$ 0.047 \\
GB   & 0.711 $\pm$ 0.058 & 0.675 $\pm$ 0.054 & 0.702 $\pm$ 0.083 & 0.691 $\pm$ 0.064 & 0.674 $\pm$ 0.067 \\
ET   & 0.807 $\pm$ 0.037 & 0.693 $\pm$ 0.040 & 0.743 $\pm$ 0.057 & 0.707 $\pm$ 0.051 & 0.702 $\pm$ 0.045 \\
AdaB & 0.774 $\pm$ 0.048 & 0.710 $\pm$ 0.051 & 0.773 $\pm$ 0.073 & 0.712 $\pm$ 0.060 & 0.720 $\pm$ 0.058 \\
DT   & 0.656 $\pm$ 0.059 & 0.632 $\pm$ 0.055 & 0.640 $\pm$ 0.092 & 0.650 $\pm$ 0.071 & 0.620 $\pm$ 0.071 \\
MLP  & 0.750 $\pm$ 0.052 & 0.674 $\pm$ 0.052 & 0.685 $\pm$ 0.070 & 0.694 $\pm$ 0.068 & 0.668 $\pm$ 0.056 \\
KNN  & 0.771 $\pm$ 0.045 & 0.685 $\pm$ 0.045 & 0.731 $\pm$ 0.058 & 0.699 $\pm$ 0.053 & 0.694 $\pm$ 0.048 \\
GNB  & 0.802 $\pm$ 0.030 & 0.697 $\pm$ 0.025 & 0.815 $\pm$ 0.042 & 0.682 $\pm$ 0.026 & 0.729 $\pm$ 0.029 \\
LDA  & 0.834 $\pm$ 0.035 & 0.721 $\pm$ 0.036 & 0.771 $\pm$ 0.047 & 0.731 $\pm$ 0.044 & 0.732 $\pm$ 0.039 \\
QDA  & 0.792 $\pm$ 0.049 & 0.692 $\pm$ 0.044 & 0.807 $\pm$ 0.065 & 0.683 $\pm$ 0.045 & 0.722 $\pm$ 0.047 \\
\bottomrule
\end{tabular}}
\caption[Appendix table: Reward sensitivity ML classification performance with white matter integrity]{Decoding reward sensitivity is classifier-independent: further evidence from multiple machine learning metrics. Performance of 12 machine learning classifiers in predicting reward sensitivity from white matter microstructural features, including fractional anisotropy and mean diffusivity. Reported values reflect the mean ± standard deviation across 1,000 iterations of stratified 5-fold cross-validation with randomized data partitioning. Classifier abbreviations: LR, logistic regression; SVM, support vector machine; RF, random forest; GB, gradient boosting; ET, extra trees; AdaB, AdaBoost; DT, decision tree; MLP, multi-layer perceptron; KNN, k-nearest neighbors; GNB, Gaussian naïve Bayes; LDA, linear discriminant analysis; QDA, quadratic discriminant analysis. Performance metrics include area under the receiver operating characteristic curve (AUC), accuracy, recall (sensitivity), precision (positive predictive value), and F1 score (harmonic mean of precision and recall). Across models, LR and LDA yielded the highest AUC values (0.837 ± 0.034, highlighted in the table and 0.834 ± 0.035, respectively), indicating strong overall discriminative ability. GNB and QDA also performed competitively (AUCs of 0.802 ± 0.030 and 0.792 ± 0.049, respectively), with GNB achieving the highest recall (0.815 ± 0.042), suggesting strong sensitivity to effort-related signal in the input features. In contrast, tree-based methods such as DT and GB exhibited comparatively lower performance across most metrics, highlighting limitations in their generalizability to the underlying data structure. While MLP and KNN classifiers offered moderate performance, ensemble methods like AdaB and ET demonstrated a balance between precision and recall, though their AUCs fell below those of the linear models. These results suggest that linear decision boundaries may be particularly well suited for classifying reward sensitivity from microstructural white matter features, potentially reflecting a low-dimensional organization of the relevant neural representations. Further, despite differences in classifier type and complexity, most models achieved consistent above-chance accuracy, demonstrating that reward sensitivity is robustly decodable. These results again indicate that reward sensitivity can be reliably predicted from white matter features across a wide range of machine learning approaches.}
  \label{table:chap4_reward_prediction}
\end{table}

\begin{figure}[h]
    \centering
    \includegraphics[width=1\linewidth]{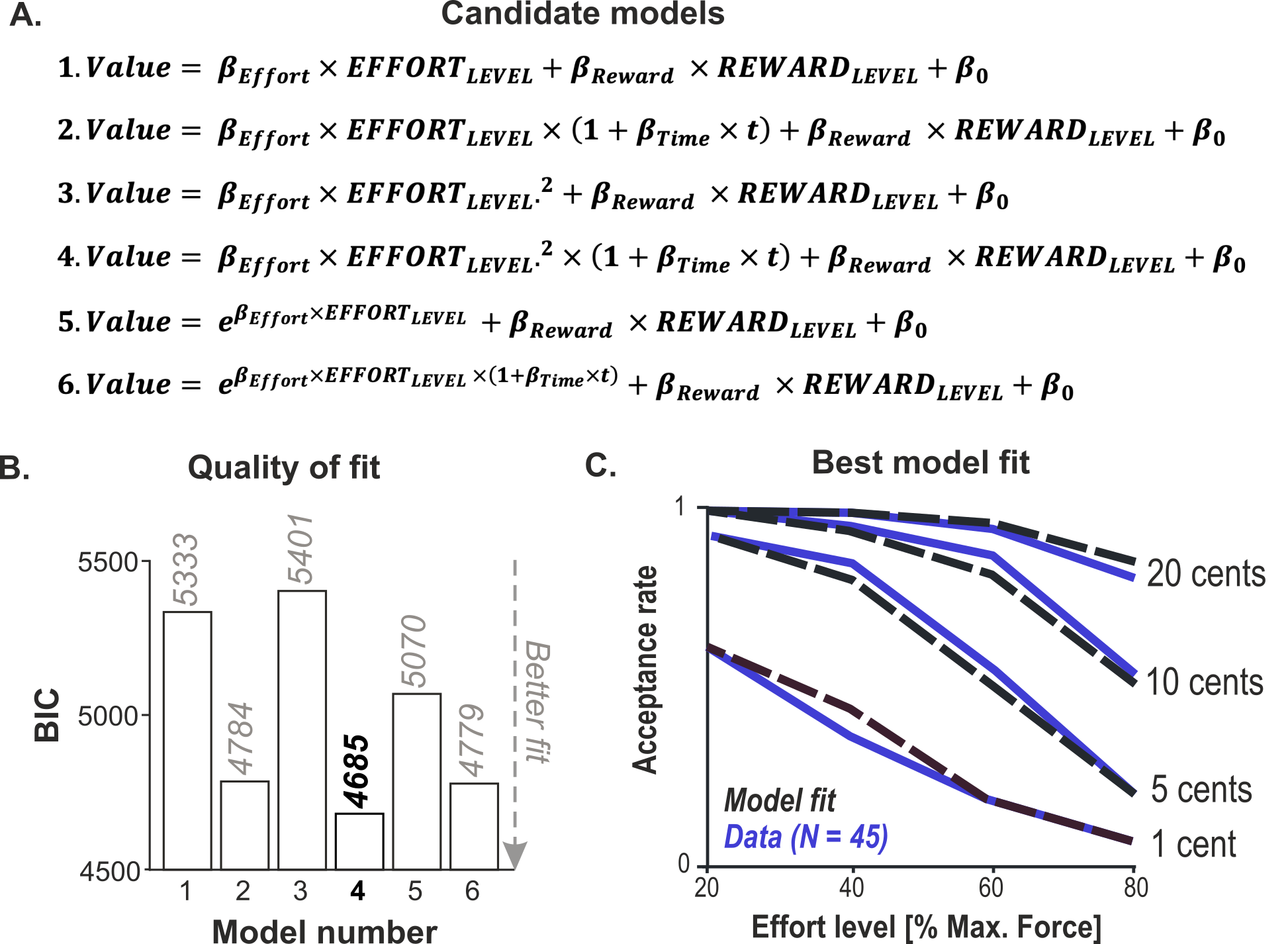}
    \caption[Computational modelling of effort-based decision-making.]{\textbf{A. Candidate models.} Candidate models of subjective value computation were tested to quantify individual differences in effort and reward sensitivity. Models differed in cost function (linear, quadratic, or exponential) and in whether they included a time-dependent modulation term ($\beta_{\text{Time}}$), following procedures established in prior work (e.g., \cite{le2018anatomy}). Value was modelled as a function of effort level ($EFFORT_\text{LEVEL}$), reward level ($REWARD_\text{LEVEL}$), and trial number ($t$), with free parameters capturing individual sensitivities ($\beta_{\text{Effort}}$, $\beta_{\text{Reward}}$) and baseline bias ($\beta_{0}$). \textbf{B. Model fit comparison based on Bayesian information criterion (BIC).} Model~4, which combined a quadratic cost function with a time-modulated cost term, showed the best fit (lowest BIC), consistent with prior findings (\cite{le2018anatomy, pessiglione2018not, prevost2010separate}.  \textbf{C. Best-fitting model vs. observed behavior.} Model-derived acceptance rate (black dashed lines) closely captured observed behavioral data (blue lines; $N = 45$), across varying effort levels and reward magnitudes.
}
    \label{fig:chapter4-suppfigure1}
\end{figure}

\begin{figure}[h]
    \centering
    \includegraphics[width=1\linewidth]{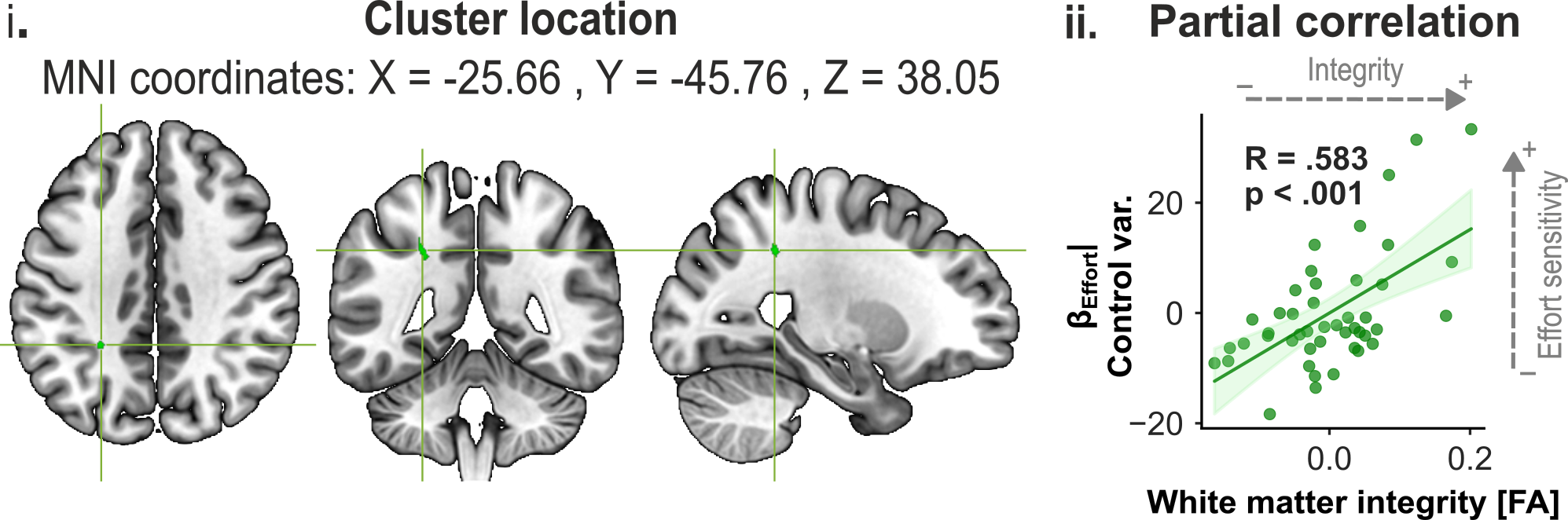}
    \caption[Superficial cluster showing a positive association between fractional anisotropy (FA) and effort sensitivity]{\textbf{Superficial cluster showing a positive association between fractional anisotropy (FA) and $\beta_{\text{Effort}}$.} i. A significant cluster was identified outside canonical white matter tracts, located in the left parietal cortex (MNI coordinates: X = $-25.66$, Y = $-45.76$, Z = $38.05$; volume: 62 mm$^{3}$). ii. Partial correlation analysis confirmed this association, revealing a robust positive correlation between FA values extracted from the cluster and $\beta_{\text{Effort}}$ ($R = .583$, $p < .001$). These findings suggest that, in addition to the major tracts identified using a probabilistic white matter atlas, more superficial white matter regions may also contribute to individual variability in effort sensitivity. This further highlights the added value of whole-brain, data-driven approaches for uncovering relevant structural substrates beyond predefined tracts-of-interest.
}
    \label{fig:chapter4-suppfigure2}
\end{figure}

\begin{figure}
    \centering
    \includegraphics[width=1\linewidth]{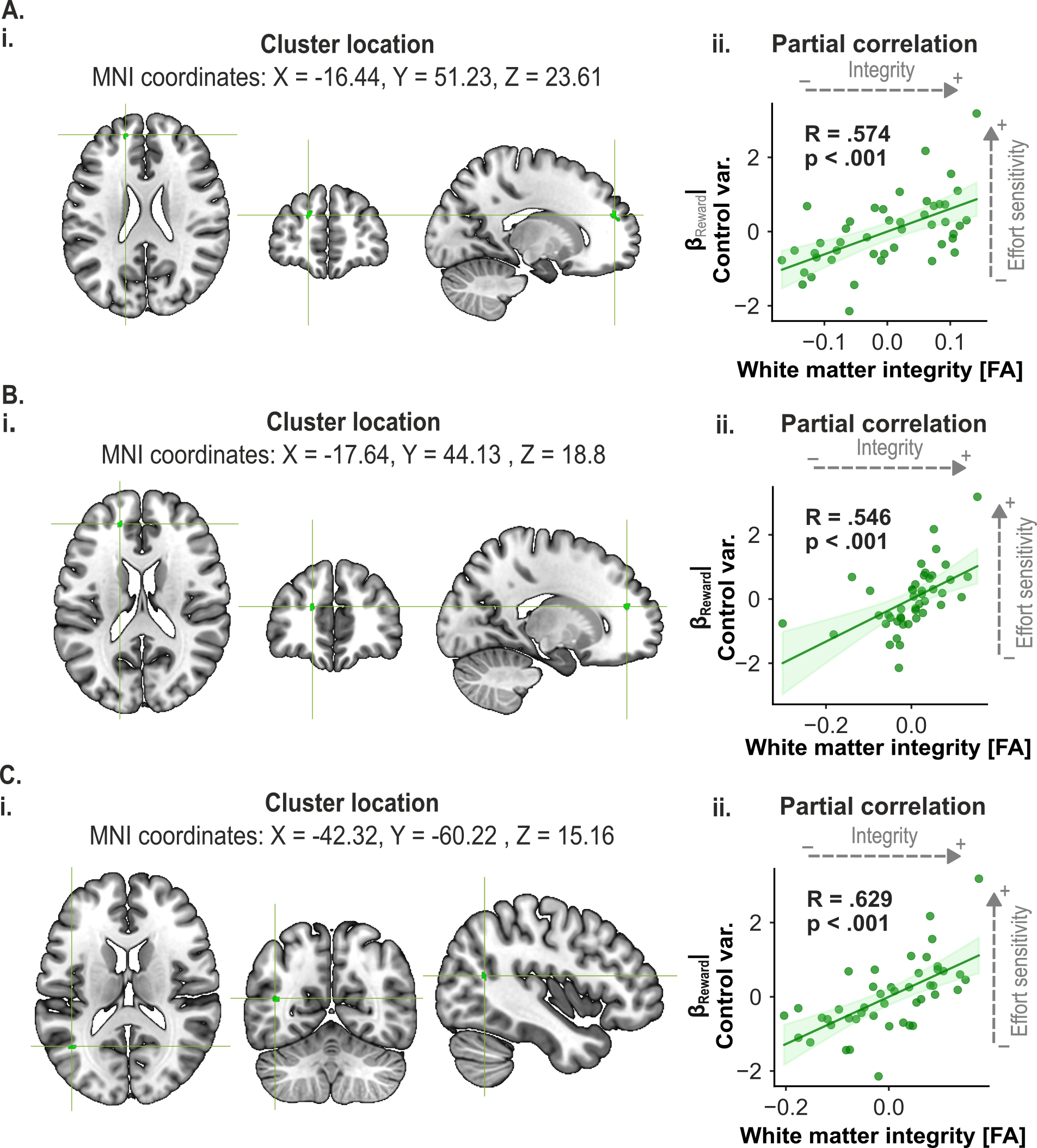}
    \caption[]{\textbf{Three clusters with positive $\beta_{\text{Reward}}$ associations in superficial white matter of the left frontal and parietal cortex outside major tracts.} Three additional clusters exhibiting significant positive associations between white matter integrity and reward sensitivity ($\beta_{\text{Reward}}$) were identified in superficial regions of the left frontal and parietal cortex, outside canonical major white matter tracts. A. i. Cluster (green) located at MNI coordinates: X = $-16.44$, Y = $51.23$, Z = $23.61$; volume: 88 mm$^{3}$. ii. Partial correlation between fractional anisotropy (FA) in this cluster and $\beta_{\text{Reward}}$: $R = .574$, $p < .001$. B. i. Cluster (green) located at MNI coordinates: X = $-17.64$, Y = $44.13$, Z = $18.88$; volume: 64 mm$^{3}$. ii. Partial correlation with $\beta_{\text{Reward}}$: $R = .546$, $p < .001$. C. i. Cluster (green) located at MNI coordinates: X = $-42.32$, Y = $-60.22$, Z = $15.16$; volume: 64 mm$^{3}$). ii. Partial correlation with $\beta_{\text{Reward}}$: $R = .629$, $p < .001$.
}
    \label{fig:chapter4-suppfigure3} 
\end{figure}

\begin{figure}
    \centering
    \includegraphics[width=.7\linewidth]{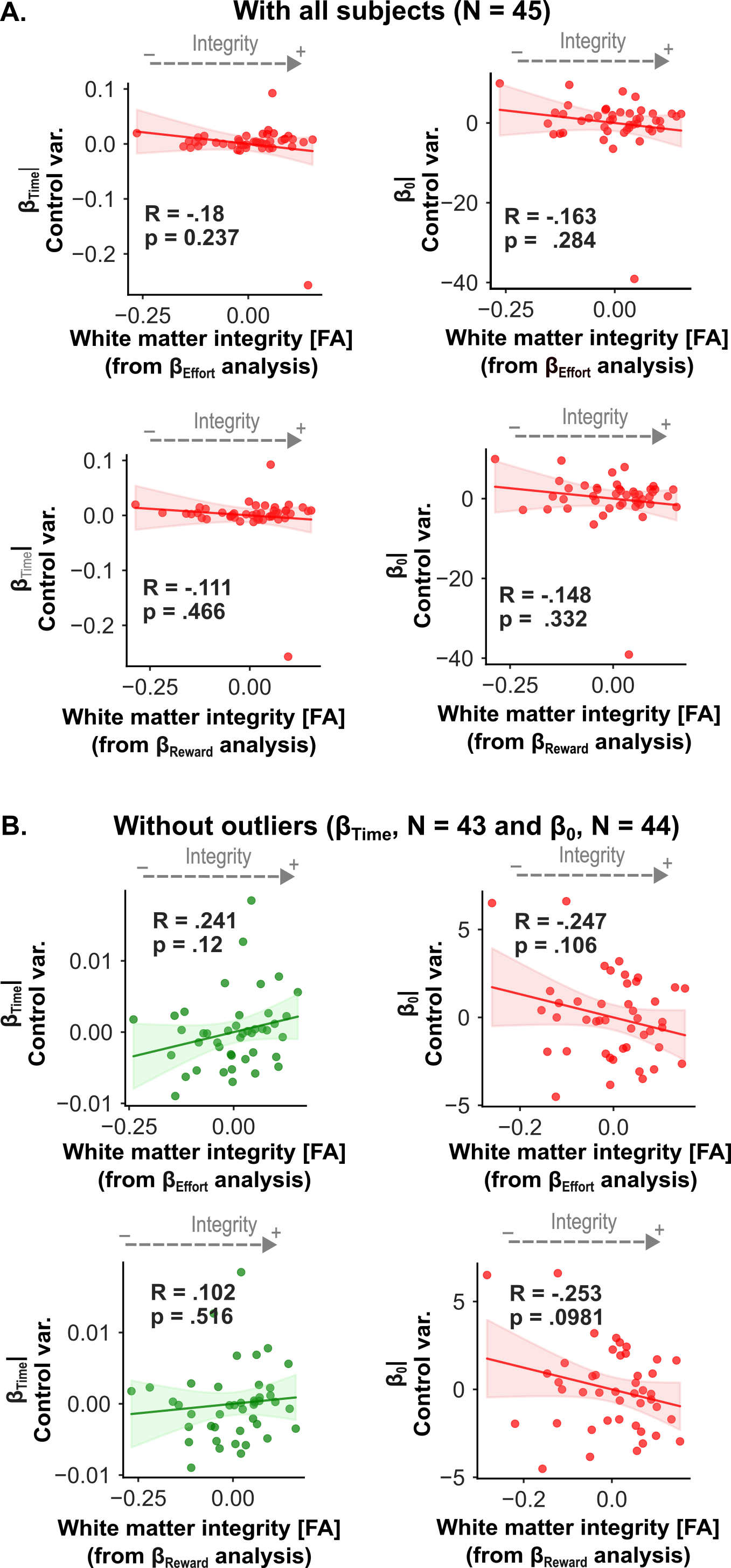}
    \caption[The SMA cluster associated with both effort and reward sensitivities is specific and unrelated to the two other model parameters]{\textbf{The SMA cluster associated with both $\beta_{\text{Effort}}$ and $\beta_{\text{Reward}}$ is specific and unrelated to the two other model parameters.} (Figure caption continued on next page.) 
}
    \label{fig:chapter4-suppfigure4}
\end{figure}

\begin{figure}
    \ContinuedFloat
    \caption[]{To confirm that the dual association with $\beta_{\text{Effort}}$ and $\beta_{\text{Reward}}$ was not trivially driven by their derivation from the same acceptance rates, we tested whether FA values in these SMA clusters correlated with the two other model-derived parameters, $\beta_{\text{Time}}$ and $\beta_{0}$. A. Correlations including all participants ($n = 45$). No significant correlations were observed for either the effort-related cluster ($\beta_{\text{Time}}$: $R = -.180$, $p = .237$; $\beta_{0}$: $R = -.163$, $p = .284$) or the reward-related cluster ($\beta_{\text{Time}}$: $R = -.111$, $p = .466$; $\beta_{0}$: $R = -.148$, $p = .332$), underscoring the specificity of this SMA white matter locus for effort and reward sensitivities. B. Correlations excluding outliers in $\beta_{\text{Time}}$ and $\beta_{0}$. Results remained consistent, with no significant correlations for either cluster (effort-related: $\beta_{\text{Time}}$: $R = .241$, $p = .120$; $\beta_{0}$: $R = -.247$, $p = .106$; reward-related: $\beta_{\text{Time}}$: $R = .102$, $p = .516$; $\beta_{0}$: $R = -.253$, $p = .0981$). Across analyses, FA values in these clusters did not relate to $\beta_{\text{Time}}$ or $\beta_{0}$, confirming that this SMA cluster specifically covaries with individual differences in effort- and reward-related parameters.}
\end{figure}

\clearpage
\section{Appendix materials for chapter 5}

\begin{table}[h]
\centering
\small % or \footnotesize, \scriptsize

  \resizebox{\textwidth}{!}{
\begin{tabular}{llllll}
\toprule
Model & AUC & Accuracy & Recall & Precision & F1 \\
\midrule
LR & 0.815 ± 0.037 & 0.722 ± 0.041 & 0.735 ± 0.053 & 0.746 ± 0.053 & 0.721 ± 0.045 \\
SVM & 0.786 ± 0.052 & 0.704 ± 0.047 & 0.641 ± 0.066 & 0.769 ± 0.072 & 0.674 ± 0.058 \\
RF & 0.739 ± 0.044 & 0.685 ± 0.039 & 0.681 ± 0.049 & 0.717 ± 0.054 & 0.678 ± 0.043 \\
GB & 0.656 ± 0.058 & 0.616 ± 0.051 & 0.614 ± 0.076 & 0.639 ± 0.070 & 0.605 ± 0.062 \\
ET & 0.755 ± 0.043 & 0.691 ± 0.042 & 0.664 ± 0.053 & 0.736 ± 0.061 & 0.676 ± 0.046 \\
AdaB & 0.714 ± 0.047 & 0.647 ± 0.043 & 0.595 ± 0.063 & 0.697 ± 0.072 & 0.614 ± 0.055 \\
DT & 0.630 ± 0.057 & 0.611 ± 0.055 & 0.585 ± 0.082 & 0.642 ± 0.079 & 0.590 ± 0.068 \\
MLP & 0.730 ± 0.050 & 0.658 ± 0.047 & 0.644 ± 0.061 & 0.693 ± 0.068 & 0.644 ± 0.053 \\
KNN & 0.701 ± 0.050 & 0.653 ± 0.047 & 0.528 ± 0.061 & 0.738 ± 0.088 & 0.590 ± 0.061 \\
GNB & 0.785 ± 0.036 & 0.701 ± 0.032 & 0.725 ± 0.035 & 0.718 ± 0.046 & 0.704 ± 0.032 \\
LDA & 0.811 ± 0.037 & 0.716 ± 0.039 & 0.735 ± 0.050 & 0.736 ± 0.051 & 0.718 ± 0.041 \\
QDA & 0.734 ± 0.053 & 0.657 ± 0.048 & 0.700 ± 0.062 & 0.670 ± 0.058 & 0.666 ± 0.051 \\
\bottomrule
\end{tabular}
}

\caption[Appendix table: Apathy scores ML classification performance with grey matter volumes]{Appendix table: Apathy scores ML classification performance with grey matter volumes}
  \label{table:chap5_lars_prediction}
\end{table}

\begin{table}[h]
\centering
\small % or \footnotesize, \scriptsize

  \resizebox{\textwidth}{!}{
\begin{tabular}{llllll}
\toprule
Model & AUC & Accuracy & Recall & Precision & F1 \\
\midrule
LR & 0.562 ± 0.055 & 0.631 ± 0.036 & 0.100 ± 0.074 & 0.158 ± 0.132 & 0.114 ± 0.085 \\
SVM & 0.470 ± 0.077 & 0.617 ± 0.038 & 0.053 ± 0.055 & 0.084 ± 0.096 & 0.060 ± 0.062 \\
RF & 0.568 ± 0.058 & 0.638 ± 0.045 & 0.317 ± 0.079 & 0.426 ± 0.129 & 0.340 ± 0.083 \\
GB & 0.541 ± 0.068 & 0.608 ± 0.051 & 0.259 ± 0.098 & 0.333 ± 0.141 & 0.271 ± 0.100 \\
ET & 0.553 ± 0.060 & 0.616 ± 0.041 & 0.130 ± 0.073 & 0.223 ± 0.143 & 0.153 ± 0.086 \\
AdaB & 0.564 ± 0.057 & 0.588 ± 0.049 & 0.240 ± 0.099 & 0.271 ± 0.125 & 0.236 ± 0.095 \\
DT & 0.545 ± 0.071 & 0.597 ± 0.060 & 0.368 ± 0.115 & 0.383 ± 0.129 & 0.351 ± 0.103 \\
MLP & 0.556 ± 0.065 & 0.575 ± 0.054 & 0.331 ± 0.086 & 0.356 ± 0.112 & 0.322 ± 0.082 \\
KNN & 0.515 ± 0.076 & 0.596 ± 0.048 & 0.143 ± 0.080 & 0.230 ± 0.145 & 0.165 ± 0.092 \\
GNB & 0.590 ± 0.048 & 0.578 ± 0.041 & 0.300 ± 0.070 & 0.356 ± 0.109 & 0.304 ± 0.072 \\
LDA & 0.650 ± 0.043 & 0.611 ± 0.037 & 0.309 ± 0.063 & 0.396 ± 0.111 & 0.322 ± 0.065 \\
QDA & 0.540 ± 0.056 & 0.570 ± 0.045 & 0.218 ± 0.075 & 0.282 ± 0.121 & 0.229 ± 0.079 \\
\bottomrule
\end{tabular}
}

\caption[Appendix table: Effort sensitivity ML classification performance with grey matter volumes]{Appendix table: Effort sensitivity ML classification performance with grey matter volumes}
  \label{table:chap5_effort_prediction}
\end{table}

\begin{table}[h]
\centering
\small % or \footnotesize, \scriptsize

  \resizebox{\textwidth}{!}{
\begin{tabular}{llllll}
\toprule
Model & AUC & Accuracy & Recall & Precision & F1 \\
\midrule
LR & 0.649 ± 0.050 & 0.605 ± 0.038 & 0.405 ± 0.080 & 0.563 ± 0.126 & 0.445 ± 0.085 \\
SVM & 0.487 ± 0.101 & 0.591 ± 0.046 & 0.298 ± 0.078 & 0.556 ± 0.153 & 0.359 ± 0.084 \\
RF & 0.646 ± 0.052 & 0.591 ± 0.046 & 0.481 ± 0.071 & 0.556 ± 0.093 & 0.490 ± 0.066 \\
GB & 0.670 ± 0.060 & 0.627 ± 0.050 & 0.509 ± 0.081 & 0.624 ± 0.105 & 0.526 ± 0.069 \\
ET & 0.630 ± 0.048 & 0.580 ± 0.044 & 0.400 ± 0.066 & 0.544 ± 0.106 & 0.433 ± 0.064 \\
AdaB & 0.685 ± 0.051 & 0.673 ± 0.044 & 0.471 ± 0.081 & 0.727 ± 0.124 & 0.529 ± 0.072 \\
DT & 0.643 ± 0.061 & 0.638 ± 0.056 & 0.550 ± 0.082 & 0.625 ± 0.099 & 0.555 ± 0.071 \\
MLP & 0.583 ± 0.065 & 0.548 ± 0.050 & 0.450 ± 0.064 & 0.504 ± 0.090 & 0.453 ± 0.061 \\
KNN & 0.601 ± 0.052 & 0.598 ± 0.046 & 0.492 ± 0.073 & 0.565 ± 0.089 & 0.502 ± 0.068 \\
GNB & 0.638 ± 0.039 & 0.588 ± 0.028 & 0.427 ± 0.040 & 0.567 ± 0.080 & 0.463 ± 0.041 \\
LDA & 0.674 ± 0.033 & 0.605 ± 0.028 & 0.502 ± 0.040 & 0.587 ± 0.071 & 0.516 ± 0.041 \\
QDA & 0.621 ± 0.043 & 0.584 ± 0.032 & 0.418 ± 0.041 & 0.561 ± 0.084 & 0.455 ± 0.043 \\
\bottomrule
\end{tabular}
}

\caption[Appendix table: Reward sensitivity ML classification performance with grey matter volumes]{Appendix table: Reward sensitivity ML classification performance with grey matter volumes}
  \label{table:chap5_reward_prediction}
\end{table}

\end{document}